\documentclass[12pt]{iopart}
\usepackage{amsmath}
\usepackage{amssymb}
\usepackage{mathtools}
\usepackage{amsthm}
\usepackage{algorithm}
\usepackage{algorithmic}
\usepackage{url}
\usepackage{hhline}
\usepackage{subcaption}

\newtheorem{theorem}{Theorem}[section]
\newtheorem{lemma}[theorem]{Lemma}

\usepackage{xcolor}

\def\bi#1{\mbox{\boldmath{$#1$}}}
\newcommand{\norm}[1]{\left\lVert#1\right\rVert}

\begin{document}

\title[Sequential Kalman Tuning]{Sequential Kalman Tuning of the $t$-preconditioned Crank-Nicolson algorithm: efficient, adaptive and gradient-free inference for Bayesian inverse problems}

\author{Richard D.P. Grumitt$^1$, Minas Karamanis$^{2, 3}$, Uro\v{s} Seljak$^{2,3}$}

\address{$^1$Department of Astronomy, Tsinghua University, Beijing 100084, China}
\address{$^2$Berkeley Center for Cosmological Physics and Department of Physics, University of California, Berkeley, CA 94720}
\address{$^3$Physics Department, Lawrence Berkeley National Laboratory, Cyclotron Rd, Berkeley, CA
94720}
\ead{rgrumitt@mail.tsinghua.edu.cn}
\vspace{10pt}
\begin{indented}
\item[]June 2024
\end{indented}

\begin{abstract}
Ensemble Kalman Inversion (EKI) has been proposed as an efficient method for the approximate solution of Bayesian inverse problems with expensive forward models. However, when applied to the Bayesian inverse problem EKI is only exact in the regime of Gaussian target measures and linear forward models. In this work we propose embedding EKI and Flow Annealed Kalman Inversion (FAKI), its normalizing flow (NF) preconditioned variant, within a Bayesian annealing scheme as part of an adaptive implementation of the $t$-preconditioned Crank-Nicolson (tpCN) sampler. The tpCN sampler differs from standard pCN in that its proposal is reversible with respect to the multivariate $t$-distribution. The more flexible tail behaviour allows for better adaptation to sampling from non-Gaussian targets. Within our Sequential Kalman Tuning (SKT) adaptation scheme, EKI is used to initialize and precondition the tpCN sampler for each annealed target. The subsequent tpCN iterations ensure particles are correctly distributed according to each annealed target, avoiding the accumulation of errors that would otherwise impact EKI. We demonstrate the performance of SKT for tpCN on three challenging numerical benchmarks, showing significant improvements in the rate of convergence compared to adaptation within standard SMC with importance weighted resampling at each temperature level, and compared to similar adaptive implementations of standard pCN. The SKT scheme applied to tpCN offers an efficient, practical solution for solving the Bayesian inverse problem when gradients of the forward model are not available. Code implementing the SKT schemes for tpCN is available at \url{https://github.com/RichardGrumitt/KalmanMC}.
\end{abstract}

%
\vspace{2pc}
\\
%
\submitto{\IP}
%
%
%

\section{Introduction}

Many scientific inference tasks can be viewed within the Bayesian inverse problem framework. In the Gaussian inverse problem setting, we can write the forward problem as
\begin{equation}
    \bi{y} = \mathcal{F}(\bi{x}) + \bi{\eta},
    \label{eq:forward problem}
\end{equation}
where $\bi{y}\in\mathbb{R}^{n_y}$ is the data vector, $\mathcal{F}$ is a forward model that maps the parameters $\bi{x}\in\mathbb{R}^{d}$ to our observables, and $\bi{\eta}\sim\mathcal{N}(0, \Gamma)$ is additive Gaussian noise with fixed noise covariance $\Gamma\in\mathbb{R}^{n_y\times n_y}$. For the Bayesian inverse problem we assign some prior over the parameters $\bi{x}\sim\pi_0(\bi{x})$, with the goal then being to recover the posterior distribution
\begin{equation}
    \pi(\bi{x}|\bi{y})= \frac{\pi(\bi{y}|\bi{x})\pi_0(\bi{x})}{\mathcal{Z}},
\end{equation}
where $\mathcal{Z}$ is some generally unknown normalizing constant and $\pi(\bi{y}|\bi{x})=\mathcal{N}(\bi{y}|\mathcal{F}(\bi{x}), \Gamma)$ \cite{kaipio2005inverse, mackay2003information}.

The particular regime we are concerned with for this work is where we do not have access to gradients of a typically expensive forward model. This is a common setting for scientific inverse problems, where evaluating the forward model often involves running some black-box solver for which gradients cannot be easily and/or accurately obtained e.g., cosmological Boltzmann solvers \cite{lewis2002cosmological, blas2011cosmic}, computational fluid dynamics simulators \cite{jasak2007openfoam}, etc. Non-differentiable forward models can also be a result of inherently discontinuous physics, e.g., in cloud modelling \cite{tan2018eddy}. Given the forward problem definition in Equation \ref{eq:forward problem}, we are restricted to Bayesian inference tasks with Gaussian likelihoods. However, this still encompasses a large number of scientific inverse problems, and is the regime for which the Ensemble Kalman methods we exploit in this work have been developed \cite{iglesias2013ensemble, iglesias2016regularizing, iglesias2018bayesian, chada2018parameterizations, kovachki2019ensemble, chada2020tikhonov,iglesias2021adaptive, zhiyan2021kalman, huang2022iterated, huang2022efficient, chada2022convergence, grumitt2024flow}. In addition to this, we are concerned with developing methods that can reliably obtain low bias estimates of posterior moments. This is critical for many scientific inference tasks where we require accurate uncertainty quantification on any model parameter constraints.  

A typical approach to solving Bayesian inverse problems involves exploiting some form of sampling algorithm. This covers a wide range of methods, e.g., Markov Chain Monte Carlo (MCMC) algorithms \cite{geyer1992MCMC, gelman1997weak, neal2011mcmc, cotter2013pCN}, simulating interacting particle systems \cite{vrugt2008diffevo, foreman2013emcee, leimkuhler2018ensemble, garbuno2020interacting, karamanis2021ensemble, grumitt2022deterministic} etc. In MCMC algorithms, we seek to construct some transition, $T(\bi{x}^\prime,\bi{x})$ that preserves the target, $\pi(\bi{x}|\bi{y})$ as an invariant distribution, i.e., 
\begin{equation}
        \pi(\bi{x}|\bi{y})=\int T(\bi{x}^\prime,\bi{x})\pi(\bi{x}^\prime|\bi{y})\,\mathrm{d}\bi{x}^\prime.
\end{equation}
Appropriately constructed, such methods enjoy target invariance and ergodicity properties. However, especially in the gradient-free regime we consider in this work, this often comes at the cost of requiring $\gtrsim\mathcal{O}(10^4)$ serial model evaluations \cite{huang2022efficient}, quickly rendering such algorithms intractable for expensive and high dimensional models.

An alternative class of method involves constructing some coupling scheme, where we have a transition $C(\bi{x}^\prime, \bi{x})$, that moves us from the prior, $\pi_0(\bi{x})$ to the target $\pi(\bi{x}|\bi{y})$, i.e.,
\begin{equation}
    \pi(\bi{x}|\bi{y})=\int C(\bi{x}^\prime, \bi{x})\pi_0(\bi{x})\,\mathrm{d}\bi{x}.
\end{equation}
Examples of coupling methods include the Ensemble Kalman Filter (EKF) \cite{geir2006ekf, geir2009ekf, schillings2017analysis}, Ensemble Kalman Inversion (EKI) \cite{iglesias2013ensemble, chada2018parameterizations, kovachki2019ensemble, chada2020tikhonov, iglesias2021adaptive, chada2022convergence}, and Sequential Monte Carlo (SMC) \cite{moral2006SMC, wu2022ensemble, dau2022waste, karamanis2022accelerating}. 

In addition to developing methods for solving Bayesian inverse problems, it is crucial to consider adaptation strategies that allow for efficient, tuning-free implementations of these methods that can used by practitioners. Extensive work has been done on the development of tuning-free implementations of gradient-based algorithms such as Hamiltonian Monte Carlo (HMC) \cite{hoffman2014no, hoffman2021chees, sountsov2021focusing, hoffman2022meads, riou2023adaptive}, including in the context of SMC \cite{buchholz2021adaptive}. For gradient-free algorithms a notable adaptive method is Preconditioned Monte Carlo (PMC), implemented in the \textsc{PocoMC} library \cite{karamanis2022accelerating, karamanis2022_pocoMC}, which uses normalizing flow (NF) \cite{dinh2016density, papamakarios2017masked, kingma2018glow, dai2021sliced} preconditioning within sequential Monte Carlo (SMC) to accelerate gradient-free sampling. 

\subsection{Our Contributions}

\begin{itemize}
    \item We develop an adaptive, tuning-free implementation of the $t$-preconditioned Crank-Nicolson (tpCN)  algorithm, designed for performing efficient gradient-free inference in Bayesian inverse problems. The tpCN algorithm preserves the exact target distribution as its invariant measure, allowing for accurate posterior moment estimation when faced with non-Gaussian targets and nonlinear forward models, which is critical for scientific inference tasks. Compared to the standard pCN algorithm, the tpCN algorithm is found to have significantly improved performance on non-Gaussian targets.
    \item Our adaptive scheme exploits the natural connection between EKI and Bayesian annealing approaches, by using EKI within an SMC sampling scheme. Controlling the transition between temperature levels as we move from the prior to the posterior in SMC allows us to apply EKI updates treating the target at the previous temperature level as an effective prior. EKI then provides a highly effective initialization and preconditioner for the tpCN updates. The tpCN updates help to ensure we correctly converge on the target at each temperature level, preventing the accumulation of errors that would result from applying EKI alone within an annealing scheme to the Bayesian inverse problem.
    \item We demonstrate the empirical performance of our adaptive sampling scheme on three challenging inverse problem benchmarks. We show that tpCN significantly outperforms standard pCN. We also show that the use of EKI as an initialization and preconditioner within a Bayesian annealing scheme for the exact tpCN updates yields significant performance improvements compared to using the standard importance resampling step in SMC.
\end{itemize}

The structure of the paper is as follows: In Section \ref{sec: p2CN} we describe the tpCN sampling algorithm we propose developing an adaptive implementation of in this work. In Section \ref{sec: background} we describe essential background regarding the methods used for adapting the tpCN sampler. In Section \ref{sec: FAKISMC} we describe the Sequential Kalman Tuning (SKT) adaptation scheme for the tpCN algorithm, and its NF preconditioned variant NF-SKT, proposed in this work for rapid gradient-free Bayesian inference. In Section \ref{sec: experiments} we present numerical results comparing the performance of the adaptive SKT samplers against adaptation in standard SMC using importance weighted resampling, and we conclude in Section \ref{sec: conclusions}. Code implementing the adaptive SKT samplers presented in this work is available at \url{https://github.com/RichardGrumitt/KalmanMC}.

\section{$t$-preconditioned Crank-Nicolson Algorithm}\label{sec: p2CN}

In this work we consider the adaptation of the tpCN sampling algorithm within SMC, which has been implemented in the context of NF preconditioned SMC in the \textsc{pocoMC} sampling package\footnote{\url{https://github.com/minaskar/pocomc/}}. At its core, tpCN modifies the standard pCN proposal such that it is reversible with respect to the multivariate $t$-distribution, as opposed to the multivariate Gaussian distribution for the pCN proposal. In \cite{kamatani2018pCN} the mixed preconditioned Crank-Nicolson (MpCN) algorithm was proposed, which uses a proposal that is reversible with respect to the $\sigma$-finite measure $\bar{p}(\mathrm{d}\bi{x})=\norm{\bi{x}}_2^{-d}\mathrm{d}\bi{x}$. Detailed theoretical studies of the MpCN algorithm were performed in \cite{kamatani2017ergodicity, kamatani2018pCN}, which showed improved convergence results for MpCN on heavy tailed targets compared to pCN. However, in our own numerical studies we found that the MpCN algorithm could not be easily adapted for sampling on the non-Gaussian targets we consider in this work. Whilst the base $t$-distribution in tpCN can be adapted for each target, adjusting the corresponding tail behaviour of the proposal, the base distribution of MpCN is not so readily adaptable. Even after pre-whitening of the target, we found the MpCN acceptance rate was typically close to zero. We therefore do not consider it further as a numerical benchmark in this work. A similarly detailed theoretical study of the tpCN algorithm as in \cite{kamatani2017ergodicity, kamatani2018pCN} is beyond the scope of this work, where we focus on its practicable adaptive implementation. However, we do show that the tpCN algorithm has superior empirical performance compared to standard pCN on a range of challenging benchmarks, when allowing for similar adaptation in their sampling hyper-parameters. In the remainder of this section we describe the pCN and tpCN algorithms.

\subsection{pCN algorithm}

Consider some target measure with probability density function (PDF) $p(\bi{x})$. The standard pCN algorithm generates samples from the target by iterating over the procedure described in Algorithm \ref{alg: pCN}. We denote the PDF of the multivariate Gaussian distribution at some location $\bi{x}$ as $\varphi_\mathcal{N}(\bi{x};\bi{\mu}, \mathcal{C})$, where $\bi{\mu}$ is the Gaussian mean and $\mathcal{C}$ is the Gaussian covariance. 
\begin{algorithm}
   \caption{pCN update}
\begin{algorithmic}[1]
   \STATE {\bfseries Input:} Current particle location $\bi{x}_{m-1}$, pCN proposal reference mean $\bi{\mu}$, pCN proposal reference covariance $\mathcal{C}$, pCN step size $\rho$, target density $p(\bi{x})$.
   \STATE Draw $\bi{W}_m\sim\mathcal{N}(0, \mathcal{C})$.
   \STATE $\bi{x}_{m}^\prime = \bi{\mu} + \sqrt{1-\rho^2}(\bi{x}_{m-1} - \bi{\mu})+\rho \bi{W}_m$.
   \STATE Update particle location,
   \begin{equation}
       \bi{x}_m=\begin{cases}
			\bi{x}_m^\prime, & \text{with probability}\, \alpha(\bi{x}_m^\prime, \bi{x}_{m-1}),\\
            \bi{x}_{m-1}, & \text{with probability}\, 1-\alpha(\bi{x}_m^\prime, \bi{x}_{m-1}),
		 \end{cases}
   \end{equation}
   where
   \begin{equation}
       \alpha(\bi{x}_m^\prime, \bi{x}_{m-1})=\mathrm{min}\left\{1, \frac{p(\bi{x}_m^\prime)\varphi_{\mathcal{N}}(\bi{x}_{m-1}; \bi{\mu},\mathcal{C})}{p(\bi{x}_{m-1})\varphi_{\mathcal{N}}(\bi{x}_m^\prime; \bi{\mu},\mathcal{C})}\right\}.
   \end{equation}
   \STATE {\bfseries Output}: New particle location $\bi{x}_m$.
\end{algorithmic}
\label{alg: pCN}
\end{algorithm}

The pCN step size parameter, $\rho$ controls the extent to which a proposal sample is correlated with the previous sample. In the limit where $\rho\rightarrow 1$, the pCN proposal reduces to an independent proposal drawn from $\mathcal{N}(\bi{\mu}, \mathcal{C})$. The proposal kernel for pCN as defined in Algorithm \ref{alg: pCN} is given by
\begin{equation}
    \mathcal{K}(\bi{x}, \mathrm{d}\bi{x}^\prime) = \mathcal{N}(\bi{\mu}+\sqrt{1-\rho^2}(\bi{x}-\bi{\mu}), \rho^2\mathcal{C}),
\end{equation}
which is reversible with respect to the Gaussian distribution $\mathcal{N}(\bi{\mu}, \mathcal{C})$ i.e.,
\begin{equation}
    \varphi_{\mathcal{N}}(\bi{x}; \bi{\mu}, \mathcal{C})\mathrm{d}\bi{x}\mathcal{K}(\bi{x}, \mathrm{d}\bi{x}^\prime)=\varphi_{\mathcal{N}}(\bi{x}^\prime; \bi{\mu}, \mathcal{C})\mathrm{d}\bi{x}^\prime \mathcal{K}(\bi{x}^\prime, \mathrm{d}\bi{x}).
\end{equation}
The pCN algorithm has been shown to exhibit a dimension independent spectral gap for large class of target measures which are the finite dimensional approximations of densities defined with respect to some Gaussian reference measure i.e., for some target posterior measure $\pi$ we have the Radon-Nikodym derivative
\begin{equation}
    \frac{\mathrm{d}\pi}{\mathrm{d}\pi_{0}}(\bi{x})\propto\exp(-\Phi(\bi{x})),
\end{equation}
where the reference prior measure $\pi_0$ is taken to be the Gaussian $\mathcal{N}(\bi{\mu}, \mathcal{C})$ and $\Phi(\bi{x})$ is the likelihood potential \cite{cotter2013pCN, hairer2014spectral}. The pCN algorithm performs well when the target measure is close to Gaussian. However, for non-Gaussian targets and targets with heavy tails the performance of the algorithm can be severely degraded. Indeed, in \cite{kamatani2018pCN} it was shown that pCN performs worse than Random Walk Metropolis-Hastings (RWHM) on a family of heavy tailed targets. This presents a problem for many scientific inference tasks where the target distribution can be expected to show some degree of non-Gaussianity.

\subsection{tpCN algorithm}

To develop an adaptive sampling scheme that will perform well against non-Gaussian targets we consider the tpCN algorithm. Instead of using a Gaussian base distribution to generate a proposal, as with standard pCN, the tpCN algorithm uses a multivariate $t$-distribution $t_{\nu_s}(\bi{\mu}_s,\mathcal{C}_s)$, where $\nu_s>0$ denotes the degrees of freedom, $\bi{\mu}_s$ is the mean and $\mathcal{C}_s$ is the scale matrix. A simple non-adaptive variant of tpCN has previously been used in the estimation of drift and diffusion parameters for stochastic differential equations in \cite{kamataniSDE2015}. The tpCN algorithm generates samples by iterating over the procedure in Algorithm \ref{alg: tpCN}. For brevity in the discussion below we use the inner product notation $\langle\bi{x}_1, \bi{x}_2\rangle_s=(\bi{x}_1-\bi{\mu}_s)^\intercal\mathcal{C}_s^{-1}(\bi{x}_2-\bi{\mu}_s)$.

\begin{algorithm}
   \caption{tpCN update}
\begin{algorithmic}[1]
   \STATE {\bfseries Input:} Current particle location $\bi{x}_{m-1}$, $t$-distribution mean $\bi{\mu}_s$, $t$-distribution scale matrix $\mathcal{C}_s$, $t$-distribution degrees of freedom $\nu_s$, pCN step size $\rho$, target density $p(\bi{x})$.
   \STATE Draw $Z_m^{-1}\sim\mathrm{Gamma}(k=\frac{1}{2}(d+\nu_s), \theta=2/(\nu_s+\langle\bi{x}_{m-1},\bi{x}_{m-1}\rangle_s))$ and $\bi{W}_m\sim\mathcal{N}(0, \mathcal{C}_s)$.
   \STATE $\bi{x}_{m}^\prime=\bi{\mu}_s+\sqrt{1-\rho^2}(\bi{x}_{m-1}-\bi{\mu}_s)+\rho\sqrt{Z_m}\bi{W}_m$.
   \STATE Update particle location,
   \begin{equation}
       \bi{x}_m=\begin{cases}
			\bi{x}_m^\prime, & \text{with probability}\, \alpha(\bi{x}_m^\prime, \bi{x}_{m-1}),\\
            \bi{x}_{m-1}, & \text{with probability}\, 1-\alpha(\bi{x}_m^\prime, \bi{x}_{m-1}),
		 \end{cases}
   \end{equation}
   where
   \begin{equation}
       \alpha(\bi{x}_m^\prime, \bi{x}_{m-1})=\mathrm{min}\left\{1, \frac{p(\bi{x}_m^\prime)(1+\langle\bi{x}_{m-1},\bi{x}_{m-1}\rangle_s/\nu_s)^{-(d+\nu_s)/2}}{p(\bi{x}_{m-1})(1+\langle\bi{x}_{m}^\prime,\bi{x}_{m}^\prime\rangle_s/\nu_s)^{-(d+\nu_s)/2}}\right\}.
       \label{eqn:tpcn alpha}
   \end{equation}
   \STATE {\bfseries Output}: New particle location $\bi{x}_m$.
\end{algorithmic}
\label{alg: tpCN}
\end{algorithm}

It can be shown that the tpCN proposal is reversible with respect to the multivariate $t$-distribution $t_{\nu_s}(\bi{\mu}_s, \mathcal{C}_s)$. The reversibility and acceptance rate properties of the tpCN algorithm are stated in Lemma \ref{lemma:tpCN}, with the corresponding proof given in \ref{sec: tpCN proofs}.
\begin{lemma}
The proposal transition kernel of the tpCN algorithm is reversible with respect to the multivariate $t$-distribution $t_{\nu_s}(\bi{\mu}_s, \mathcal{C}_s)$ and the proposal acceptance probability is given by Equation \ref{eqn:tpcn alpha}.
\label{lemma:tpCN}
\end{lemma}
Similarly to the MpCN algorithm, the tpCN proposal is reversible with respect to a distribution that will generally have heavier tails than the standard pCN algorithm. One may therefore expect that it will show similarly improved performance on heavy tailed targets. A key difference between the tpCN and MpCN algorithms is the ability to tune the degrees of freedom $\nu_s$, which controls the tail behaviour of the tpCN proposal. However, it is worth emphasising that the benefits of using the $t$-distribution as a base distribution extend beyond heavy tailed targets to non-Gaussian targets more generally. The ability to tune the parameters of the more flexible $t$-distribution to the target allows for improved sampling of non-Gaussian targets, as observed in our numerical experiments in Section \ref{sec: experiments}. The multivariate $t$-distribution has been exploited in the development of adaptive elliptical slice sampling implementations in \cite{nishihara2014parallel}, where the increased flexibility of the multivariate $t$-distribution in approximating the target distribution was found to yield significant performance gains compared to standard elliptical slice sampling. 

For targets with strong non-Gaussianity, the performance of tpCN can be futher improved through NF preconditioning. In this case the tpCN updates are performed on the NF latent space particles, $\bi{z}=f(\bi{x})$ with the corresponding latent space acceptance probability being given by
\begin{equation}
    \alpha(\bi{z}, \bi{z}^\prime)=\mathrm{min}\left\{1, \frac{\pi_0(f_n^{-1}(\bi{z}^\prime))\pi(\bi{y}|f^{-1}(\bi{z}^\prime))^{\beta_{n+1}}|\mathrm{det}Df^{-1}(\bi{z}^\prime)|(1+\langle\bi{z},\bi{z}\rangle_s/\nu_s)^{-(d+\nu_s)/2}}{\pi_0(f^{-1}(\bi{z}))\pi(\bi{y}|f^{-1}(\bi{z}))^{\beta_{n+1}}|\mathrm{det}Df^{-1}(\bi{z})|(1+\langle\bi{z}^\prime,\bi{z}^\prime\rangle_s/\nu_s)^{-(d+\nu_s)/2}}\right\},
\label{eqn: tpCN latent alpha}
\end{equation}
where $Df^{-1}(\bi{z})=\partial f^{-1}(\bi{z})/\partial \bi{z}$ is the Jacobian of the inverse NF transformation. The use of NF preconditioning within our adaptation scheme is discussed in detail in Section \ref{subsec: nf preconditioning}.

\section{Background Methods}\label{sec: background}

In this section we introduce essential background regarding the methods we use for implementing an adaptive tpCN sampler, targeted at solving the Bayesian inverse problem. In Section \ref{subsec: eki} we give a brief description of EKI as applied to the Bayesian inverse problem and its connection to Bayesian annealing, in order to motivate its use within our adaptation scheme. In Section \ref{subsec: smc} we describe the SMC sampling scheme within which we embed our adaptation procedures. In Section \ref{subsec: temperature adaptation} we describe the temperature schedule adaptation, and in Section \ref{subsec: nf preconditioning} we describe the use of NFs for additional preconditioning. 

\subsection{Ensemble Kalman Inversion}\label{subsec: eki}

EKI is a coupling-based algorithm that leverages ideas from EKF to construct iterative particle ensemble updates for the solution of inverse problems \cite{iglesias2013ensemble, iglesias2016regularizing, iglesias2018bayesian, chada2018parameterizations, kovachki2019ensemble, chada2020tikhonov,iglesias2021adaptive, chada2022convergence}. In its standard setting, EKI seeks to solve the variational inverse problem i.e., finding parameter values that minimize the misfit functional
\begin{equation}
    \Phi(\bi{x}) = \frac{1}{2}\norm{\Gamma^{-1/2}(\bi{y}-\mathcal{F}(\bi{x}))}_2^2.
    \label{eqn:eki loss}
\end{equation}
In this optimization setting, the EKI ensemble will collapse to a single solution \cite{iglesias2013ensemble}, and can also be applied rigorously to problems with nonlinear forward models \cite{chada2022convergence}. However, EKI has also been applied to the solution of the Bayesian inverse problem, where we seek to obtain an approximation to the full target posterior \cite{iglesias2021adaptive, grumitt2024flow}. In this setting, EKI is only exact when applied to linear forward models with Gaussian targets, providing an uncontrolled posterior approximation outside this regime. 

In this work we consider leveraging the EKI target approximation within the context of Bayesian annealing to initialize and precondition tpCN sampling iterations. It is worth noting that EKI has previously been used in the context of preconditioning for sampling in \cite{garbuno2020interacting}, where the Ensemble Kalman Sampler (EKS) was developed based on the EKI ensemble structure. However, this sampling scheme is only exact for linear forward models, and can otherwise give highly biased posterior moment estimates. A key difference in our work is in the direct use of EKI updates to intialize and precondition the tpCN sampling scheme, which preserves the exact target as its invariant measure. For adaptive MCMC schemes based within Bayesian annealing this achieves significant convergence acceleration whilst allowing for accurate posterior moment estimation outside the regime of linear forward models and Gaussian targets. In \ref{sec: eks} we demonstrate the performance of EKS on our numerical experiments, showing that it fails to recover accurate posterior moment estimates.

Following \cite{iglesias2021adaptive}, we can motivate EKI applied to the solution of Bayesian inverse problems within the context of a Bayesian annealing scheme. Given a prior measure $\pi_0(\bi{x})$, EKI proceeds by constructing a sequence of Gaussian ensemble approximations to the intermediate measures
\begin{equation}
    \pi_{n}(\mathrm{d}\bi{x})\propto \pi_0(\mathrm{d}\bi{x})\mathcal{N}(\bi{y}|\mathcal{F}(\bi{x}), \beta_n^{-1}\Gamma),
    \label{eq:annealed targets}
\end{equation}
where the inverse temperatures satisfy $0\equiv\beta_0<\beta_1<\ldots<\beta_N<\beta_{N+1}\equiv 1$. From Equation \ref{eq:annealed targets} we can obtain the recursion
\begin{equation}
    \frac{\pi_{n+1}(\mathrm{d}\bi{x})}{\pi_n(\mathrm{d}\bi{x})}\propto \mathcal{N}(\bi{y}|\mathcal{F}(\bi{x}),\alpha_n\Gamma),
    \label{eq:basic recursion}
\end{equation}
where the annealing step size $\alpha_n=(\beta_{n+1}-\beta_n)^{-1}$. The step size can be viewed as a regularization parameter \cite{iglesias2016regularizing, chada2020tikhonov, iglesias2021adaptive}, which can be selected such that we make a gradual transition from the prior to the posterior.

The ensemble updates for EKI can be derived by assuming we have some Gaussian approximation to the prior measure $\tilde{\pi}_{0}=\mathcal{N}(\bi{m}_0, C_0)$, proceeding to move through a sequence of Gaussian approximations, $\tilde{\pi}_{n}=\mathcal{N}(\bi{m}_n, C_n)$ using the recursion
\begin{equation}
    \frac{\tilde{\pi}_{n+1}(\mathrm{d}\bi{x})}{\tilde{\pi}_{n}(\mathrm{d}\bi{x})}\propto \mathcal{N}\left(\bi{y}|\mathcal{F}(\bi{m}_n) + \frac{\partial\mathcal{F}_n}{\partial\bi{x}}(\bi{x}-\bi{m}_n), \alpha_n\Gamma\right).
    \label{eq:approx recursion}
\end{equation}
The forward model has been linearized around the approximation mean, $\mathcal{F}(\bi{x})\approx\mathcal{F}(\bi{m}_n) -\partial\mathcal{F}_n/\partial\bi{x}(\bi{x}-\bi{m}_n)$, where $\partial\mathcal{F}_n/\partial\bi{x}=\partial\mathcal{F}/\partial\bi{x}|_{\bi{x}=\bi{m}_n}$. From Equation \ref{eq:approx recursion} we can obtain recursions for the approximation means and covariances,
\begin{align}
    \bi{m}_{n+1}&=\bi{m}_n+C_n\frac{\partial\mathcal{F}_n^{*}}{\partial\bi{x}}\left(\frac{\partial\mathcal{F}_n}{\partial\bi{x}}C_n\frac{\partial\mathcal{F}_n^*}{\partial\bi{x}}+\alpha_n\Gamma\right)^{-1}(\bi{y}-\mathcal{F}(\bi{m}_n)), \label{eqn: eki mean recusion}\\
    C_{n+1} &= C_n - C_n\frac{\partial\mathcal{F}_n^*}{\partial\bi{x}}\left(\frac{\partial\mathcal{F}_n}{\partial\bi{x}}C_n\frac{\partial\mathcal{F}_n^*}{\partial\bi{x}}+\alpha_n\Gamma\right)^{-1}\frac{\partial\mathcal{F}_n}{\partial\bi{x}}C_n \label{eqn: eki cov recursion},
\end{align}
where $\partial\mathcal{F}_n^*/\partial\bi{x}$ is the adjoint of $\partial\mathcal{F}_n/\partial\bi{x}$. Using the linearized forward model, the terms involving derivatives can be approximated as
\begin{align}
    C_n\frac{\partial\mathcal{F}_n^*}{\partial\bi{x}}&\approx\mathbb{E}_n[(\bi{x}_n-\bi{m}_n)\otimes (\mathcal{F}(\bi{x}_n)-\mathbb{E}_n[\mathcal{F}(\bi{x}_n)])],\\
    \frac{\partial\mathcal{F}_n}{\partial\bi{x}}C_n\frac{\partial\mathcal{F}_n^*}{\partial\bi{x}}&\approx \mathbb{E}_n[(\mathcal{F}(\bi{x}_n)-\mathbb{E}_n[\mathcal{F}(\bi{x}_n)])\otimes (\mathcal{F}(\bi{x}_n)-\mathbb{E}_n[\mathcal{F}(\bi{x}_n)])],
\end{align}
where $\mathbb{E}_n$ denotes the expectation with respect to $\tilde{\pi}_n$. These expectations cannot be computed in closed form. To overcome this, EKI exploits an ensemble approximation. Given an initial particle ensemble $\{\bi{x}_0^i\sim\pi_0(\bi{x})\}_{i=1}^J$, EKI applies embarrassingly parallel recursive updates using the expression
\begin{equation}
    \bi{x}^{i}_{n+1}=\bi{x}^{i}_n + C_n^{\bi{x}\mathcal{F}}\left(C_n^{\mathcal{F}\mathcal{F}}+\alpha_n\Gamma\right)^{-1}\left(\bi{y}-\mathcal{F}(\bi{x}_n^i)+\sqrt{\alpha_n}\bi{\xi}_n^i\right),
    \label{eqn:eki update}
\end{equation}
where $\bi{\xi}_n^i\sim\mathcal{N}(0,\Gamma)$ is a Gaussian noise vector \cite{iglesias2021adaptive}. The empirical covariances are given by
\begin{align}
    C_n^{\bi{x}\mathcal{F}} &= \frac{1}{J-1}\sum_{i=1}^J (\bi{x}_n^i-\langle \bi{x}_n\rangle)\otimes(\mathcal{F}(\bi{x}_n^i)-\langle\mathcal{F}_n\rangle),\label{eqn: xF cov}\\
    C_n^{\mathcal{F}\mathcal{F}} &= \frac{1}{J-1}\sum_{i=1}^J (\mathcal{F}(\bi{x}_n^i)-\langle\mathcal{F}_n\rangle)\otimes(\mathcal{F}(\bi{x}_n^i)-\langle\mathcal{F}_n\rangle),
    \label{eqn: FF cov}
\end{align}
where $\langle\bi{x}_n\rangle=\frac{1}{J}\sum_{i=1}^{J}\bi{x}_n^i$ and $\langle\mathcal{F}_n\rangle=\frac{1}{J}\sum_{i=1}^{J}\mathcal{F}(\bi{x}_n^i)$. It can be shown that the ensemble means and covariances obtained through the EKI updates approximate those in Equations \ref{eqn: eki mean recusion} and \ref{eqn: eki cov recursion} as $J\rightarrow\infty$ \cite{mandel2011convergence, iglesias2021adaptive}.

Applied to the Bayesian inverse problem, EKI enjoys rapid convergence properties, typically converging in $\mathcal{O}(10)$ iterations \cite{huang2022efficient}. For the case of Gaussian targets with linear forward models the particle ensemble will be distributed according to the target posterior as the ensemble size $J\rightarrow\infty$ \cite{iglesias2021adaptive}, otherwise giving an uncontrolled approximation. In \cite{grumitt2024flow}, NF maps were learned at each temperature level in the EKI iterations. By learning an NF map, one can map the particle distribution at a given temperature level to a Gaussian latent space and perform the EKI update in this latent space. Whilst this can improve the stability of EKI when faced with non-Gaussian targets, it does not address the linearity assumptions used in deriving EKI. Further, the NF map can introduce additional nonlinearity in the forward model evaluation due to the need to apply the inverse transformation when evaluating the forward model at each latent space location. These problems can result in the converged particle ensemble being a poor approximation to the true posterior, which poses a major drawback for scientific inference tasks where we desire accurate estimation of the first and second moments of the target posterior.

Despite the limitations of EKI when applied to the solution of the Bayesian inverse problem alone, we can exploit its natural connection with Bayesian annealing to form part of an adaptive SMC sampling scheme. The core idea here is that an EKI update can be used at each temperature level to target the next temperature level. The updated ensemble provides both an initialization and preconditioner for the subsequent sampling iterations.

\subsection{Sequential Monte Carlo}\label{subsec: smc}

SMC encompasses a class of sampling methods that move through a sequence of probability measures $\{\pi_n(\bi{x})\}_{n=1}^{N+1}$ in order to sample from the final target measure $\pi_{N+1}(\bi{x})$ \cite{moral2006SMC}. The method has seen extensive applications in sequential Bayesian inference where one has a set of sequential observations $\{\bi{y}_t\}_{t=1}^T$ e.g., time series data \cite{chopin2002sequential}. In this case SMC moves through targets $\pi_n(\bi{x})=p(\bi{x}|\bi{y}_1,...,\bi{y}_{n})$, where at each iteration an additional observation is added. Ensemble Kalman methods have previously been exploited for SMC in sequential Bayesian inference in \cite{wu2022ensemble}, where the EKF update was used to construct an efficient importance sampling proposal for SMC.

An alternative setting involves moving from some tractable density $\pi_0(\bi{x})$, through a sequence of intermediate measures towards the final target. In this work, we consider the situation where $\pi_0(\bi{x})$ is the prior and we move through a sequence of temperature annealed targets $\pi_n(\bi{x})\propto\pi_0(\bi{x})\pi(\bi{y}|\bi{x})^{\beta_{n}}$ where $\pi(\bi{y}|\bi{x})$ is the likelihood. As with EKI, the inverse temperatures satisfy $0\equiv\beta_0<\beta_1<\ldots<\beta_N<\beta_{N+1}\equiv 1$, with $\beta_0=0$ corresponding to the prior and $\beta_{N+1}=1$ corresponding to the full posterior.

Consider a particle ensemble at the inverse temperature $\beta_n$. Assuming the ensemble is distributed according to the annealed target $\pi_n(\bi{x})\propto\pi_0(\bi{x})\pi(\bi{y}|\bi{x})^{\beta_{n}}$, we can calculate the unnormalized importance weights corresponding to the next temperature level,
\begin{equation}
    w_{n}(\bi{x}_n^i)=\frac{\pi(\bi{y}|\bi{x}_n^i)^{\beta_{n+1}}}{\pi(\bi{y}|\bi{x}_n^i)^{\beta_{n}}}.
\end{equation}
From this we obtain an estimator for expectations of test functions $f(\bi{x})$ with respect to the subsequent annealed target given by
\begin{equation}
    \mathbb{E}_{\pi_{n+1}}[f(\bi{x})] = \frac{\sum_{i=1}^{J} f(\bi{x}_n^i)w_n(\bi{x}_n^i)}{\sum_{i=1}^{J}w_n(\bi{x}_n^i)}.
\end{equation}
If the importance sampling proposal distribution, in this case $\pi_n(\bi{x})$, is not close to the target, the importance sampling estimator can have very high variance, scaling approximately with the variance of the importance weights \cite{moral2006SMC}. 

Direct application of importance weighting through the annealed targets in SMC can quickly result in weight collapse, where all the importance weight is assigned to a single particle in the ensemble. This issue can be partially addressed through resampling, where the particle ensemble is resampled according to their importance weights, duplicating particles with high weight and removing particles with low weight \cite{douc2005comparison}. This also gives an equal weight particle ensemble that is approximately distributed according to the annealed target. We discuss the exact resampling scheme used in this work in \ref{sec: resampling}.

In order to further improve the quality of the MC approximation given by the particle ensemble, one can perform sampling updates at each temperature level. If resampling has been performed, this also helps to disperse particles and remove duplicates in the ensemble. Typically, this will involve the application of some $\pi_{n+1}(\bi{x})$ invariant MCMC kernel, $K_n(\bi{x}^\prime|\bi{x})$ for several iterations such that the particle ensemble is distributed according to $\pi_{n+1}(\bi{x})$. Pseudocode for the SMC algorithms we use as benchmarks in this work is given in \ref{sec: smc implement}.

In principle, SMC can produce a particle ensemble that provides asymptotically unbiased approximations to posterior marginal moments, without the limitations of EKI in only being exact for Gaussian targets with linear forward models. However, in order to attain low bias on these moment estimates in practice we must run multiple iterations of the MCMC updates at each temperature level \cite{karamanis2022accelerating}. For scientific inverse problems with expensive forward models we would like to minimize the number of MCMC iterations required to achieve low bias. Previous works have leveraged ideas from EKF within MCMC, with examples including \cite{zhang2020ekf_mcmc}, where a proposal kernel was developed based on the analysis step in the EKF update, and \cite{drovandi2022ensemble} which used EKF to accelerate pseudo-marginal MCMC in state space models. In this work we propose using EKI as part of an adaptation scheme for the tpCN sampler, replacing the resampling step in SMC by instead using the EKI update as an initialization and preconditioner for each intermediate target.

\subsection{Temperature Adaptation}\label{subsec: temperature adaptation}

The choice of temperature schedule is crucial to both EKI and SMC. We wish to take steps in inverse temperature that are neither too small, which would unnecessarily increase the number of model evaluations, nor too large, which would render the particle ensemble obtained at the previous temperature level of limited use in adapting the MCMC kernel used for the next target temperature. This is particularly relevant when we learn NF maps for preconditioning, where we rely on the particle distribution from the previous temperature level to inform our preconditioning.

In this work we select temperature levels by estimating the effective sample size (ESS) of the particle ensemble and choosing a value of $\beta$ such that we attain some fractional ESS target. The ESS in targeting some $\beta_{n+1}$ from $\beta_n$ can be estimated by calculating the importance weights given by
\begin{equation}
    w_n^i=\exp\left(-\frac{1}{2}\left(\beta_{n+1}-\beta_n\right)\norm{\Gamma^{-1/2}\left(\bi{y}-\mathcal{F}(\bi{x}_n^i)\right)}_2^2\right).
\end{equation}
The value of the target inverse temperature can then be obtained by solving for $\beta_{n+1}$ in,
\begin{equation}
    \left(\sum_{i=1}^{J}w_n^{i}(\beta_{n+1})^2\right)^{-1}\left(\sum_{i=1}^{J}w_n^i(\beta_{n+1})\right)^2 = \tau J,
    \label{eqn:ess criterion}
\end{equation}
where $0<\tau<1$ is the fractional ESS threshold. The value of $\tau$ controls the size of the steps in $\beta$, with larger values of $\tau$ resulting in smaller steps. This method has seen extensive application in adaptive SMC and EKI implementations due to the ability to control the ensemble ESS, which is crucial for effective resampling \cite{de2018quantifying, iglesias2018bayesian}.

Alternative temperature adaption schemes can be used, for example if one was seeking to use a more aggressive temperature schedule \cite{iglesias2021adaptive}. However, more agressive temperature schedules can be unstable for standard SMC, which uses importance resampling at each temperature level, if the ESS becomes very low. In order to make more direct comparisons between SKT algorithms and standard SMC algorithms in this work, we only consider the ESS-based criterion expressed in Equation \ref{eqn:ess criterion}. It is worth noting that such temperature adaptation renders SMC a biased but consistent method. However, this bias is typically negligible, and the ability to adapt the temperature schedule to each problem offers significant advantages in avoiding the need to manually select an appropriate schedule, hence the widespread use of adaptive temperature schedules in SMC \cite{moral2012adaptive, beskos2016convergence}.

\subsection{Normalizing Flow Preconditioning}\label{subsec: nf preconditioning}

In this paper we leverage NFs in two contexts; learning a map to a Gaussian latent space at each temperature level to improve the fidelity of the EKI target approximation \cite{grumitt2024flow}, and to act as a preconditioner for the subsequent tpCN sampling iterations.

NFs are generative models where one learns a bijective map between some original data space, $\bi{x}\in\mathbb{R}^d$ and a simple latent space, $\bi{z}\in\mathbb{R}^d$. They can be used for highly expressive density estimation and allow efficient sampling from the learned generative model \cite{dinh2016density, papamakarios2017masked, kingma2018glow, dai2021sliced}. The full bijective map, $\bi{z}=f(\bi{x})$ proceeds through a sequence of invertible transformations $f=f_1\circ\ldots\circ f_{n_L}$, with the latent space base distribution typically chosen to be the standard Gaussian such that $\bi{z}\sim p_{\bi{z}}(\bi{z})=\mathcal{N}(0, I_d)$, where $I_d$ denotes the $d\times d$ identity matrix. Data space samples can be obtained from the NF distribution by drawing samples from the latent space base distribution and evaluating the inverse transformation $\bi{x}=f^{-1}(\bi{z})$.

The learned NF density, $q(\bi{x})$ can be evaluated using the standard change of variables formula,
\begin{equation}
    q(\bi{x}) = p_{\bi{z}}(f(\bi{x}))\left|\mathrm{det}\,Df(\bi{x})\right| = p_{\bi{z}}(f(\bi{x}))\prod_{l=1}^{n_L} \left|\mathrm{det}\,Df_l(\bi{x})\right|,
\end{equation}
where $Df(\bi{x})=\partial f(\bi{x})/\partial \bi{x}$ is the Jacobian for the NF transformation. In this work we use neural spline flows (NSF) \cite{durkan2019neural} as implemented in the \textsc{FlowMC} package \cite{gabrie2022adaptive, wong2022flowmc}, which have been found to be highly expressive flow architectures able to capture complex target geometries. In the numerical experiments performed in this work we were able use a single set of default configurations across the test models without the need for extensive NSF hyper-parameter searches.

The impact of the NF in the annealing schemes considered in this work can be seen by considering the recursive expression for the target with inverse temperature $\beta_{n+1}$,
\begin{equation}
    \pi_{n+1}(\bi{x})\propto\pi_{n}(\bi{x})\mathcal{N}(\bi{y}|\mathcal{F}(\bi{x}), \alpha_n\Gamma).
    \label{eqn: recursive target}
\end{equation}
We can view $\pi_{n}(\bi{x})$ as a pseudo-prior for $\pi_{{n+1}}(\bi{x})$, with the likelihood contribution being controlled by $\alpha_n$. By fitting an NF to the particle ensemble obtained for $\beta_n$, and assuming the particle ensemble is correctly distributed as $\pi_{{n}}(\bi{x})$, we can map the pseudo-prior to an approximately Gaussian space. In the NF latent space, the $\beta_{n+1}$ target is given by
\begin{equation}
    \pi_{n+1}(\bi{z})\propto\pi_{n}(f_n^{-1}(\bi{z}))|\mathrm{det}Df_n^{-1}(\bi{z})|\mathcal{N}(\bi{y}|\mathcal{F}(f_n^{-1}(\bi{z})), \alpha_n\Gamma).
    \label{eqn:latent target recursion}
\end{equation}
The latent space pseudo-prior is approximately the standard Gaussian. For EKI updates performed in the NF latent space, we can view this as single step EKI with prior $\pi_{n}(\bi{z})=\pi_{n}(f_n^{-1}(\bi{z}))|\mathrm{det}Df_n^{-1}(\bi{z})|$ and target posterior $\pi_{n+1}(\bi{z})$. Provided the particle ensemble for $\beta_n$ was correctly distributed according $\pi_{n}(\bi{x})$, by performing the EKI update in the latent space we have a Gaussian prior ensemble. If the value of $\alpha_n$ is chosen to be sufficiently large (i.e., small step size in $\beta$) such that $\pi_{n}(\bi{z})$ is prior dominated, we are able to effectively relax the Gaussian ansatz of EKI. 

Whilst the use to NFs in EKI has been found to improve robustness against non-Gaussianity \cite{grumitt2024flow}, the NF maps do not address the assumption that the forward model is linear. If the forward model is nonlinear the EKI update will not be exact, even when performed in the Gaussian latent space. This means the particle ensemble will not be correctly distributed according to the subsequent tempered target. When this is then treated as the pseudo-prior for the next temperature level, the NF will map the incorrect particle ensemble to a Gaussian latent space which does not correspond to the correct pseudo-prior distribution. These errors can accumulate as one progresses from the prior to the posterior, resulting in a low fidelity ensemble approximation to the final posterior.

Nonetheless, within our adaptive sampling scheme NF preconditioning has the benefit of helping to stabilize EKI/Flow Annealed Kalman Inversion (FAKI) updates, and acting as a nonlinear preconditioner for the tpCN updates. Considering again Equation \ref{eqn:latent target recursion}, if the particle ensemble obtained for $\beta_n$ can be mapped to a Gaussian latent space, and the value of $\alpha_n$ is chosen such that the $\beta_{n+1}$ target is dominated by the pseudo-prior, the NF provides a highly effective preconditioner that is able to account for local variations in the target geometry. Mapping to a Gaussian latent space has the additional advantage of allowing us to use scaling relations derived for samplers with Gaussian targets when selecting sampling hyper-parameters \cite{gelman1997weak, roberts2001optimal, beskos2013optimal}. The use of NFs as preconditioners has seen several applications for sampling, including in MCMC \cite{hoffman2019neutra, gabrie2022adaptive}, with interacting particle systems \cite{grumitt2022deterministic} and in SMC \cite{karamanis2022accelerating, karamanis2022_pocoMC}.

\section{Adaptation of tpCN with Ensemble Kalman Inversion in Sequential Monte Carlo}\label{sec: FAKISMC}

In this section we describe the SKT adaptation procedure we propose for tpCN. In essence, EKI is used within a Bayesian annealing scheme, with the core performance improvements arising from the ability of the EKI updates to provide an effective initialization and preconditioner for the tpCN sampling iterations at each temperature level. The tpCN iterations are then able to efficiently converge on the target at each temperature level, allowing for accurate posterior moment estimation. In Section \ref{subsec: tpCN kernel adapt} we describe the procedure for adapting the tpCN kernel parameters, and a method for selecting the number of tpCN iterations to perform at each temperature level. We then provide pseudocode outlining the full adaptive sampling algorithm in Section \ref{subsec: faki+smc}.

\subsection{Adaptation of tpCN kernel parameters}\label{subsec: tpCN kernel adapt}

At each temperature level we need to select values of the tpCN kernel parameters such that the base $t$-distribution approximates the target distribution well. In \cite{nishihara2014parallel}, a parallel adaptation scheme for elliptical slice sampling was used where the parameters of the proposal $t$-distribution were obtained by dividing a particle ensemble into two groups. The parameters for the proposal in one group were then obtained by fitting a $t$-distribution to the particles in the other group. For the adaptive SMC scheme we consider in this work we do not need to divide the ensemble into groups, instead using the particle ensemble prior to sampling at each temperature level to fit for the $t$-distribution parameters. This also means we do not need to alternate the tpCN updates between groups as in \cite{nishihara2014parallel}, instead relying on control of the transition between temperature levels and the quality of the EKI target approximation to provide an effective tpCN kernel.

To select the $t$-distribution parameters at each temperature level we use the expectation maximization (EM) algorithm \cite{Meng1997TheEA, liu1995ml}, described in Algorithm 4 of \cite{nishihara2014parallel}. This is a stable choice provided the size of the particle ensemble $J\geq 2d$, where $d$ is the target dimension. As noted in \cite{nishihara2014parallel}, more sophisticated procedures could be used in high dimensions, although the structure of our adaptation scheme would be largely the same. The $t$-distribution parameters are fitted to the ensemble after applying the EKI update, such that it approximates the target distribution. Similarly, for the standard SMC benchmarks we fit the $t$-distribution parameters to the resampled particle ensemble.

In addition to selecting appropriate parameters for the tpCN kernel, it is also important to run a sufficient number of tpCN iterations at each temperature level to ensure particles are distributed according to each intermediate target. In this work we study both the sampling performance using a fixed number of sampling iterations at each temperature level, in order to more directly assess the impact of the EKI adaptation step compared to resampling, and also present numerical results when selecting the number of tpCN iterations based on autocorrelation statistics. Such an approach has previously been used in the context of adaptive SMC, for example in \cite{buchholz2021adaptive} the number of MCMC iterations was selected by monitoring the component-wise first-order autocorrelation of the statistic $\bi{x}_{n,m}^i(j)+\bi{x}_{n,m}^i(j)^2$, where $i$ denotes the ensemble member, $n$ denotes the temperature level, $m$ denotes the MCMC iteration number and $j$ denotes the component of $\bi{x}_{n,m}^i$.  This statistic monitors the correlation of the first and second moments of the ensemble. In this work we present numerical results where MCMC iterations are performed at each temperature level until the product of the first-order autocorrelations falls below some threshold $\tau_{\mathrm{corr}}$ for all dimensions i.e.,
\begin{equation}
    \prod_{m=1}^{M}\hat{\rho}_{m}(j)<\tau_{\mathrm{corr}}\:\forall j,
\end{equation}
where $\hat{\rho}_m(j)$ is the autocorrelation statistic calculated over the ensemble for successive states $\{\bi{x}_{n,m-1}^i\}_{i=1}^J$ and $\{\bi{x}_{n,m}^i\}_{i=1}^J$. 

An important difference to note between the adaptive SKT samplers we propose here and standard SMC is that, for a fixed number of MCMC iterations at each temperature level, the SMC particle ensemble will converge asymptotically to the target posterior as the ensemble size $J\rightarrow\infty$ \cite{moral2006SMC}. In contrast, the SKT algorithm requires sufficient MCMC iterations to be performed at each temperature level such that we converge on each intermediate target. However, in the practical settings we consider in this work, where the ensemble size is some multiple of the target dimension, standard SMC still requires a large number of MCMC iterations at each temperature level to ensure we obtain low bias posterior moment estimates. This is required to ensure the particle ensemble does not collapse through the repeated resampling steps, and to remove duplicates from the resampled ensembles that can otherwise provide high variance posterior moment estimates. In Section \ref{sec: experiments} we provide numerical results demonstrating both the need for a large number of MCMC iterations at each temperature level in SMC, and the ability of the EKI adaptation step to accelerate the convergence of the MCMC iterations.

\subsection{Sequential Kalman Tuning for tpCN}\label{subsec: faki+smc}

All of the procedures we have outlined thus far can be combined to produce the SKT adaptive sampling scheme for tpCN.  The core of the SKT approach lies in replacing the importance resampling step of SMC with an EKI update. That is, given an ensemble of particles $\{\bi{x}_n^{i}\}_{i=1}^J$ distributed according to the target at inverse temperature $\beta_n$, we apply the EKI update in Equation \ref{eqn:eki update} to obtain a particle ensemble that approximates the target at $\beta_{n+1}$, which acts as an initialization and preconditioner for subsequent sampling updates. We can then fit for the parameters of the $t$-distribution reference measure before performing tpCN updates to correctly distribute the particle ensemble according to the target at $\beta_{n+1}$. In addition to performing the EKI update at each temperature level, we can also use NF preconditioning to improve the stability of EKI when approximating non-Gaussian measures, and to act as a nonlinear preconditioner for the tpCN sampling. Pseudocode for SKT with NF preconditioning is given in Algorithm \ref{alg: nf SKT}. For SKT without NF preconditioning, the structure of the algorithm is largely identical, without any NF fits being performed such that EKI and tpCN updates are performed in the original data space. For completeness, we provide pseudocode describing the benchmark SMC implementations used in this work in \ref{sec: smc implement}.

\begin{algorithm}
   \caption{Flow Preconditioned Sequential Kalman Tuning}
\begin{algorithmic}[1]
   \STATE {\bfseries Input:} Set of $J$ samples from the prior $\{\bi{x}_0^i\sim\pi_0(\bi{x})\}_{i=1}^J$, data $\bi{y}$, observation covariance $\Gamma$, target fractional ESS $\tau$, maximum number of tpCN iterations to perform at each temperature level $M$, initial tpCN step size $\rho$, target tpCN acceptance rate $\alpha^{\star}$, tpCN autocorrelation threshold $\tau_\mathrm{corr}$.
   \STATE Set $\beta_0=0$ and iteration counter $n=0$.
   \WHILE{$\beta_n<1$ \do}
        \STATE Solve for target inverse temperature $\beta_{n+1}$ in Equation \ref{eqn:ess criterion}.
        \IF{$\beta_{n+1}=1$}
            \STATE $n^* \leftarrow n+1$
        \ENDIF
        \STATE $\alpha_n \leftarrow \beta_{n+1} - \beta_n$
        \STATE Fit NF map, $\bi{z}=f_n(\bi{x})$ to current particle locations $\{\bi{x}_n^i\}_{i=1}^{J}$.
        \STATE Obtain latent space particle locations $\{\bi{z}_n^i=f_n(\bi{x}_n^i)\}_{i=1}^J$.
        \FOR{$i=1, \ldots, J$}
        \STATE Update latent space particle ensemble with
                \begin{equation}
                    \bi{z}^{i}_{n+1}=\bi{z}^{i}_n + C_n^{\bi{z}\mathcal{F}}\left(C_n^{\mathcal{F}\mathcal{F}}+\alpha_n\Gamma\right)^{-1}\left(\bi{y}-\mathcal{F}(f_n^{-1}(\bi{z}_n^i))+\sqrt{\alpha_n}\bi{\xi}_n^i\right),
                \end{equation}
                where $C_n^{\bi{z}\mathcal{F}}$ and $C_n^{\mathcal{F}\mathcal{F}}$ are defined analogously to Equations \ref{eqn: xF cov} and \ref{eqn: FF cov} respectively in the NF latent space, and $\bi{\xi}_n^i\sim\mathcal{N}(0, \Gamma)$.
            \ENDFOR
        \STATE Fit the multivariate $t$-distribution, $t_{\nu_s}(\mu_s, \mathcal{C}_s)$ to the latent space particle ensemble $\{\bi{z}^{i}_{n+1}\}_{i=1}^J$ with an EM algorithm. Set component-wise autocorrelations $\hat{\rho}_0(j)=1\:\forall j$.
        \FOR{$m=1,\ldots,M$}
                \FOR{$i=1, \ldots , J$}
                \STATE Update particle state $\bi{z}_{n+1}^i$ using Algorithm \ref{alg: tpCN} in NF latent space.
            \ENDFOR
            \STATE $\log\rho \leftarrow \log\rho + (\langle\alpha\rangle-\alpha^\star)/m$
            \STATE $\bi{\mu}_s \leftarrow \bi{\mu}_s + (\langle\bi{z}_{n+1}\rangle - \bi{\mu_s}) / m$
            \STATE Calculate component-wise autocorrelations $\hat{\rho}_m(j)$.
            \IF{$\prod_{l=1}^m\hat{\rho}_l(j)<\tau_{\mathrm{corr}}\:\forall j$}
                \STATE End tpCN iterations.
            \ENDIF
      \ENDFOR
      \STATE Map particle ensemble back to the original data space $\{\bi{x}_{n+1}^i=f^{-1}_n(\bi{z}_{n+1}^i)\}_{i=1}^{J}$
      \STATE $n \leftarrow n+1$
    \ENDWHILE
   \STATE {\bfseries Output}: Converged particle ensemble $\{\bi{x}_{n^*}^i\}_{i=1}^J$.
\end{algorithmic}
\label{alg: nf SKT}
\end{algorithm}

The use of EKI as an adaptation step within an annealed sampling scheme has two core benefits. Compared to applying EKI directly to solving the Bayesian inverse problem, using it as an adaptation step for the tpCN sampler that preserves the target measure as its invariant measure, we are able to obtain low bias posterior moment estimates outside the linear, Gaussian setting where standard EKI can otherwise give highly biased results. Beyond allowing us to correct the errors in direct EKI and FAKI, the EKI adaptation significantly accelerates tpCN sampling. The EKI update distributes particles approximately according to the target measure. When used to fit the reference $t$-distribution for tpCN, we are able to more closely capture the target geometry. Coupled with NF preconditioning, we obtain a doubly preconditioned sampler, with the NF mapping us to an approximately Gaussian latent target space, and the $t$-preconditioning in tpCN giving improved performance in sampling any residual non-Gaussianity in the target.


For the tpCN implementations, both with the SKT adaptation and the benchmark SMC adaptation, we perform diminishing adaptation \cite{roberts2007diminish} of the tpCN step size $\rho$, and the reference $t$-distribution mean $\bi{\mu}_s$. For some sampling iteration $m$, the tpCN kernel parameters at iteration $m+1$ are given by
\begin{align}
    \log\rho^{m+1}&=\log\rho^m + \frac{\langle\alpha^m\rangle-\alpha^\star}{m},\\
    \bi{\mu}_s^{m+1}&=\bi{\mu}_s^{m}+\frac{\langle\bi{x}^m\rangle-\bi{\mu}_s^m}{m},
\end{align}
where $\langle\alpha^m\rangle$ is the mean tpCN acceptance probability at iteration $m$, $\alpha^\star$ is some target acceptance probability and $\langle\bi{x}^m\rangle$ is the mean of the particle ensemble at iteration $m$. Performing diminishing adaptation in this way helps to ensure the robust performance of the tpCN algorithm across all the adaptive sampling schemes, with similar adaptation previously being implemented in the \textsc{pocoMC} package for NF preconditioned SMC \cite{karamanis2022accelerating, karamanis2022_pocoMC}.

\section{Numerical Experiments}\label{sec: experiments}

In this section we present the results from three numerical experiments. In Section \ref{subsec: heat}, we study the recovery of an initial temperature field evolving under the heat equation. In Section \ref{subsec: gravity}, we study the recovery of an underlying density field from surface measurements of the gravitational field. Finally, in Section \ref{subsec: reaction diffusion}, we study the recovery of a source term from observations of a signal evolving under the reaction-diffusion equation.

We compare the performance of the adaptive SKT scheme against adaptation with importance resampling SMC, both with and without NF preconditioning. For the purposes of labelling the results from each adaptation algorithm we use the following acronyms:
\begin{enumerate}
    \item \textbf{SKT}: The SKT algorithm without NF preconditioning, analogous to Algorithm \ref{alg: nf SKT} without the NF steps.
    \item \textbf{NF-SKT}: The SKT algorithm with NF preconditioning, as described in Algorithm \ref{alg: nf SKT}.
    \item \textbf{SMC}: Importance resampling SMC without NF preconditioning, analogous to Algorithm \ref{alg: nf smc tpcn} without the NF steps.
    \item \textbf{NF-SMC}: Importance resampling SMC with NF preconditioning, as described in Algorithm \ref{alg: nf smc tpcn}.
\end{enumerate}
Alongside testing the performance of the adaptation algorithms for the tpCN sampler, we also provide results for standard pCN. When adapting the pCN sampler, we fit for the mean and covariance of the Gaussian base distribution using the empirical mean and covariance of the particle ensemble prior to sampling at each temperature level. Learning the NF maps at each temperature level took approximately 10 seconds for each of the experiments we consider here. This could likely be further improved by implementing e.g., early stopping based on the NF validation loss \cite{karamanis2022_pocoMC}. However, in general, we expect that learning each NF map will take of order seconds of wall time at each temperature level up to $\mathcal{O}(100)$ dimensions. For problems where the cost of a single forward model evaluation is comparable, the NF training cost becomes negligible compared to the need to apply repeated sampling updates at each temperature level. However, we find that obtaining high quality NF fits requires scaling both the number of particles and the complexity of the NF maps with dimension, which will quickly render NF training prohibitive as we move to $\mathcal{O}(10^3)$ dimensions.

To quantify the performance of the samplers we compare the squared bias on the estimated first and second moments of the target posterior, averaged over the target dimensions, which has previously been used in studying the rate of convergence of MCMC algorithms \cite{hoffman2019neutra, hoffman2022meads}. The dimension averaged squared bias, normalized by the posterior variance, on the estimate for some quantity $g(\bi{x})$ is given by
\begin{equation}
    \langle b_g^2\rangle = \left\langle\frac{\left(\mathbb{E}_{\beta=1}[g_k(\bi{x})]-\mathbb{E}_\pi[g_k(\bi{x})]\right)^2}{\sigma_{g,k}^2}\right\rangle_{k\in d},
\end{equation}
where $\mathbb{E}_{\beta=1}[g_k(\bi{x})]=J^{-1}\sum_{i=1}^J g_k(\bi{x}^i_{n^\star})$ is the mean of $g(\bi{x})$ for the dimension $k$, evaluated over the final particle ensemble $\{\bi{x}_{n^\star}^i\}_{i=1}^J$, $\mathbb{E}_\pi[g_k(\bi{x})]$ is the expectation value of $g(\bi{x})$ for the dimension $k$, evaluated with respect to the true target posterior, $\sigma_{g,k}^2$ is the true posterior variance of $g(\bi{x})$ for the dimension $k$, and $\langle \cdot\rangle_{k\in d}$ denotes the average over the dimensions. We estimate $\mathbb{E}_\pi[g_k(\bi{x})]$ and $\sigma_{g,k}^2$ for each problem from long runs of Hamiltonian Monte Carlo (HMC), using the No-U-Turn Sampler implementation in the \textsc{numpyro} library \cite{phan2019composable, bingham2019pyro}. We denote the dimension averaged squared bias on the first moment ($g_k(\bi{x})=\bi{x}_k$) as $\langle b_1^2\rangle$ and on the second moment ($g_k(\bi{x})=\bi{x}_k^2$) as $\langle b_2^2\rangle$, where $\bi{x}_k$ is the element of $\bi{x}$ corresponding to the dimension $k$.

Given $N$ independent samples from the posterior $\{\hat{\bi{x}}_i\}_{i=1}^N$, we have the estimator $\mathbb{E}_\pi[g_k(\bi{x})]=N^{-1}\sum_{i=1}^N g_k(\hat{\bi{x}}_i)$. Invoking the central limit theorem, the squared error on this estimator will be of the order $\sim\sigma_{g,k}^2/N$ \cite{hoffman2022meads}. Whilst we do not have independent samples from the posterior from HMC, we ensure that we run chains sufficiently long such that the estimated effective sample size (ESS) $\gtrsim 10^3$. These heuristics can also be used to define a regime for low bias where $\langle b_g^2\rangle<10^{-2}$, which corresponds approximately with a dimension averaged squared bias less than one hundredth of the posterior variance. 

In comparing the performance of each adaptive algorithm, we consider two sets of tests. In the first we run the SKT and NF-SKT algorithms with 10 tpCN (pCN) iterations at each temperature level, and the SMC and NF-SMC algorithms with 11 tpCN (pCN) iterations at each temperature level. The fixed computational budget at each temperature level allows for a more direct assessment of the performance of the EKI adaptation step. The additional tpCN (pCN) iteration for SMC and NF-SMC is to account for the additional set of forward model evaluations  used for the EKI updates in SKT and NF-SKT. We report results for ensemble sizes $J\in\{2d, 4d, 6d, 8d, 10d\}$, where $d$ is the target dimension of each model. In the second set of tests we compare the performance of each algorithm using an adaptive number of sampling iterations at each temperature level, with a correlation threshold of $\tau_\mathrm{corr}=0.1$ and an ensemble size $J=10d$. To avoid excessive computation, we set the maximum number of sampling iterations at each temperature level to 50 for SKT and NF-SKT, and 51 for SMC and NF-SMC. For the results presented in this section we use a target tpCN (pCN) acceptance rate of $\alpha^\star=0.234$, an initial tpCN (pCN) step size of $\rho=1$ and a fractional target ESS of $\tau=0.5$ when adapting the annealing schedule. Each algorithm is run over ten different random seeds to estimate the corresponding variation in performance. For completeness, we also provide corner plots comparing the converged particle ensembles obtained with each adaptation method against reference HMC samples in \ref{sec: corner plots}.

In \ref{subsec: faki and eki}, we demonstrate the performance of EKI and FAKI on our numerical benchmarks without embedding them as part of an annealed sampling scheme. Each of the numerical experiments we consider in this work show varying degrees of non-Gaussianity, which results in highly biased posterior inferences for EKI and FAKI. Whilst both methods converge on their final ensembles with $\sim 30-50$ embarrassingly parallel model evaluations, the highly biased posterior moment estimates mean such methods are unsuitable for many scientific inference tasks when applied alone.

In addition to the bias on posterior moment estimates, one may also be interested in the field reconstructions obtained with each method. In \ref{sec:field recon} we show the relevant field and source term reconstructions for each experiment obtained with each algorithm. For FAKI and EKI there are clear discrepancies between the reconstructed fields and source terms and the reference reconstruction from HMC. This is to be expected given the highly biased final ensembles obtained with these methods. For each of the adaptive sampling schemes considered in this work the qualitative reconstruction of the initial fields and source terms was largely comparable to the recovery with HMC, both applied to tpCN and pCN. However, the ability of each method to obtain comparable field and source term recoveries does not fully reflect the ability of the various algorithms in accurately approximating marginal posterior moments, which are the key object of study for many scientific inference tasks.

\subsection{Heat Equation}\label{subsec: heat}

The heat equation is a partial differential equation (PDE) describing the evolution of some field $u(\bi{x}, t)$ over time. For our experiment we consider the case of a two-dimensional temperature field evolving according to
\begin{equation}
    \frac{\partial u(\bi{x}, t)}{\partial t}=D\nabla^2 u(\bi{x}, t)=D\left(\frac{\partial^2 u(\bi{x}, t)}{\partial x^2_1}+\frac{\partial^2 u(\bi{x}, t)}{\partial x^2_2}\right),
    \label{eqn: heat equation}
\end{equation}
where we set the thermal diffusivity constant $D=0.5$. The temperature field is taken to be on a square plate, with the length of a side set to $L=10$. We impose Dirichlet boundary conditions on the domain $\Omega\subset\mathbb{R}^2$, such that $u(\bi{x}, t)=0, \forall\bi{x}\in\partial\Omega$. The forward model consists in solving Equation \ref{eqn: heat equation} for the evolution of some initial temperature field $u(\bi{x}, t=0)$ up to a time $t_f=1$.  We solve the heat equation using the forward time centered space (FTCS) method \cite{anderson1988cfd} on a $64\times 64$ grid, with 1000 time steps.

\begin{figure}[ht]
    \centering
    \includegraphics[width=\textwidth]{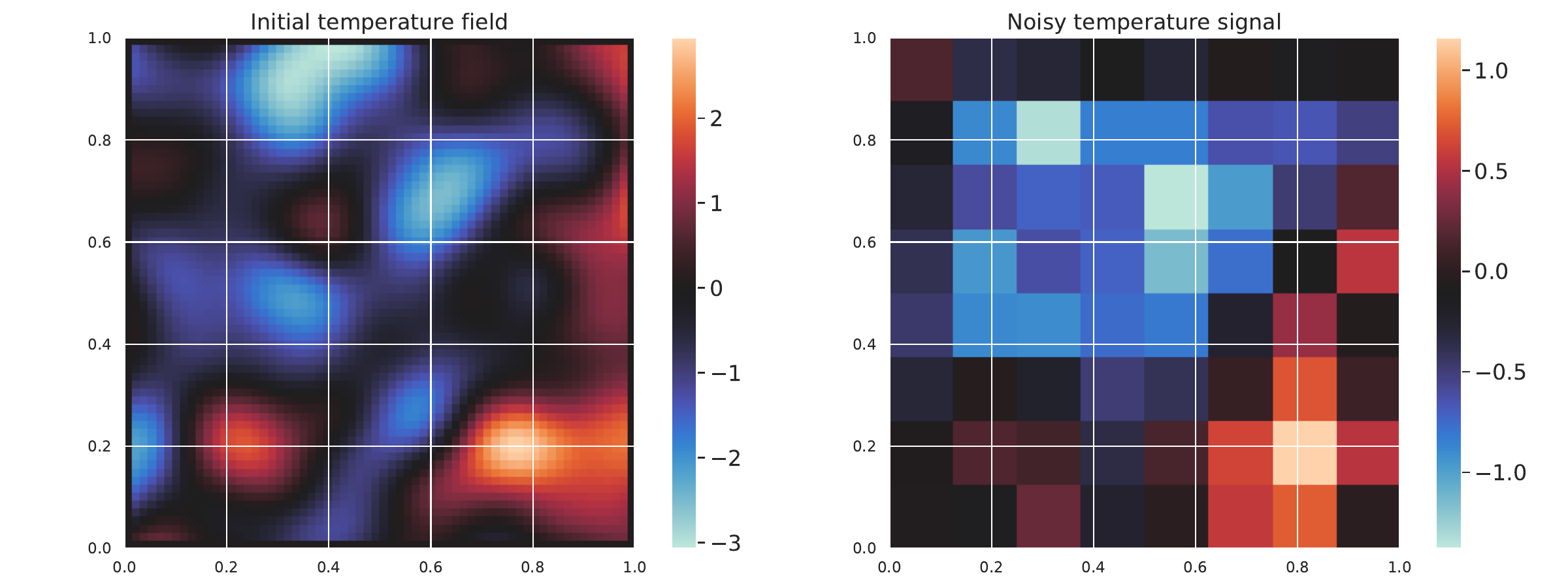}
    \caption{The initial temperature field $u(\bi{x}, t=0)$ (left panel), shown alongside the low resolution observed temperature field at time $t_f=1$ (right panel).}
    \label{fig:heat initial signal field}
\end{figure}

For our simulated data, we consider the situation where measurements of the temperature field are made at time $t_f$ on a low resolution $8\times 8$ grid, with the signal in a low resolution grid pixel being the average of the temperature signal from the $64\times 64$ grid pixels contained within it. The observation noise was taken to be independent in each pixel, with a Gaussian noise standard deviation of $\sigma_\eta=0.2$. The true initial temperature field was generated from the Karhunen-Loeve (KL) expansion of a Gaussian random field (GRF) with a squared exponential covariance kernel
\begin{equation}
    C(\bi{x}, \bi{x}^\prime) = \exp\left(-\frac{\norm{\bi{x}-\bi{x}^\prime}_2^2}{2\ell^2}\right),
    \label{eqn:sq expon kernel}
\end{equation}
where $\ell$ is the GRF length scale. The KL expansion, up to some order $R$, for the GRF is given by
\begin{equation}
    u(\bi{x}, t=0) = \mu_K+\sigma_K\sum_{k=1}^R \sqrt{\lambda_k}\phi_k(\bi{x})\theta_k,
    \label{eqn: heat KL}
\end{equation}
where $\mu_K$ is the GRF mean, $\sigma^2_K$ is the GRF variance, $\{\lambda_k\}_{k=1}^R$ is a sequence of strictly decreasing, real and positive eigenvalues for the covariance kernel in Equation \ref{eqn:sq expon kernel}, $\{\phi_k(\bi{x})\}_{k=1}^R$ are the set of corresponding eigenfunctions of the covariance kernel, and $\{\theta_k\sim\mathcal{N}(0, 1)\}_{k=1}^R$ are a set of standard Gaussian random variables. To generate the true field for this numerical study we set $\mu=0$, $\sigma^2=1$, $\ell=0.1$ and generate $R=200$ standard Gaussian random variables $\{\theta_k\}_{k=1}^{200}$. The simulated data are then generated by solving for the time evolution of $u(\bi{x}, t)$ up to time $t_f=1$, averaging the signal onto the low resolution grid and adding Gaussian noise realizations to each pixel. The true initial temperature field and the low resolution observed field are shown in Figure \ref{fig:heat initial signal field}.

For the test model, we consider recovering the $R=100$ leading modes of the KL expansion, along with the thermal diffusivity constant. Defining $\bi{\theta}=(\theta_1,\ldots,\theta_{100})^\intercal$, the full model is given by
\begin{align}
    D&\sim|\mathcal{N}(\mu=0, \sigma^2=0.5^2)|,\\
    \mu_K &\sim \mathcal{N}(\mu=0, \sigma=0.1),\\
    \sigma_K &\sim |\mathcal{N}(\mu=0, \sigma^2=1.0)|,\\
    \bi{\theta} &\sim \mathcal{N}(0, I_{100}),\\
    \bi{y} &\sim\mathcal{N}(F_H(D,\mu_K,\sigma_K,\bi{\theta}), \sigma_\eta^2 I_{64}),
\end{align}
where $|\mathcal{N}(\mu=0, \sigma^2)|$ is the Half-Normal distribution with scale $\sigma$ and $F_H(D,\mu_K,\sigma_K,\bi{\theta})$ denotes the forward model for the heat equation, mapping from the initial temperature field to the low resolution observations $\bi{y}$ at time $t_f=1$. For performing inference we apply log-transformations to $D$ and $\sigma_K$ to map the all the parameters to an unconstrained space, modifying the the target distribution with the corresponding Jacobian factors. The target dimension for this problem is $d=103$.

\begin{figure}[ht]
    \centering
    \includegraphics[width=\textwidth]{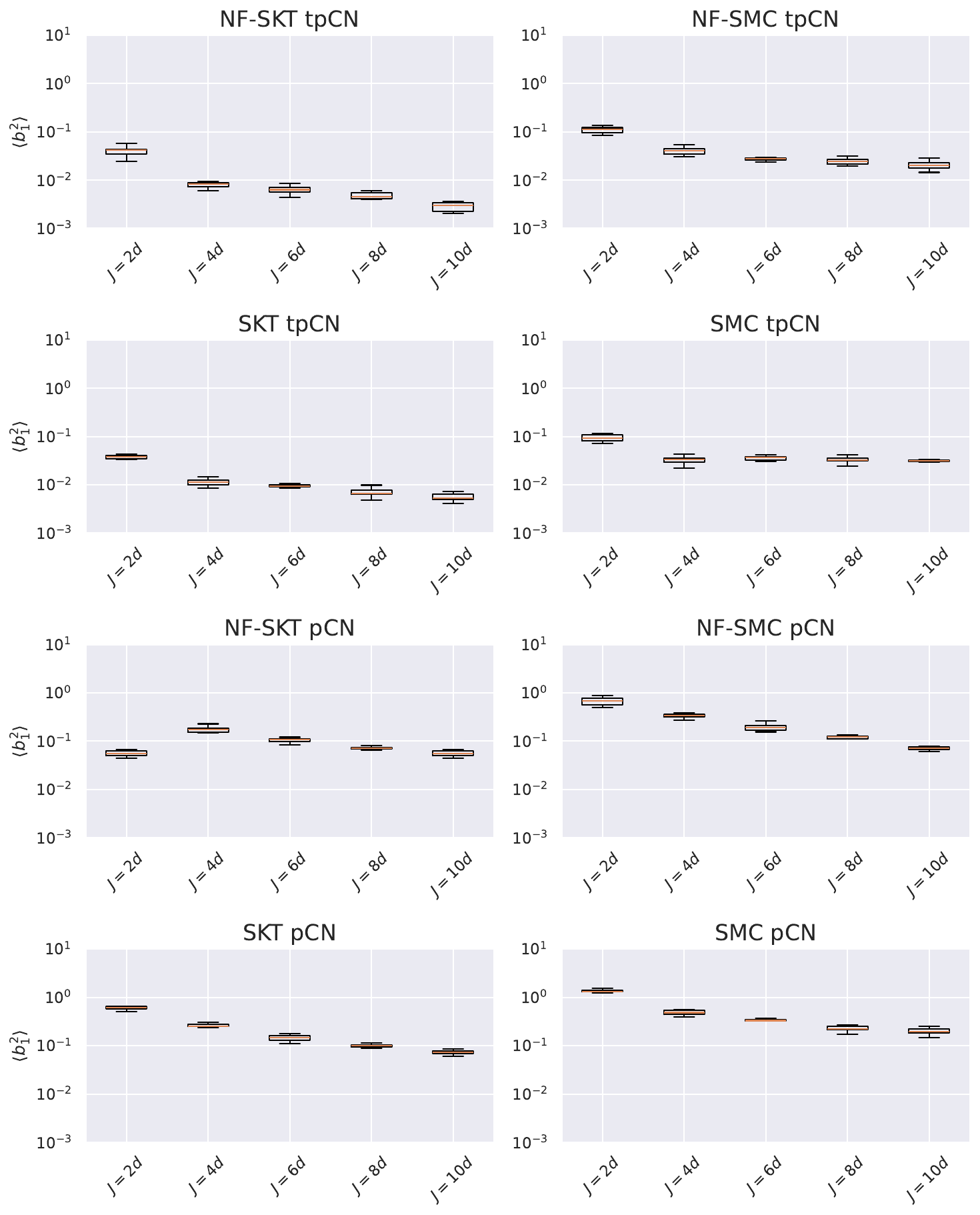}
    \caption{Final dimension averaged squared bias on the first moment for the heat equation model, plotted against the ensemble size $J$ expressed as a multiple of the target dimension $d=103$. Results are shown for each adaptation algorithm applied to the tpCN and pCN samplers.}
    \label{fig:heat b1 10tpCN}
\end{figure}
\begin{figure}[ht]
    \centering
    \includegraphics[width=\textwidth]{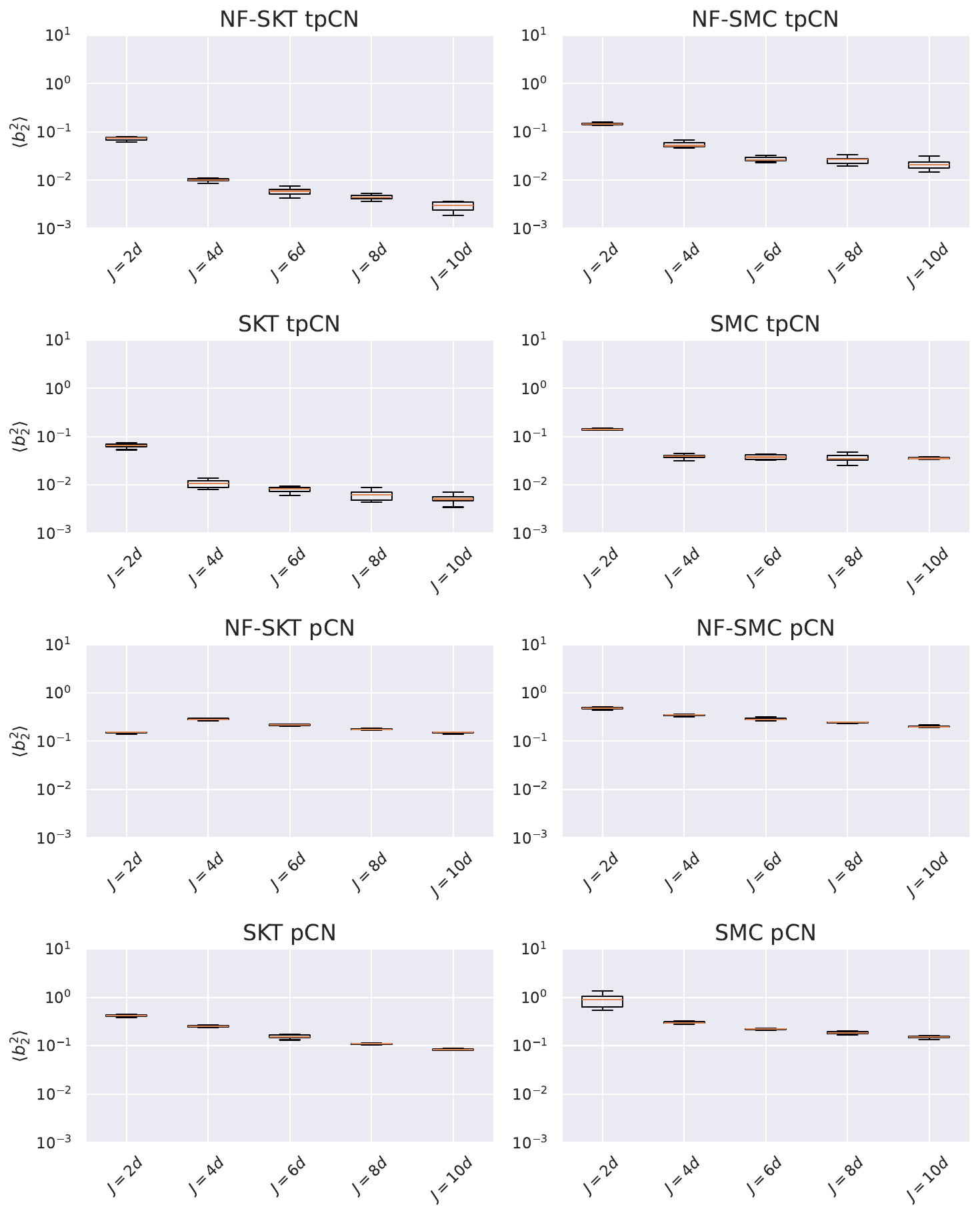}
    \caption{Final dimension averaged squared bias on the second moment for the heat equation model, plotted against the ensemble size $J$ expressed as a multiple of the target dimension $d=103$. Results are shown for each adaptation algorithm applied to the tpCN and pCN samplers.}
    \label{fig:heat b2 10tpCN}
\end{figure}
The first set of results are provided when running each adaptive algorithm with a fixed computational budget at each temperature level. In Figure \ref{fig:heat b1 10tpCN} we show the results for $\langle b_1^2\rangle$ obtained with the final particle ensembles for each algorithm, with the box plots showing the variation over the 10 runs. Similarly, Figure \ref{fig:heat b2 10tpCN} shows the results for $\langle b_2^2\rangle$ obtained for each algorithm with a fixed computational budget at each temperature level. The mean and standard deviation of $\langle b_1^2\rangle$ and $\langle b_2^2\rangle$ for each algorithm over their 10 runs are reported in Table \ref{tab:heat eqn tpCN bias table} for tpCN and in Table \ref{tab:heat eqn pCN bias table} for pCN, alongside the number of temperature levels, $N_\beta$ used by each algorithm. For the second set of results, where the number of sampling iterations at each temperature level is chosen by monitoring sample autocorrelations, we report the mean and standard deviation of $\langle b_1^2\rangle$ and $\langle b_2^2\rangle$ for each algorithm over their 10 runs in Table \ref{tab: heat eqn rho0p1 bias}, alongside the number of forward model evaluations divided by the ensemble size $N_\mathrm{eval}/J$, which corresponds to the number of embarrassingly parallel forward model evaluations. 

Beginning with comparisons where we use a fixed computational budget at each temperature level, we find that tpCN achieves lower squared bias on the first and second moments compared to pCN. Given the same adaptive SKT and SMC schemes, the greater flexibility in the base $t$-distribution means tpCN is able to better adapt to the non-Gaussian targets, resulting in more rapid convergence to the target at each temperature level. 

Comparing SKT and SMC adaptation schemes, we can see that SKT obtains lower values for $\langle b_1^2\rangle$ and $\langle b_2^2\rangle$ for all ensemble sizes. The number of temperature levels used by both algorithms is comparable. It is apparent from these tests using SKT adaptation we are able to converge more rapidly at each temperature level, indicating that the EKI update provides a better initialization for the tpCN updates than the importance resampling used in SMC. The EKI update also helps to provide improved preconditioning for the tpCN updates, with the $t$-distribution being fitted to the annealed target particle approximation obtained via the EKI update.

A similar pattern is observed when comparing NF-SKT and NF-SMC. In comparing each with SKT and SMC respectively, we see that the improvement from NF preconditioning becomes more pronounced as the ensemble size is increased. For $J=2d$ there are not enough particles for the NF to learn a useful map between between the original data space and a Gaussian latent space. Indeed, in this regime the NF can degrade performance by failing to map to a latent space where the target is effectively Gaussianized, and in the case of NF-SKT introducing additional non-linearity in the forward model evaluation for the EKI update. For larger ensemble sizes ($J\geq 6d$) the use of NF transformations reduces the final bias for both the NF-SKT and NF-SMC algorithms. The effect is more pronounced for NF-SKT, where the NF acts to both relax the Gaussian ansatz of EKI and provide nonlinear preconditioning for the tpCN updates. However, it is worth noting that SKT was able to achieve low bias without NF preconditioning, demonstrating the potential of the EKI update as an adaptation step for tpCN by both initializing and preconditioning the sampling updates within an annealing scheme.
\begin{table}[ht]
    \centering
    \begin{tabular}{c|c|c|c|c} 
    \hhline{= = = = =}
         Algorithm&   $J$&  $N_\beta$ &$\langle b_1^2\rangle$ &$\langle b_2^2\rangle$ \\ 
         \hhline{=|=|=|=|=}
         NF-SKT tpCN& 
     $10d$&  $15.0\pm 0.0$&$0.0029\pm 0.0006$&$0.0029\pm 0.0007$\\ 
 NF-SKT tpCN& $8d$& $15.0\pm 0.0$& $0.0048\pm 0.0008$&$0.0045\pm 0.0005$\\ 
 NF-SKT tpCN& $6d$& $15.0\pm 0.0$& $0.0065\pm 0.001$&$0.0060\pm 0.001$\\ 
 NF-SKT tpCN& $4d$& $15.0\pm 0.0$& $0.0084\pm 0.001$&$0.010\pm 0.007$\\ 
 NF-SKT tpCN& $2d$& $14.8\pm 0.4$& $0.040\pm 0.009$&$0.072\pm 0.006$\\ 
 \hhline{=|=|=|=|=}
 SKT tpCN& $10d$& $15.0\pm 0.0$& $0.0056\pm 0.001$&$0.0053\pm 0.001$\\ 
 SKT tpCN& $8d$& $15.0\pm 0.0$& $0.0071\pm 0.001$&$0.0062\pm 0.001$\\ 
 SKT tpCN& $6d$& $15.0\pm 0.0$& $0.0094\pm 0.001$&$0.0080\pm 0.001$\\ 
 SKT tpCN& $4d$& $15.0\pm 0.0$& $0.011\pm 0.002$&$0.011\pm 0.002$\\ 
 SKT tpCN& $2d$& $15.0\pm0.0$& $0.038\pm 0.003$&$0.064\pm 0.007$\\ 
 \hhline{=|=|=|=|=}
 NF-SMC tpCN& $10d$& $15.2\pm 0.4$& $0.021\pm 0.004$&$0.021\pm 0.005$\\ 
 NF-SMC tpCN& $8d$& $15.7\pm 0.5$& $0.025\pm 0.004$&$0.026\pm 0.004$\\ 
 NF-SMC tpCN& $6d$& $15.4\pm 0.5$& $0.029\pm 0.005$&$0.028\pm 0.005$\\ 
 NF-SMC tpCN& $4d$& $15.5\pm 0.5$& $0.041\pm 0.007$&$0.055\pm 0.008$\\ 
 NF-SMC tpCN& $2d$&$14.9\pm 0.3$ & $0.11\pm 0.02$&$0.15\pm 0.009$\\ 
 \hhline{=|=|=|=|=}
 SMC tpCN& $10d$& $15.3\pm 0.5$& $0.032\pm 0.004$&$0.036\pm 0.004$\\ 
 SMC tpCN& $8d$&$15.5\pm 0.5$& $0.034\pm 0.005$&$0.037\pm 0.007$\\ 
 SMC tpCN& $6d$& $15.7\pm 0.5$& $0.036\pm 0.004$&$0.038\pm 0.004$\\ 
 SMC tpCN& $4d$& $15.6\pm 0.5$& $0.033\pm 0.006$&$0.039\pm 0.004$\\ 
 SMC tpCN& $2d$& $15.1\pm 0.3$& $0.094\pm 0.02$&$0.14\pm 0.009$\\ 
 \hhline{= = = = =}\end{tabular}
    \caption{Results for the number of temperature levels used by each algorithm $N_\beta$ and squared bias results, $\langle b_1^2\rangle$ and $\langle b_2^2\rangle$, obtained with the final particle ensemble for each algorithm when performing inference on the heat equation example, adapting the tpCN sampler. We report the mean and standard deviation for each statistic over the 10 algorithm runs, and show results for each of the tested ensemble sizes $J$. The number of parallelized model evaluations is given by $11N_\beta$ for each algorithm, with the total number of model evaluations being given by $11JN_\beta$. The target dimension is $d=103$.}
    \label{tab:heat eqn tpCN bias table}
\end{table}
\begin{table}[ht]
    \centering
    \begin{tabular}{c|c|c|c|c} 
    \hhline{= = = = =}
         Algorithm&   $J$&  $N_\beta$&$\langle b_1^2\rangle$&$\langle b_2^2\rangle$ \\ 
         \hhline{=|=|=|=|=}
         NF-SKT pCN& 
     $10d$&  $13.2\pm 0.4$&$0.056\pm 0.008$&$0.15\pm 0.006$\\ 
 NF-SKT pCN& $8d$& $13.0\pm 0.0$& $0.072\pm 0.006$&$0.18\pm 0.0004$\\ 
 NF-SKT pCN& $6d$& $13.0\pm 0.0$& $0.10\pm 0.01$&$0.21\pm 0.007$\\ 
 NF-SKT pCN& $4d$& $13.0\pm 0.0$& $0.18\pm 0.02$&$0.28\pm 0.01$\\ 
 NF-SKT pCN& $2d$& $13.2\pm 0.4$& $0.056\pm 0.008$&$0.15\pm 0.006$\\ 
 \hhline{=|=|=|=|=}
 SKT pCN& $10d$& $15.9\pm 0.3$& $0.074\pm 0.008$&$0.083\pm 0.005$\\ 
 SKT pCN& $8d$& $15.9\pm 0.3$& $0.097\pm 0.001$&$0.11\pm 0.006$\\ 
 SKT pCN& $6d$& $15.9\pm 0.3$& $0.15\pm 0.02$&$0.16\pm 0.01$\\ 
 SKT pCN& $4d$& $15.7\pm 0.5$& $0.26\pm 0.02$&$0.25\pm 0.01$\\ 
 SKT pCN& $2d$& $15.2\pm 0.4$& $0.61\pm 0.05$&$0.41\pm 0.02$\\ 
 \hhline{=|=|=|=|=}
 NF-SMC pCN& $10d$& $15.0\pm 0.0$& $0.073\pm 0.009$&$0.20\pm 0.006$\\ 
 NF-SMC pCN& $8d$& $15.1\pm 0.3$& $0.12\pm 0.01$&$0.24\pm 0.006$\\ 
 NF-SMC pCN& $6d$& $15.9\pm 0.3$& $0.20\pm 0.01$&$0.29\pm 0.02$\\ 
 NF-SMC pCN& $4d$& $17.4\pm 0.5$& $0.33\pm 0.03$&$0.34\pm 0.02$\\ 
 NF-SMC pCN& $2d$& $20.6\pm 0.4$& $0.67\pm 0.12$&$0.48\pm 0.03$\\ 
 \hhline{=|=|=|=|=}
 SMC pCN& $10d$& $18.2\pm 0.4$& $0.21\pm 0.03$&$0.15\pm 0.008$\\ 
 SMC pCN& $8d$&  $19.2\pm 0.4$& $0.23\pm 0.03$&$0.19\pm 0.01$\\ 
 SMC pCN& $6d$& $17.4\pm 0.5$& $0.33\pm 0.04$&$0.22\pm 0.01$\\ 
 SMC pCN& $4d$& $21.2\pm 0.4$& $0.48\pm 0.06$&$0.30\pm 0.01$\\ 
 SMC pCN& $2d$& $14.6\pm 0.5$& $1.36\pm 0.16$&$0.88\pm 0.26$\\ 
 \hhline{= = = = =}\end{tabular}
    \caption{Results for the number of temperature levels used by each algorithm $N_\beta$ and squared bias results, $\langle b_1^2\rangle$ and $\langle b_2^2\rangle$, obtained with the final particle ensemble for each algorithm when performing inference on the heat equation example, adapting the pCN sampler. We report the mean and standard deviation for each statistic over the 10 algorithm runs, and show results for each of the tested ensemble sizes $J$. The number of parallelized model evaluations is given by $11N_\beta$ for each algorithm, with the total number of model evaluations being given by $11JN_\beta$. The target dimension is $d=103$.}
    \label{tab:heat eqn pCN bias table}
\end{table}

When we select the number of sampling iterations at each temperature level based on the first order autocorrelations, we see that for the NF-SKT and SKT adaptation schemes, the tpCN sampler is able to reach the low bias regime ($\langle b_g^2\rangle < 10^{-2}$). This is not the case for the NF-SMC and SMC samplers. From the tests using a fixed computational budget at each temperature level, we expect these adaptation schemes to require more tpCN iterations at each temperature level. Despite this, the tpCN sampling updates are terminated earlier for the NF-SMC and SMC adaptation schemes, indicating that a lower value of $\tau_\mathrm{corr}$ is required. For pCN, we do not reach the low bias regime for all adaptation schemes, again indicating a more stringent requirement on the number of sampling iterations is necessary.
\begin{table}
    \centering
    \begin{tabular}{c|c|c|c}
    \hhline{= = = =}
         Algorithm ($\tau_\mathrm{corr}=0.1$)&  $N_\mathrm{eval} / J$&  $\langle b_1^2 \rangle$& $\langle b_2^2 \rangle$\\
         \hhline{=|=|=|=}
         NF-SKT tpCN&  $210\pm 140$&  $0.0029\pm 0.0005$& $0.0028\pm 0.0004$\\
         NF-SKT pCN&  $320\pm 210$&  $0.056\pm 0.008$& $0.15\pm 0.006$\\
         \hhline{=|=|=|=}
         SKT tpCN&  $390\pm 180$&  $0.0031\pm 0.0007$& $0.0034\pm 0.0009$\\
         SKT pCN&  $790\pm 24$&  $0.069\pm 0.008$& $0.079\pm0.005$\\
         \hhline{=|=|=|=}
         NF-SMC tpCN&  $165\pm 31$&  $0.030\pm 0.005$& $0.033\pm 0.006$\\
         NF-SMC pCN&  $170\pm 22$&  $0.093\pm 0.012$& $0.19\pm 0.009$\\
         \hhline{=|=|=|=}
         SMC tpCN&  $180\pm 22$&  $0.039\pm 0.003$& $0.045\pm 0.004$\\
         SMC pCN&  $930\pm 24$&  $0.19\pm 0.03$& $0.15\pm 0.0007$\\
         \hhline{=|=|=|=}
    \end{tabular}
    \caption{Results for the number of embarrassingly parallel model evaluations $N_\mathrm{eval}/J$, and squared bias results, $\langle b_1^2\rangle$ and $\langle b_2^2\rangle$, obtained with the final particle ensemble for each algorithm when adapting the number of sampling iterations at each temperature level using $\tau_\mathrm{corr}=0.1$, as applied to the heat equation example. We show results when adapting both the tpCN and pCN samplers with an ensemble size of $J=10d$, where the target dimension is $d=103$.}
    \label{tab: heat eqn rho0p1 bias}
\end{table}

\subsection{Gravity Survey}\label{subsec: gravity}

For this problem we adapt the two-dimensional gravity surveying problem presented in \cite{lykkegaard2023mlda}. We have some mass density field $\varrho(\bi{x})$, located at a depth $\delta$ from the surface at which measurements of the vertical component of the gravitational field are made. The vertical component of the gravitational field at some point $\bi{s}$ at the surface is given by
\begin{equation}
    \zeta(\bi{s})=\iint_X \frac{\delta}{\norm{\bi{s}-\bi{x}}_2^3}\varrho(\bi{x})\mathrm{d}\bi{x},
    \label{eqn: gravity integral}
\end{equation}
where $X=[0, 1]^2$ is the domain $\varrho(\bi{x})$. The forward model therefore consists in solving the integral in Equation \ref{eqn: gravity integral}. We follow \cite{lykkegaard2023mlda} in evaluating this integral using midpoint quadrature. Using $Q$ quadrature points along each dimension, the integral expression becomes
\begin{equation}
    \zeta(\bi{s}_i) = \sum_{l=1}^Q\omega_l\sum_{k=1}^{Q}\omega_k\frac{\delta}{\norm{\bi{s}_i-\bi{x}_{k,l}}_2^3}\hat{\varrho}(\bi{x}_{k,l})=\sum_{j=1}^{Q^2}\omega_j\frac{\delta}{\norm{\bi{s}_i-\bi{x}_{j}}_2^3}\hat{\varrho}(\bi{x}_{j}),
    \label{eqn: gravity fwd}
\end{equation}
where $\omega_j=1/Q^2, \forall j$ are the quadrature weights, $\hat{\varrho}(\bi{x}_{j})$ is the approximate subsurface density at the quadrature point $\bi{x}_j$, and $\zeta(\bi{s}_i$) is the vertical component of the gravitational field at the collocation point on the surface $\bi{s}_i, i\in\{1,\ldots,N^2\}$. 

\begin{figure}[ht]
    \centering
    \includegraphics[width=\textwidth]{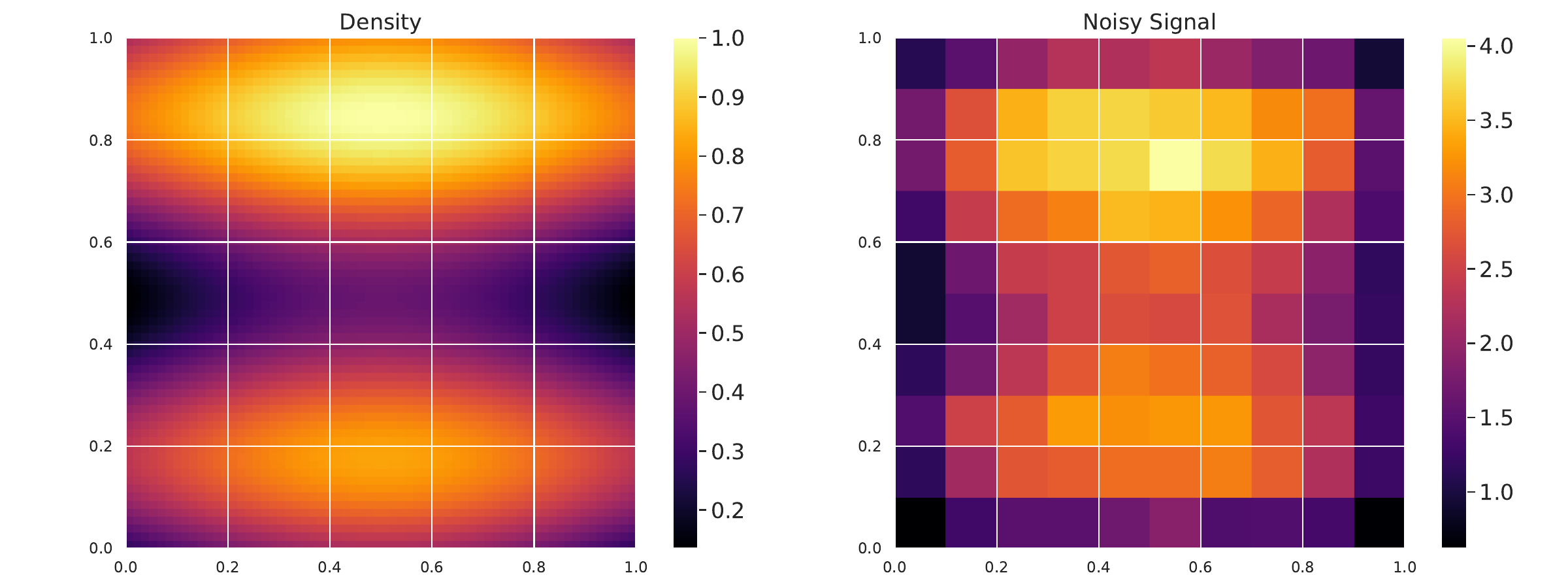}
    \caption{The true subsurface mass density field $\varrho(\bi{x})$ (left panel), shown alongside the low resolution measurements of the gravitational field at the surface $\zeta(\bi{x})$ (right panel).}
    \label{fig:gravity density signal field}
\end{figure}

The simulated data was obtained by generating a ground truth subsurface density field with profile given by
\begin{equation}
    \varrho(\bi{x})\propto \sin(\pi x_1)+\sin(3\pi x_2) + x_2 + 1,\quad x_1,x_2\in [0, 1]
\end{equation}
normalized to have a maximum value of 1. This signal was projected onto a $64\times 64$ grid. The surface signal was evaluated using Equation \ref{eqn: gravity fwd} on a $10\times 10$ grid, with Gaussian white noise with standard deviation $\sigma_{\eta}=0.1$ being added to each surface pixel to give the simulated data. The true subsurface mass density, and the corresponding surface gravitational field measurements used in this example are shown in Figure \ref{fig:gravity density signal field}.

For the inference task we model the subsurface density as a GRF with a Mat\'ern  3/2 covariance kernel,
\begin{equation}
    C(\bi{x}, \bi{x}^\prime)=\left(1+\frac{\sqrt{3}\norm{\bi{x}-\bi{x}^\prime}_2}{\ell}\right)\exp\left(-\frac{\sqrt{3}\norm{\bi{x}-\bi{x}^\prime}_2}{\ell}\right),
    \label{eqn: matern 3/2 cov}
\end{equation}
where $\ell$ is the correlation length scale. The subsurface density field is parameterized using a KL expansion of the $R=60$ leading eigenmodes,
\begin{equation}
\varrho(\bi{x})=\mu_K+\sigma_K\sum_{k=1}^{R=60}\sqrt{\lambda_k}\phi_k(\bi{x})\theta_k,
\label{eqn: density KL}
\end{equation}
where $\mu_K$ and $\sigma_K^2$ are the field mean and variance respectively, $\{\lambda_k\}_{k=1}^{R=60}$ is a sequence of strictly decreasing, real and positive eigenvalues of the covariance kernel in Equation \ref{eqn: matern 3/2 cov}, $\phi_k(\bi{x})$ are the corresponding eigenfunctions of the covariance kernel and $\{\theta_k\sim\mathcal{N}(0, 1)\}_{k=1}^{R=60}$ are a set of standard Gaussian random variables. Defining $\bi{\theta}=(\theta_1,\ldots,\theta_{60})^\intercal$, the full model for this example is given by
\begin{align}
    \mu_K &\sim\mathcal{N}(\mu=0, \sigma^2=1^2),\\
    \sigma_K &\sim |\mathcal{N}(\mu=0, \sigma^2=0.2^2)|,\\
    \bi{\theta} &\sim\mathcal{N}(0, I_{60}),\\
    \bi{y}&\sim\mathcal{N}(F_\zeta(\mu_K, \sigma_K, \bi{\theta}), \sigma_\eta^2 I_{100})
\end{align}
where $F_\zeta(\mu_K, \sigma_K, \bi{\theta})$ denotes the full gravity survey forward model, mapping from the subsurface mass density field to the low resolution surface measurements of the gravitational field $\bi{y}$. When performing inference a log-transformation is applied to $\sigma_K$ such that all parameters are in an unconstrained space, with the target distribution being modified by the corresponding Jacobian. The target dimension for this problem is $d=62$.

\begin{figure}[ht]
    \centering
    \includegraphics[width=\textwidth]{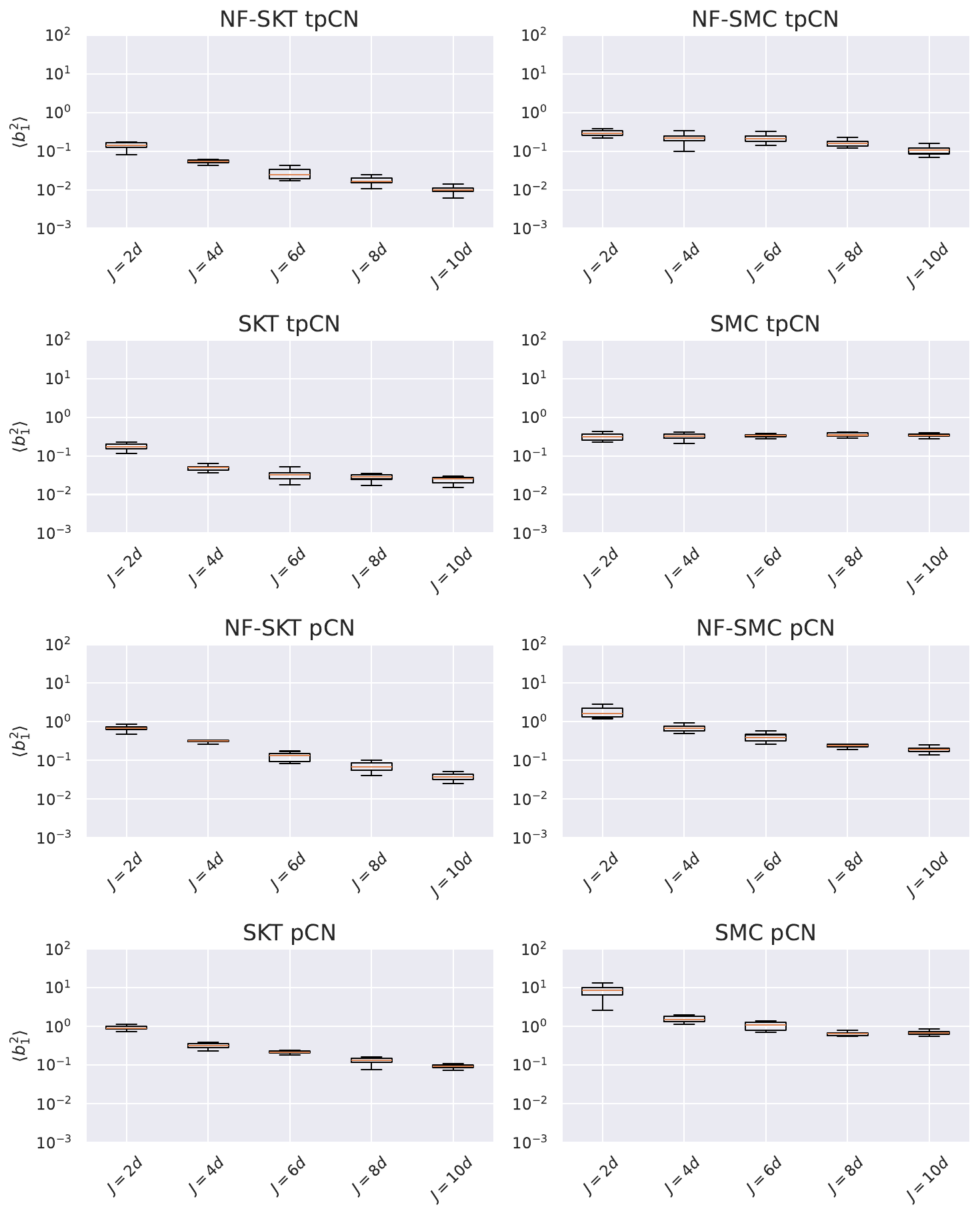}
    \caption{Final dimension averaged squared bias on the first moment for the gravity survey model, plotted against the ensemble size $J$ expressed as a multiple of the target dimension $d=62$. Results are shown for each adaptation algorithm applied to the tpCN and pCN samplers.}
    \label{fig:gravity b1 10tpCN}
\end{figure}

\begin{figure}[ht]
    \centering
    \includegraphics[width=\textwidth]{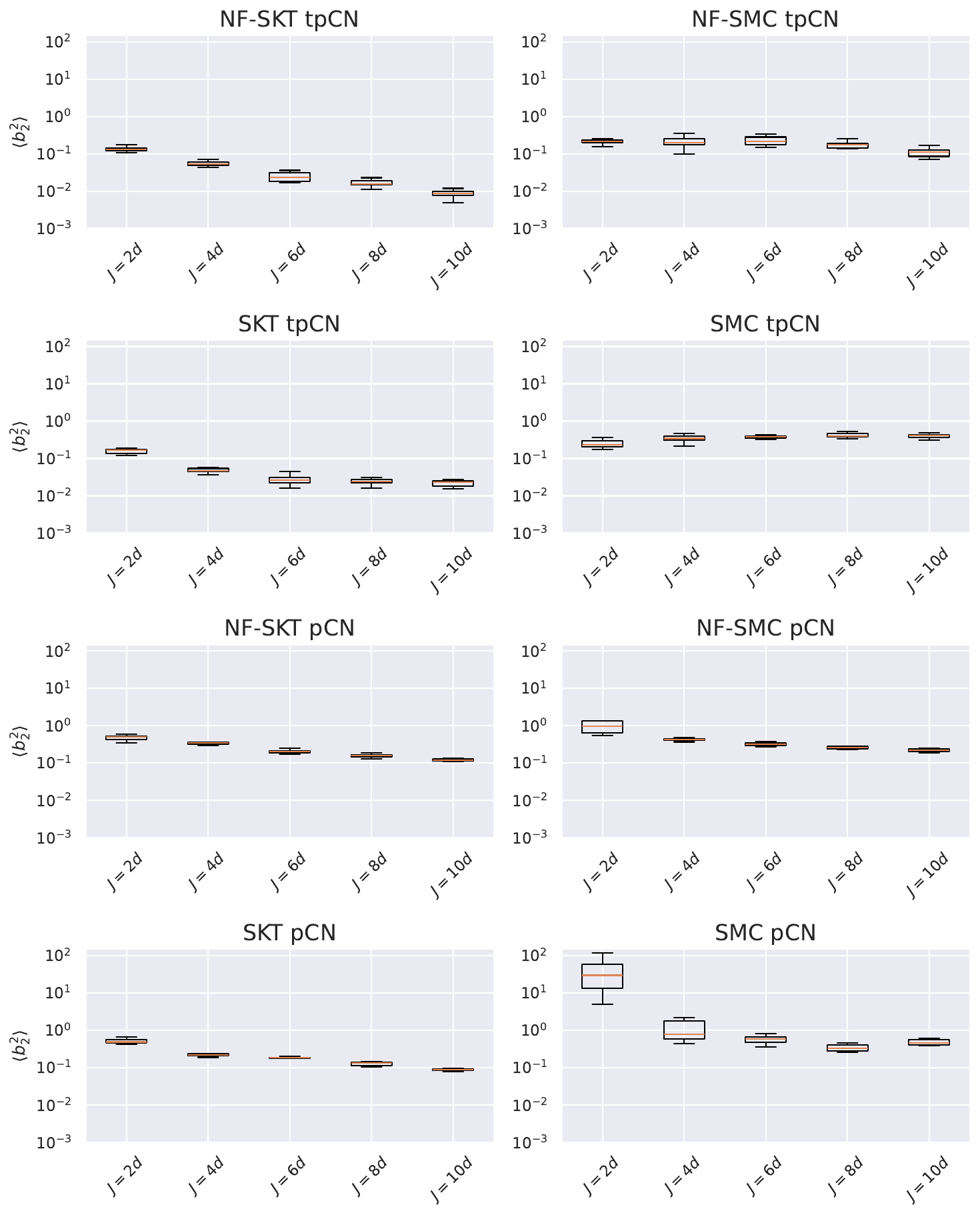}
    \caption{Final dimension averaged squared bias on the second moment for the gravity survey model, plotted against the ensemble size $J$ expressed as a multiple of the target dimension $d=62$. Results are shown for each adaptation algorithm applied to the tpCN and pCN samplers.}
    \label{fig:gravity b2 10tpCN}
\end{figure}

We start again with tests where we enforce a fixed computational budget at each temperature level. In Figures \ref{fig:gravity b1 10tpCN} and \ref{fig:gravity b2 10tpCN} we show the recovered estimates for $\langle b_1^2\rangle$ and $\langle b_2^2\rangle$ respectively, with box plots again showing the variation over the 10 runs for each algorithm and ensemble size, using a fixed computational budget at each temperature level. The mean and standard deviation for $\langle b_1^2\rangle$ and $\langle b_2^2\rangle$, along with the number of temperature levels used by each each algorithm over the 10 runs are reported in Table \ref{tab:gravity tpCN bias table} for tpCN and Table \ref{tab:gravity pCN bias table} for pCN. We report the mean and standard deviation for $\langle b_1^2\rangle$ and $\langle b_2^2\rangle$, along with the number of embarrassingly parallel model evaluations $N_\mathrm{eval} / J$, for the second set of tests, where we select the number of sampling iterations adaptively, in Table \ref{tab: gravity rho0p1 bias}.

Given a fixed computational budget at each temperature level, we see again that tpCN is able to achieve a lower squared bias on the final ensembles for all adaptation methods. For SMC applied to tpCN, we find that the values for $\langle b_1^2\rangle$ and $\langle b_2^2\rangle$ are high and largely independent of the ensemble size. In this case, SMC requires significantly more tpCN iterations at each temperature level in order to correctly distribute the particle ensemble after the importance resampling step. In comparison, SKT is able to achieve a lower bias with the same computational budget being used at each temperature level. The particle ensemble obtained by the EKI update provides a better initialization and preconditioner for the tpCN updates compared to the importance resampled particle ensemble, achieving lower bias with fewer model evaluations. 

A similar pattern is again observed when comparing the NF-SKT and NF-SMC algorithms. For larger ensemble sizes the NF is able to map the effective prior at each temperature level to a Gaussian latent space, where the target is approximately Gaussian. For the same computational budget at each temperature level, the NF preconditioning more rapidly distributes particles according to the given target, with the low bias threshold being reached for an ensemble size of $J=10d$ for the NF-SKT algorithm. For smaller ensemble sizes, the NF is unable to learn useful non-Gaussian features in the geometry of the effective prior, meaning we do not obtain an improvement from NF preconditioning.
    \begin{table}[ht]
     \centering
    \begin{tabular}{c|c|c|c|c} 
    \hhline{= = = = =}
         Algorithm&   $J$&  $N_\beta$&$\langle b_1^2\rangle$&$\langle b_2^2\rangle$\\ 
         \hhline{=|=|=|=|=}
         NF-SKT tpCN& 
     $10d$&  $22.6\pm 0.5$&$0.0098\pm 0.0024$&$0.0089\pm 0.0025$\\ 
 NF-SKT tpCN& $8d$& $23.0\pm 0.0$& $0.017\pm 0.004$&$0.017\pm 0.003$\\ 
 NF-SKT tpCN& $6d$& $23.0\pm 0.0$& $0.027\pm 0.009$&$0.025\pm 0.007$\\ 
 NF-SKT tpCN& $4d$& $23.0\pm 0.0$& $0.058\pm 0.011$&$0.056\pm 0.008$\\ 
 NF-SKT tpCN& $2d$& $23.2\pm 0.4$& $0.14\pm 0.03$&$0.13\pm 0.02$\\ 
 \hhline{=|=|=|=|=}
 SKT tpCN& $10d$& $23.1\pm 0.3$& $0.024\pm 0.005$&$0.022\pm 0.004$\\ 
 SKT tpCN& $8d$& $23.0\pm 0.0$& $0.028\pm 0.006$&$0.024\pm 0.004$\\ 
 SKT tpCN& $6d$& $23.2\pm 0.4$& $0.032\pm 0.009$&$0.027\pm 0.008$\\ 
 SKT tpCN& $4d$& $23.3\pm 0.5$& $0.050\pm 0.010$&$0.049\pm 0.009$\\ 
 SKT tpCN& $2d$& $23.4\pm 0.5$& $0.17\pm 0.03$&$0.16\pm 0.02$\\ 
 \hhline{=|=|=|=|=}
 NF-SMC tpCN& $10d$& $23.0\pm 0.0$& $0.11\pm 0.03$&$0.11\pm 0.03$\\ 
 NF-SMC tpCN& $8d$& $23.8\pm 0.4$& $0.17\pm 0.05$&$0.19\pm 0.05$\\ 
 NF-SMC tpCN& $6d$& $24.0\pm 0.0$& $0.22\pm 0.05$&$0.23\pm 0.06$\\ 
 NF-SMC tpCN& $4d$& $24.1\pm 0.3$& $0.22\pm 0.07$&$0.21\pm 0.07$\\ 
 NF-SMC tpCN& $2d$& $23.2\pm 0.3$& $0.31\pm 0.07$&$0.23\pm 0.08$\\ 
 \hhline{=|=|=|=|=}
 SMC tpCN& $10d$& $23.9\pm 0.3$& $0.35\pm 0.05$&$0.42\pm 0.07$\\ 
 SMC tpCN& $8d$& $24.0\pm 0.0$& $0.35\pm 0.05$&$0.42\pm 0.06$\\ 
 SMC tpCN& $6d$& $24.2\pm 0.4$& $0.34\pm 0.07$&$0.39\pm 0.10$\\ 
 SMC tpCN& $4d$& $24.0\pm 0.6$& $0.31\pm 0.06$&$0.34\pm 0.08$\\ 
 SMC tpCN& $2d$& $23.5\pm 0.5$& $0.32\pm 0.07$&$0.25\pm 0.06$\\ 
 \hhline{= = = = =}\end{tabular}
    \caption{Results for the number of temperature levels used by each algorithm $N_\beta$ and squared bias results, $\langle b_1^2\rangle$ and $\langle b_2^2\rangle$, obtained with the final particle ensemble for each algorithm, when performing inference on the gravity survey example. We report the mean and standard deviation for each statistic over the 10 algorithm runs, and show results for each of the tested ensemble sizes $J$. The number of parallelized model evaluations is given by $11N_\beta$ for each algorithm, with the total number of model evaluations being given by $11JN_\beta$. The target dimension is $d=62$.}
    \label{tab:gravity tpCN bias table}
    \end{table}
    \begin{table}[ht]
     \centering
    \begin{tabular}{c|c|c|c|c} 
    \hhline{= = = = =}
         Algorithm&   $J$&  $N_\beta$ &$\langle b_1^2\rangle$ &$\langle b_2^2\rangle$ \\ 
         \hhline{=|=|=|=|=}
         NF-SKT pCN& 
     $10d$&  $20.0\pm 0.0$&$0.040\pm 0.010$&$0.12\pm 0.01$\\ 
 NF-SKT pCN& $8d$& $19.7\pm 0.5$& $0.070\pm 0.020$&$0.16\pm 0.02$\\ 
 NF-SKT pCN& $6d$& $19.3\pm 0.5$& $0.13\pm 0.03$&$0.20\pm 0.02$\\ 
 NF-SKT pCN& $4d$& $19.3\pm 0.5$& $0.33\pm 0.06$&$0.35\pm 0.04$\\ 
 NF-SKT pCN& $2d$& $18.4\pm 0.5$& $0.69\pm 0.11$&$0.48\pm 0.08$\\ 
 \hhline{=|=|=|=|=}
 SKT pCN& $10d$& $24.6\pm 0.5$& $0.095\pm 0.014$&$0.16\pm 0.02$\\ 
 SKT pCN& $8d$& $24.4\pm 0.5$& $0.13\pm 0.03$&$0.13\pm 0.02$\\ 
 SKT pCN& $6d$& $23.6\pm 0.5$& $0.22\pm 0.03$&$0.18\pm 0.01$\\ 
 SKT pCN& $4d$& $23.5\pm 0.5$& $0.34\pm 0.08$&$0.23\pm 0.03$\\ 
 SKT pCN& $2d$& $22.0\pm 0.4$& $0.94\pm 0.15$&$0.52\pm 0.08$\\ 
 \hhline{=|=|=|=|=}
 NF-SMC pCN& $10d$& $21.0\pm 0.0$& $0.19\pm 0.04$&$0.22\pm 0.02$\\ 
 NF-SMC pCN& $8d$& $21.5\pm 0.5$& $0.25\pm 0.04$&$0.26\pm 0.02$\\ 
 NF-SMC pCN& $6d$& $22.3\pm 0.5$& $0.40\pm 0.10$&$0.32\pm 0.03$\\ 
 NF-SMC pCN& $4d$& $23.7\pm 0.5$& $0.69\pm 0.14$&$0.44\pm 0.07$\\ 
 NF-SMC pCN& $2d$& $29.2\pm 1.1$& $1.82\pm 0.58$&$1.27\pm 0.82$\\ 
 \hhline{=|=|=|=|=}
 SMC pCN& $10d$& $27.3\pm 0.6$& $0.69\pm 0.09$&$0.48\pm 0.09$\\ 
 SMC pCN& $8d$& $28.2\pm 0.4$& $0.65\pm 0.07$&$0.34\pm 0.07$\\ 
 SMC pCN& $6d$& $29.1\pm 0.5$& $1.05\pm 0.24$&$0.58\pm 0.15$\\ 
 SMC pCN& $4d$& $31.4\pm1.2$& $1.63\pm 0.45$&$1.11\pm 0.65$\\ 
 SMC pCN& $2d$& $9.4\pm 1.6$& $8.28\pm 2.86$&$50.8\pm 51.0$\\ 
 \hhline{= = = = =}\end{tabular}
    \caption{Results for the number of temperature levels used by each algorithm $N_\beta$ and squared bias results, $\langle b_1^2\rangle$ and $\langle b_2^2\rangle$, obtained with the final particle ensemble for each algorithm, when performing inference on the gravity survey example. We report the mean and standard deviation for each statistic over the 10 algorithm runs, and show results for each of the tested ensemble sizes $J$. The number of parallelized model evaluations is given by $11N_\beta$ for each algorithm, with the total number of model evaluations being given by $11JN_\beta$. The target dimension is $d=62$.}
    \label{tab:gravity pCN bias table}
    \end{table}

When we allow for adaptive selection of the number of sampling iterations we again find that we are able to reach the low bias regime with NF-SKT and SKT adaptation of the tpCN sampler. This is not the case for NF-SMC and SMC adaptation applied to tpCN, or for the pCN sampler using all adaptation algorithms. In these cases, a more stringent adaptation criterion is again required to achieve low bias estimates of posterior moments.
\begin{table}
    \centering
    \begin{tabular}{c|c|c|c}
    \hhline{= = = =}
         Algorithm ($\tau_\mathrm{corr}=0.1$)&  $N_\mathrm{eval} / J$&  $\langle b_1^2 \rangle$& $\langle b_2^2 \rangle$\\
         \hhline{=|=|=|=}
         NF-SKT tpCN&  $350\pm 100$&  $0.0078\pm 0.0032$& $0.0072\pm 0.0032$\\
         NF-SKT pCN&  $420\pm 230$&  $0.059\pm 0.011$& $0.12\pm 0.01$\\
         \hhline{=|=|=|=}
         SKT tpCN&  $810\pm 90$&  $0.0055\pm 0.0010$& $0.0060\pm 0.0012$\\
         SKT pCN&  $1200\pm 26$&  $0.15\pm 0.03$& $0.12\pm 0.01$\\
         \hhline{=|=|=|=}
         NF-SMC tpCN&  $320\pm 73$&  $0.046\pm 0.023$& $0.046\pm 0.024$\\
         NF-SMC pCN&  $190\pm 6.6$&  $0.15\pm 0.03$& $0.12\pm 0.01$\\
         \hhline{=|=|=|=}
         SMC tpCN&  $630\pm 14$&  $0.036\pm 0.008$& $0.037\pm 0.010$\\
         SMC pCN&  $1400\pm 43$&  $0.69\pm 0.11$& $0.49\pm 0.09$\\
         \hhline{=|=|=|=}
    \end{tabular}
    \caption{Results for the number of embarrassingly parallel model evaluations $N_\mathrm{eval}/J$, and squared bias results, $\langle b_1^2\rangle$ and $\langle b_2^2\rangle$, obtained with the final particle ensemble for each algorithm when adapting the number of sampling iterations at each temperature level using $\tau_\mathrm{corr}=0.1$, as applied to the gravity survey example. We show results when adapting both the tpCN and pCN samplers with an ensemble size of $J=10d$, where the target dimension is $d=62$.}
    \label{tab: gravity rho0p1 bias}
\end{table}

\subsection{Reaction-Diffusion Equation}\label{subsec: reaction diffusion}

We consider a reaction-diffusion system in one spatial dimension, where some quantity $s(x,t)$ varies with time under the action of some source term $u(x)$. This time evolution is described by a nonlinear reaction-diffusion equation of the form
\begin{equation}
    \frac{\partial s(x, t)}{\partial t} = D\frac{\partial^2 s(x, t)}{\partial x^2}+\gamma s^2(x,t)+u(x),\quad x\in\Omega=[0, 1],
    \label{eqn:reaction-diffusion}
\end{equation}
where $D=0.1$ is the diffusion constant and $\gamma=0.1$ is the reaction rate. For this problem, we study the recovery of the source function $u(x)$ from observations of $s(x, t)$. To solve Equation \ref{eqn:reaction-diffusion} we use the implicit, second-order finite difference scheme implemented in \cite{wang2021learning}. We assume Dirichlet boundary conditions such that $s(x,t)=0, \forall x\in \partial\Omega$, and the initial condition $s(x, t=0)=0$. The solution to Equation \ref{eqn:reaction-diffusion} is evaluated on a $100\times 100$ grid in $(x, t)$, up to a final time $t_f=1$.

We parameterize the source term using the Hilbert space expansion of a Gaussian Process (GP) \cite{riutort2023practical} with a squared exponential kernel,
\begin{equation}
    u(x) = \mu_H + \sum_{j=1}^{R} \left[S_{\Theta}\left(\sqrt{\lambda_j}\right)\right]^{1/2}\phi_j(x)\theta_j,
    \label{eqn:hilbert}
\end{equation}
where $\mu_H$ is the Hilbert space GP mean, $S_\Theta(\omega)=\alpha_H\sqrt{2\pi}\ell_H\exp(-\ell^2_H\omega^2/2)$ is the squared exponential kernel spectral density function, $\Theta=(\alpha_H, \ell_H)$ denotes the kernel hyperparmeters i.e., the kernel variance $\alpha_H$ and length scale $\ell_H$, $\{\lambda_j\}_{j=1}^{\infty}$ and $\{\phi_j(x)\}_{j=1}^\infty$ are the eigenvalues and eigenfunctions of the Laplacian operator on some domain $\Omega_L=[-L, L]$ respectively, and $\theta_j\sim\mathcal{N}(0, 1)$ are a set of standard Gaussian random variables. The eigenvalues and eigenfunctions of the Laplacian operator are given by
\begin{align}
    \lambda_j &= \left(\frac{j\pi}{2L}\right)^2,\\
    \phi_j(x) &= \sqrt{\frac{1}{L}}\sin\left(\sqrt{\lambda_j}(x+L)\right).
\end{align}
Without loss of generality, we can evaluate $u(x)$ on the symmetric interval $[-0.5, 0.5]$, choosing the domain for the Laplacian operator $\Omega_L=[-1, 1]$ such that it contains the full spatial domain of $u(x)$ \cite{riutort2023practical}. 

\begin{figure}[ht]
    \centering
    \includegraphics[width=\textwidth]{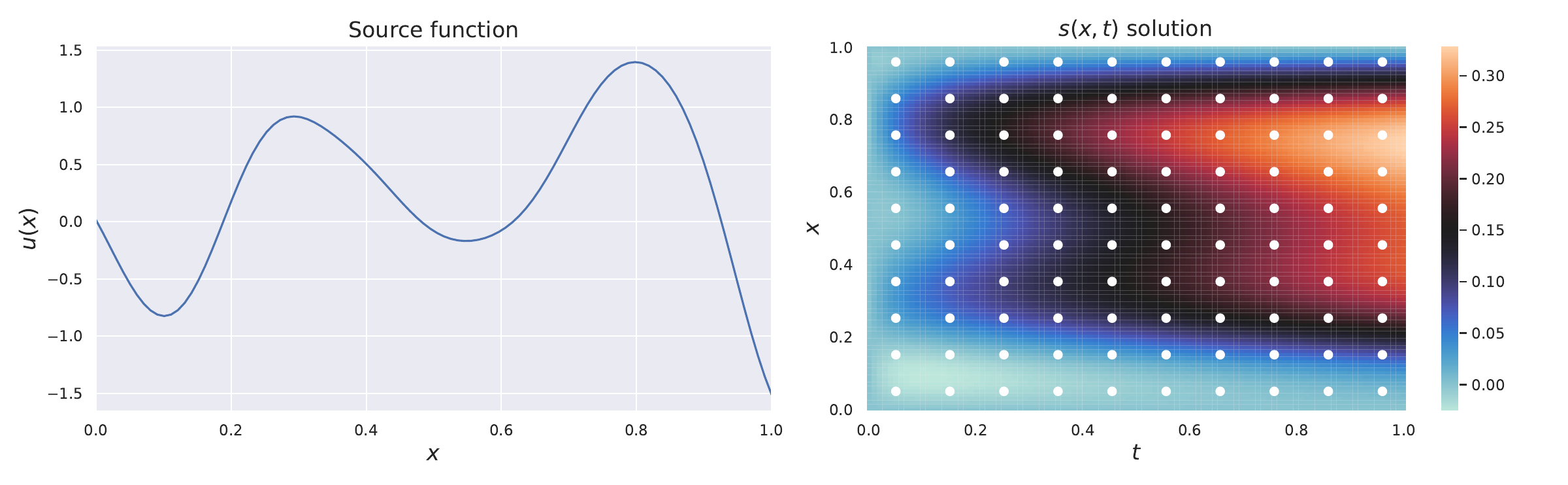}
    \caption{The true source term $u(x)$ (left panel), shown alongside the corresponding solution for $s(x, t)$ (right panel). White circles on the right panel denote the locations where measurements of $s(x, t)$ were made.}
    \label{fig:reaction diffusion signal field}
\end{figure}

To generate a simulated data set, we obtain a realisation of $u(x)$ from Equation \ref{eqn:hilbert} with $\mu_H=0$, $\alpha_H=1$ and $\ell_H=0.1$. We solve for $s(x, t)$ subject to the corresponding Dirichlet boundary conditions, up to a time $t_f=1$ on the $100\times 100$ grid in $(x, t)$. The field $s(x, t)$ is then observed at 10 equally spaced spatial locations, at 10 equally spaced times, with Gaussian observation noise corresponding to a noise standard deviation of $\sigma_\eta=0.01$. The true source function is shown in Figure \ref{fig:reaction diffusion signal field}, alongside the corresponding solution for $s(x, t)$ and the locations of the $s(x, t)$ measurements. 

For the inference task here we consider recovering the first $R=50$ terms in the Hilbert space expansion. Denoting $\bi{\theta}=(\theta_1, \ldots,\theta_{50})^\intercal$,  the full model is given by
\begin{align}
    \mu_H &\sim \mathcal{N}(\mu=0, \sigma^2=0.1^2),\\
    \alpha_H &\sim |\mathcal{N}(\mu=0, \sigma^2=1^2)|,\\
    \ell_H &\sim \mathrm{InverseGamma}(\alpha=4, \beta=0.3),\\
    \bi{\theta} &\sim \mathcal{N}(0, I_{50}),\\
    \bi{y} &\sim \mathcal{N}(F_{RD}(\mu_H, \alpha_H, \ell_H, \bi{\theta}), \sigma_\eta^2 I_{100}),
\end{align}
where $F_{RD}(\mu_H, \alpha_H, \ell_H, \bi{\theta})$ denotes the full forward model, mapping from the source function $u(x)$ to the $s(x, t)$ observations $\bi{y}$. When running our set of inference algorithms, we apply log-transformations to $\alpha_H$ and $\ell_H$ such that all parameters are mapped to an unconstrained space, making the corresponding Jacobian adjustments to the target. The target dimension for this problem is $d=53$.

\begin{figure}[ht]
    \centering
    \includegraphics[width=\textwidth]{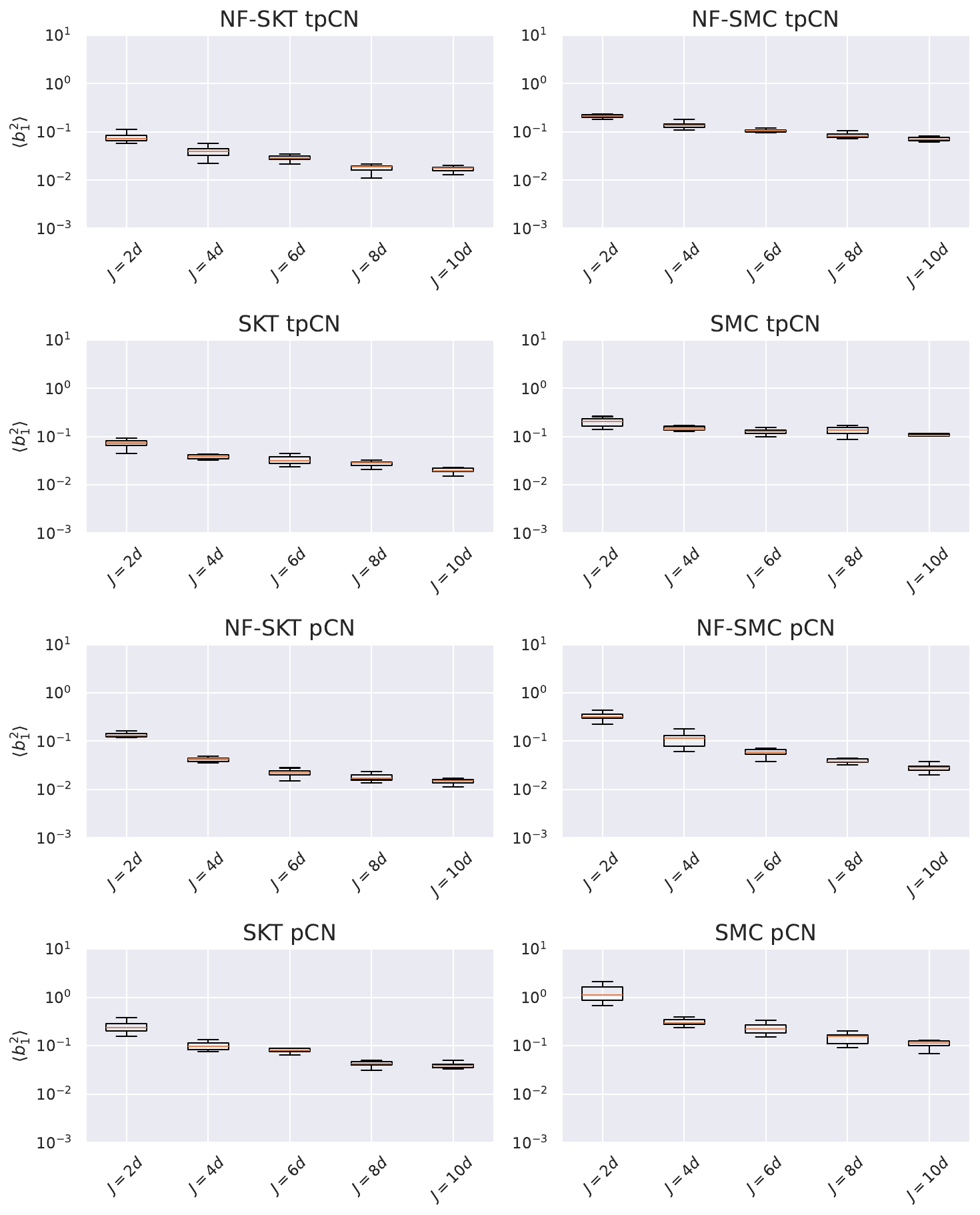}
    \caption{Final dimension averaged squared bias on the first moment for the reaction-diffusion model, plotted against the ensemble size $J$ expressed as a multiple of the target dimension $d=53$. Results are shown for each adaptation algorithm applied to the tpCN and pCN samplers.}
    \label{fig:reaction diffusion b1 10tpCN}
\end{figure}

\begin{figure}[ht]
    \centering
    \includegraphics[width=\textwidth]{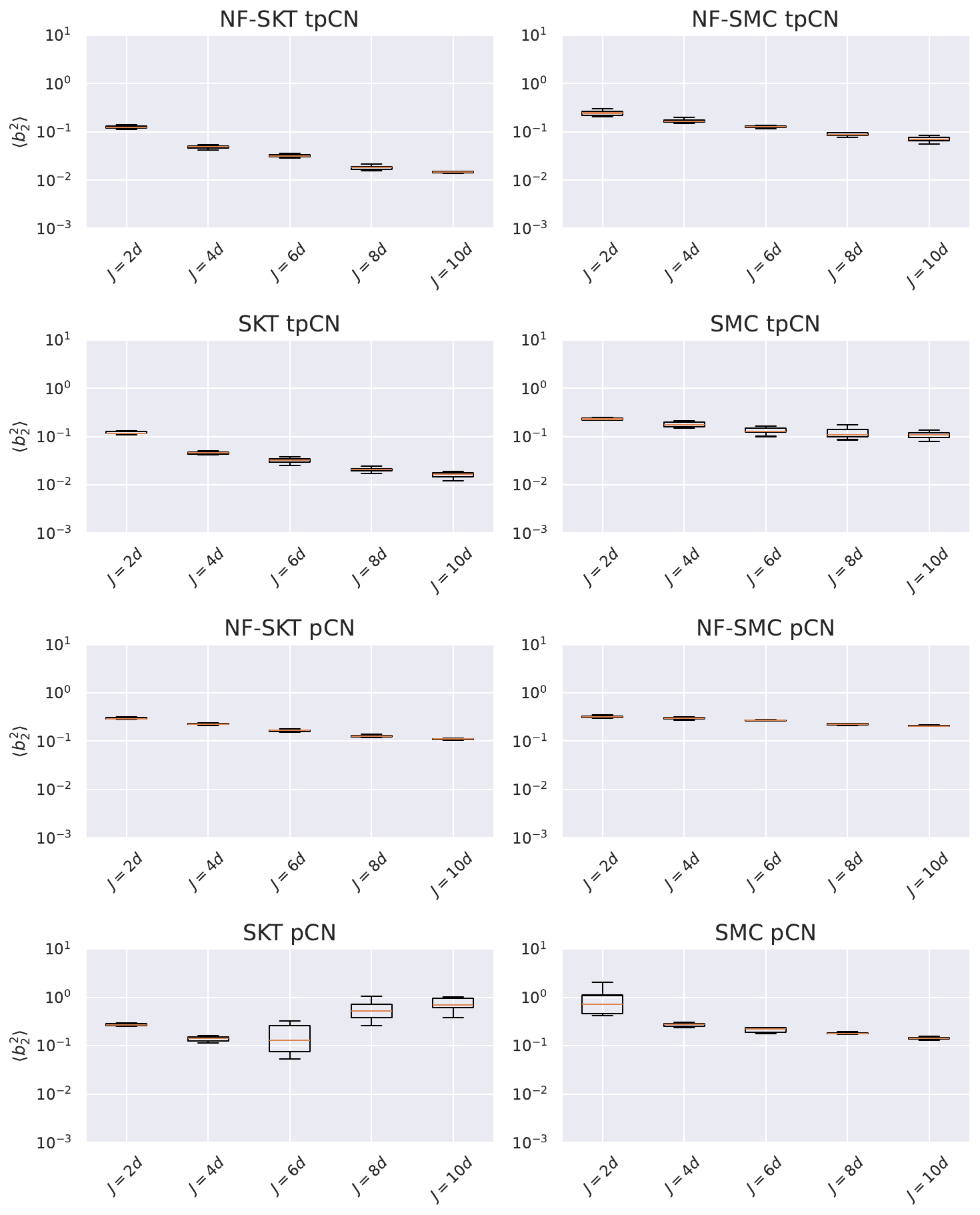}
    \caption{Final dimension averaged squared bias on the second moment for the reaction-diffusion model, plotted against the ensemble size $J$ expressed as a multiple of the target dimension $d=53$. Results are shown for each adaptation algorithm applied to the tpCN and pCN samplers.}
    \label{fig:reaction diffusion b2 10tpCN}
\end{figure}

For the first set of results we apply a fixed computational budget at each temperature level. In Figures \ref{fig:reaction diffusion b1 10tpCN} and \ref{fig:reaction diffusion b2 10tpCN} we show the recovered estimates for $\langle b_1^2\rangle$ and $\langle b_2^2\rangle$ respectively, with box plots showing the variation over the 10 runs for each algorithm and ensemble size. We report the corresponding mean and standard deviation for $\langle b_1^2\rangle$ and $\langle b_2^2\rangle$, along with the mean and standard deviation on the number of temperature levels used by each algorithm over the 10 runs in Table \ref{tab:reaction diffusion tpCN bias table} for tpCN, and in Table \ref{tab:reaction diffusion pCN bias table} for pCN. In Table \ref{tab: reaction diffusion rho0p1 bias} we report the mean and standard deviation for $\langle b_1^2\rangle$ and $\langle b_2^2\rangle$, along with the number of embarrassingly parallel model evaluations $N_\mathrm{eval} / J$, for the second set of tests where we select the number of sampling iterations at each temperature level adaptively. 
    \begin{table}[ht]
     \centering
    \begin{tabular}{c|c|c|c|c} 
    \hhline{= = = = =}
         Algorithm&   $J$&  $N_\beta$ &$\langle b_1^2\rangle$ &$\langle b_2^2\rangle$ \\ 
         \hhline{=|=|=|=|=}
         NF-SKT tpCN& 
     $10d$&  $17.0\pm 0.0$&$0.017\pm 0.002$&$0.015\pm 0.002$\\ 
 NF-SKT tpCN& $8d$& $17.0\pm 0.0$& $0.018\pm 0.003$&$0.018\pm 0.002$\\ 
 NF-SKT tpCN& $6d$& $17.8\pm 0.4$& $0.029\pm 0.004$&$0.032\pm 0.002$\\ 
 NF-SKT tpCN& $4d$& $18.0\pm 0.0$& $0.040\pm 0.010$&$0.049\pm 0.004$\\ 
 NF-SKT tpCN& $2d$& $19.1\pm 0.3$& $0.076\pm 0.016$&$0.13\pm 0.01$\\ 
 \hhline{=|=|=|=|=}
 SKT tpCN& $10d$& $17.0\pm 0.0$& $0.020\pm 0.003$&$0.016\pm 0.002$\\ 
 SKT tpCN& $8d$& $17.1\pm 0.3$& $0.028\pm 0.004$&$0.021\pm 0.002$\\ 
 SKT tpCN& $6d$& $17.6\pm 0.5$& $0.033\pm 0.007$&$0.032\pm 0.004$\\ 
 SKT tpCN& $4d$& $18.0\pm 0.0$& $0.038\pm 0.004$&$0.046\pm 0.003$\\ 
 SKT tpCN& $2d$& $19.0\pm 0.4$& $0.072\pm 0.013$&$0.12\pm 0.007$\\ 
 \hhline{=|=|=|=|=}
 NF-SMC tpCN& $10d$& $18.0\pm 0.0$& $0.072\pm 0.007$&$0.072\pm 0.008$\\ 
 NF-SMC tpCN& $8d$& $18.6\pm 0.5$& $0.084\pm 0.010$&$0.093\pm 0.015$\\ 
 NF-SMC tpCN& $6d$& $19.0\pm 0.0$& $0.11\pm 0.009$&$0.13\pm 0.01$\\ 
 NF-SMC tpCN& $4d$& $19.8\pm 0.4$& $0.14\pm 0.02$&$0.18\pm 0.03$\\ 
 NF-SMC tpCN& $2d$& $21.3\pm 0.5$& $0.21\pm 0.02$&$0.25\pm 0.03$\\ 
 \hhline{=|=|=|=|=}
 SMC tpCN& $10d$& $18.0\pm 0.0$& $0.11\pm 0.01$&$0.11\pm 0.02$\\ 
 SMC tpCN& $8d$& $18.1\pm 0.3$& $0.13\pm 0.03$&$0.12\pm 0.03$\\ 
 SMC tpCN& $6d$& $18.9\pm 0.3$& $0.13\pm 0.02$&$0.13\pm 0.02$\\ 
 SMC tpCN& $4d$& $19.5\pm 0.5$& $0.15\pm 0.01$&$0.18\pm 0.02$\\ 
 SMC tpCN& $2d$& $21.1\pm 0.5$& $0.20\pm 0.04$&$0.24\pm 0.07$\\ 
 \hhline{= = = = =}\end{tabular}
    \caption{Results for the number of temperature levels used by each algorithm $N_\beta$ and squared bias results, $\langle b_1^2\rangle$ and $\langle b_2^2\rangle$, obtained with the final particle ensemble for each algorithm, when performing inference on the reaction-diffusion example. We report the mean and standard deviation for each statistic over the 10 algorithm runs, and show results for each of the tested ensemble sizes $J$. The number of parallelized model evaluations is given by $11N_\beta$ for each algorithm, with the total number of model evaluations being given by $11JN_\beta$. The target dimension is $d=53$.}
    \label{tab:reaction diffusion tpCN bias table}
    \end{table}
        \begin{table}[ht]
     \centering
    \begin{tabular}{c|c|c|c|c} 
    \hhline{= = = = =}
         Algorithm&   $J$&  $N_\beta$ &$\langle b_1^2\rangle$ &$\langle b_2^2\rangle$ \\ 
         \hhline{=|=|=|=|=}
         NF-SKT pCN& 
     $10d$&  $16.0\pm 0.0$&$0.015\pm 0.002$&$0.11\pm 0.003$\\ 
 NF-SKT pCN& $8d$& $16.0\pm 0.0$& $0.018\pm 0.003$&$0.13\pm 0.006$\\ 
 NF-SKT pCN& $6d$& $15.9\pm 0.3$& $0.022\pm 0.004$&$0.16\pm 0.008$\\ 
 NF-SKT pCN& $4d$& $15.5\pm 0.5$& $0.042\pm 0.005$&$0.23\pm 0.006$\\ 
 NF-SKT pCN& $2d$& $14.9\pm 0.5$& $0.14\pm 0.04$&$0.30\pm 0.01$\\ 
 \hhline{=|=|=|=|=}
 SKT pCN& $10d$& $21.0\pm 0.9$& $0.038\pm 0.007$&$0.79\pm 0.32$\\ 
 SKT pCN& $8d$& $21.4\pm 2.0$& $0.041\pm 0.008$&$0.23\pm 0.01$\\ 
 SKT pCN& $6d$& $19.9\pm 1.0$& $0.082\pm 0.016$&$0.22\pm 0.25$\\ 
 SKT pCN& $4d$& $18.6\pm 0.5$& $0.10\pm 0.02$&$0.13\pm 0.04$\\ 
 SKT pCN& $2d$& $17.7\pm 0.6$& $0.25\pm 0.06$&$0.28\pm 0.02$\\ 
 \hhline{=|=|=|=|=}
 NF-SMC pCN& $10d$& $16.7\pm 0.5$& $0.018\pm 0.003$&$0.13\pm 0.006$\\ 
 NF-SMC pCN& $8d$& $17.0\pm 0.0$& $0.041\pm 0.008$&$0.23\pm 0.01$\\ 
 NF-SMC pCN& $6d$& $17.9\pm 0.5$& $0.061\pm 0.022$&$0.27\pm 0.008$\\ 
 NF-SMC pCN& $4d$& $18.9\pm 0.7$& $0.11\pm 0.04$&$0.30\pm 0.02$\\ 
 NF-SMC pCN& $2d$& $24.7\pm 0.8$& $0.35\pm 0.10$&$0.32\pm 0.03$\\ 
 \hhline{=|=|=|=|=}
 SMC pCN& $10d$& $19.8\pm 0.4$& $0.11\pm 0.02$&$0.14\pm 0.007$\\ 
 SMC pCN& $8d$& $20.9\pm 0.7$& $0.14\pm 0.03$&$0.18\pm 0.005$\\ 
 SMC pCN& $6d$& $21.5\pm 0.7$& $0.23\pm 0.06$&$0.21\pm 0.02$\\ 
 SMC pCN& $4d$& $24.2\pm 1.0$& $0.10\pm 0.02$&$0.28\pm 0.03$\\ 
 SMC pCN& $2d$& $11.1\pm 1.0$& $1.28\pm 0.49$&$0.86\pm 0.48$\\ 
 \hhline{= = = = =}\end{tabular}
    \caption{Results for the number of temperature levels used by each algorithm $N_\beta$ and squared bias results, $\langle b_1^2\rangle$ and $\langle b_2^2\rangle$, obtained with the final particle ensemble for each algorithm, when performing inference on the reaction-diffusion example. We report the mean and standard deviation for each statistic over the 10 algorithm runs, and show results for each of the tested ensemble sizes $J$. The number of parallelized model evaluations is given by $11N_\beta$ for each algorithm, with the total number of model evaluations being given by $11JN_\beta$. The target dimension is $d=53$.}
    \label{tab:reaction diffusion pCN bias table}
    \end{table}
\begin{table}[ht]
    \centering
    \begin{tabular}{c|c|c|c}
    \hhline{= = = =}
         Algorithm ($\tau_\mathrm{corr}=0.1$) &  $N_\mathrm{eval} / J$ & $\langle b_1^2 \rangle$ & $\langle b_2^2 \rangle$ \\
    \hhline{=|=|=|=}
         NF-SKT tpCN&  $510\pm43$&  $0.0089\pm 0.0010$& $0.0082\pm 0.0009$\\
         NF-SKT pCN&  $450\pm 140$&  $0.013\pm 0.004$& $0.12\pm 0.003$\\
        \hhline{=|=|=|=}
         SKT tpCN&  $600\pm 43$&  $0.0094\pm 0.0008$& $0.0095\pm 0.0013$\\
         SKT pCN&  $980\pm 15$&  $0.039\pm 0.011$& $0.76\pm 2.08$\\
         \hhline{=|=|=|=}
         NF-SMC tpCN&  $430\pm 19$&  $0.031\pm 0.005$& $0.024\pm 0.004$\\
         NF-SMC pCN&  $420\pm 62$&  $0.039\pm 0.011$& $0.22\pm 0.009$\\
         \hhline{=|=|=|=}
         SMC tpCN&  $700\pm 18$&  $0.022\pm 0.005$& $0.033\pm 0.006$\\
         SMC pCN&  $1000\pm 0.0$&  $0.12\pm 0.03$& $0.14\pm 0.008$\\
    \hhline{=|=|=|=}
    \end{tabular}
    \caption{Results for the number of embarrassingly parallel model evaluations $N_\mathrm{eval}/J$, and squared bias results, $\langle b_1^2\rangle$ and $\langle b_2^2\rangle$, obtained with the final particle ensemble for each algorithm when adapting the number of sampling iterations at each temperature level, as applied to the reaction-diffusion example. We show results when adapting both the tpCN and pCN samplers with an ensemble size of $J=10d$, where the target dimension is $d=53$.}
    \label{tab: reaction diffusion rho0p1 bias}
\end{table}

Whilst the values of $\langle b_1^2\rangle$ are comparable for tpCN and pCN on this example, the values for $\langle b_2^2\rangle$ are significantly lower for tpCN. The tpCN sampler is still able to better adapt to the targets at each temperature level, despite the target being closer to Gaussian. Comparing SKT with SMC, we can see that SKT is able to achieve significantly lower bias with the final particle ensemble using the same computational budget at each temperature level. This again indicates the the SKT update provides a better initialization and preconditioner for the subsequent tpCN iteration than the importance resampled ensemble in SMC. For this problem, we only obtain a small improvement in the final bias with NF preconditioning for larger ensemble sizes ($J\geq 6d$). The target posterior for this problem is close to Gaussian, meaning the NF map does not introduce a latent space where the target geometry is such that sampling is significantly easier.

When we make an adaptive selection of the number of sampling iterations, we find that we are able to reach the low bias regime for the tpCN sampler using the NF-SKT and SKT adaptation algorithms. This is not the case for the NF-SMC and SMC adaptation algorithms, or for the pCN sampler using any adaptation method. It is worth noting here that for pCN using SKT adaptation, for one of the random seeds the resultant pCN sampler was very poorly adapted to the target, resulting in high values for the mean and standard deviation of $\langle b_2^2\rangle$. For NF-SMC and SMC adaptation applied to tpCN, and any adaptation method applied to pCN, we would again require a more stringent criterion for selecting the number of sampling iterations at each temperature level. 

\section{Conclusions}\label{sec: conclusions}

In this work we have considered the problem of performing Bayesian inference on inverse problems where the forward model is expensive to evaluate and we do not have access to derivatives of the forward model. In such a situation, standard sampling methods such as MCMC and SMC algorithms can quickly become intractable, requiring a large number of serial model evaluations to attain low bias estimates of posterior moments \cite{huang2022efficient}. In contrast, EKI methods have been proposed that can rapidly converge on an ensemble approximation to the target posterior. However, EKI as applied to the Bayesian inverse problem is only exact in the regime of Gaussian targets and linear forward models, otherwise giving an uncontrolled approximation. This is insufficient for many scientific inference tasks where we seek accurate uncertainty estimates and hence low bias estimates for higher order posterior moments.

To address this shortcoming, we proposed integrating EKI updates within an adaptive SMC framework, replacing the standard importance resampling step at each temperature level with an EKI update. Instead of relying solely on the EKI updates to approximate the posterior, it was used to adapt the proposal kernel of the  $t$-preconditioned Crank-Nicolson (tpCN) sampler. In this way, the EKI approximation at each temperature level provides a highly effective initialization and preconditioner for the tpCN sampler. Moreover, performing tpCN sampling prevents the accumulation of errors that would result in EKI using the incorrect prior ensemble to approximate each annealed target. The tpCN proposal kernel is reversible with respect to the multivariate $t$-distribution, in contrast to the standard pCN proposal which is reversible with respect to the multivariate Gaussian. In this paper we have proposed the Sequential Kalman Tuning (SKT) adaptation scheme for tpCN, and its NF preconditioned variant NF-SKT, that provides an efficient, tuning-free sampler for the solution of Bayesian inverse problems.

We compared the performance of the SKT and NF-SKT adaptation schemes, applied to tpCN and pCN, with standard importance resampling SMC and NF-SMC, running each algorithm on three Bayesian inverse problems. Across our numerical experiments the tpCN sampler out-performed standard pCN for all the adaptation schemes we considered. The more flexible tail behaviour behaviour of the tpCN kernel means it can be more readily adapted for sampling from non-Gaussian targets. When using the same computational budget at each temperature level, the SKT and NF-SKT adaptation schemes resulted in lower bias estimates for the first and second posterior moments, compared to adapting the tpCN kernel within a standard SMC or NF-SMC scheme. When we selected the number of sampling iterations adaptively the tpCN sampler was able to rapidly reach low bias when using the SKT and NF-SKT adaptation schemes. For adaptation in SMC and NF-SMC the tpCN sampler was not able to reach low bias, and the pCN sampler failed to reach low bias for all adaptation schemes. In these cases a more stringent criterion would need to be imposed when selecting the number of sampling iterations and therefore significantly more forward model evaluations.

It is worth noting that we achieved lower bias adapting tpCN with the SKT scheme compared to using NF-SMC, demonstrating the ability of the EKI ensemble update to provide an effective initialization and preconditioner for the subsequent sampling steps. This is particularly promising for regimes where learning high fidelity NF maps becomes intractable e.g., moving beyond $\mathcal{O}(100)$ dimensions, or where one wishes to avoid the additional computational overhead from NF training. When using SKT adaptation, NF preconditioning is primarily useful for inverse problems where the cost of NF training (typically of order seconds up to $\mathcal{O}(100)$ dimensions), is insignificant compared to the cost of forward model evaluations during sampling, and where one can afford ensemble sizes that are sufficiently large to learn nonlinear features in the target geometry.

Several avenues exist for extending this work. In the first instance we plan to incorporate SKT and NF-SKT adaptation for tpCN within the \textsc{pocoMC} sampling package \cite{karamanis2022accelerating, karamanis2022_pocoMC}, which currently implements adaptive variants of SMC and NF-SMC for tpCN. It would be interesting to explore alternative NF architectures that are able to learn useful features in the target geometry with smaller ensemble sizes \cite{dai2021sliced}, and waste-free SMC methods that allow us to exploit the full sampling history in the SMC framework \cite{dau2022waste}. In this work we have only considered one variant of the EKI-type updates within SMC. It would be worth studying the performance of deterministic ensemble updates \cite{zhiyan2021kalman, huang2022efficient}, which have been shown to have superior empirical performance compared to the stochastic EKI update used in this work \cite{huang2022efficient}. It would also be useful to consider extensions allowing for parameter dependent noise covariances \cite{Botha2023iterative} and general likelihoods \cite{duffield2022ensemble}. 

One could also consider adaptation schemes for tpCN that fall outside the Bayesian annealing framework e.g., ensemble sampling schemes that directly target the full posterior \cite{leimkuhler2018ensemble}, leveraging ideas from measure transport \cite{marzouk2016introduction}, directly fitting for NF approximations to the target \cite{gabrie2022adaptive} etc. However, such adaptation schemes would require careful study to ensure stable kernel tuning during the burn-in phase where samples are far from the typical set. This problem is avoided in our sequential approach, where transitioning through a sequence of annealed targets allows for stable tuning of the tpCN kernel. This is particularly enhanced by the use of EKI updates as part of the adaptation. Moving to a non-sequential approach would require modification if we still wished to exploit Kalman-based approximations as part of the adaptation process, for example leveraging the proposed update rules in \cite{huang2022efficient}.

\section*{Acknowledgements}

This research was funded by NSFC (grant No. 12250410240) and the U.S. Department of Energy, Office of Science, Office of Advanced Scientific Computing Research under Contract No. DE-AC02-05CH11231 at Lawrence Berkeley National Laboratory to enable research for Data-intensive Machine Learning and Analysis, and by NSF grant number 2311559. RDPG was supported by a Tsinghua Shui Mu Fellowship. The authors thank Qijia Jiang and David Nabergoj for helpful discussions.

\appendix
\section{Proof of Lemma \ref{lemma:tpCN}}\label{sec: tpCN proofs}

\begin{proof}
Consider the current location $\bi{x}$ and the proposal location $\bi{x}^\prime=\bi{\mu}_s+\sqrt{1-\rho^2}(\bi{x}-\bi{\mu}_s)+\rho\sqrt{Z}\bi{W}$, where $Z^{-1}\sim\mathrm{Gamma}(k=\frac{1}{2}(d+\nu_s), \theta=2/(\nu_s+\langle\bi{x},\bi{x}\rangle_s))$ and $\bi{W}\sim\mathcal{N}(0, \mathcal{C}_s)$. 

We have that
\begin{equation}
    \sqrt{\frac{(d+\nu_s)Z}{\nu_s+\langle\bi{x}, \bi{x}\rangle_s}}\bi{W}\sim t_{d+\nu_s}(0,\mathcal{C}_s).
\end{equation}
Using the change of variables formula, we obtain the proposal transition kernel for the tpCN algorithm as
\begin{multline}
\mathcal{K}_t(\bi{x}, \mathrm{d}\bi{x}^\prime) = \frac{\gamma_1}{\rho^d}\left(\frac{d+\nu_s}{\nu_s+\langle\bi{x},\bi{x}\rangle_s}\right)^{d/2}\left[1+\right.\\
\left.(\rho^2(\nu_s+\langle\bi{x},\bi{x}\rangle_s))^{-1}(\langle\bi{x}^\prime,\bi{x}^\prime\rangle_s+(1-\rho^2)\langle\bi{x},\bi{x}\rangle_s-2\sqrt{1-\rho^2}\langle\bi{x},\bi{x}^\prime\rangle_s)\right]^{-(2d+\nu_s)/2}\mathrm{d}\bi{x}^\prime,
\end{multline}
where $\gamma_1$ is a normalizing constant. Considering the multivariate $t$-measure
\begin{equation}
    p_s(\mathrm{d}\bi{x})=\gamma_2\left[1+\frac{\langle\bi{x}, \bi{x}\rangle_s}{\nu_s}\right]^{-(d+\nu_s)/2}\mathrm{d}\bi{x},
\end{equation}
where $\gamma_2$ is a normalizing constant, we have that
\begin{multline}
p_s(\mathrm{d}\bi{x})\mathcal{K}_t(\bi{x}, \mathrm{d}\bi{x}^\prime) = \gamma_1\gamma_2\nu_s^{(d+\nu_s)/2}(d+\nu_s)^{d/2}\rho^{d+\nu_s}\left[\rho^2\nu_s+\langle\bi{x}^\prime, \bi{x}^\prime\rangle_s+\langle\bi{x}, \bi{x}\rangle_s\right.\\\left.-2\sqrt{1-\rho^2}\langle\bi{x}, \bi{x}^\prime\rangle_s\right]^{-(2d+\nu_s)/2}\mathrm{d}\bi{x}\mathrm{d}\bi{x}^\prime=p_s(\mathrm{d}\bi{x}^\prime)\mathcal{K}_t(\bi{x}^\prime, \mathrm{d}\bi{x}).
\label{eqn:p2CN reversible}
\end{multline}
The variables $\bi{x}$ and $\bi{x}^\prime$ are exchangeable in Equation \ref{eqn:p2CN reversible}. Therefore the tpCN proposal transition kernel is reversible with respect to the multivariate $t$-distribution $t_{\nu_s}(\bi{\mu}_s, \mathcal{C}_s)$.

The tpCN acceptance probability follows from the fact that for some general proposal kernel, $\mathcal{K}(\bi{x}, \mathrm{d}\bi{x}^\prime)$ with probability density function $\kappa(\bi{x}, \bi{x}^\prime)$, the Metropolis-Hastings (MH) acceptance probability is given by
\begin{equation}
    \alpha(\bi{x}, \bi{x}^\prime) = \mathrm{min}\left\{1, \frac{p(\bi{x}^\prime)\kappa(\bi{x}^\prime, \bi{x})}{p(\bi{x})\kappa(\bi{x}, \bi{x}^\prime)}\right\}.
    \label{eqn:general mh}
\end{equation}
From the reversibility expression in Equation \ref{eqn:p2CN reversible} we have that
\begin{equation}
    \frac{\kappa_t(\bi{x}^\prime, \bi{x})}{\kappa_t(\bi{x}, \bi{x}^\prime)}=\frac{p_s(\bi{x})}{p_s(\bi{x}^\prime)}=\frac{(1+\langle\bi{x},\bi{x}\rangle_s/\nu_s)^{-(d+\nu_s)/2}}{(1+\langle\bi{x}^\prime,\bi{x}^\prime\rangle_s/\nu_s)^{-(d+\nu_s)/2}},
\end{equation}
which gives the MH acceptance probability in Equation \ref{eqn:tpcn alpha}.
\end{proof}

\section{Importance Resampling}\label{sec: resampling}

Given a set of samples and associated normalized importance weights $\left\{\bi{x}^i_n, \tilde{w}_n(\bi{x}^i_n)\right\}_{i=1}^{J}$, where $\tilde{w}_n(\bi{x}_n^i)=w_n(\bi{x}_n^i)/\sum_{k=1}^{J}w_n(\bi{x}_n^k)$, we can apply a resampling algorithm to obtain a set of equal weight samples. A simple approach would be to apply multinomial resampling, where the duplication counts for each member of the ensemble $\{N^1, \ldots, N^J\}$ are obtained by sampling from the multinomial distribution $\mathrm{Mult}(J;\tilde{w}_n(\bi{x}_n^1), \ldots, \tilde{w}_n(\bi{x}_n^J))$. Whilst multinomial resampling is straightforward, lower variance methods are available \cite{douc2005comparison}. In this work we use systematic resampling for all our importance resampling SMC benchmarks. The systematic resampling algorithm pseudocode is given in Algorithm \ref{alg: systematic}.

\begin{algorithm}
   \caption{Systematic Resampling}
\begin{algorithmic}[1]
   \STATE {\bfseries Input:} Set of $J$ samples and corresponding normalized importance weights $\{\bi{x}^i, \tilde{w}^i\}_{i=1}^{J}$.
   \STATE Draw uniform random variable $U\sim\mathrm{Unif}(0, 1)$.
   \STATE Set $U^i= U+(i-1)/J$, $i\in\{1, \ldots, J\}$.
   \STATE Set $D_w=\tilde{w}^0$.
   \STATE Set index counter $k=1$
   \FOR{$i=1, \ldots, J$}
            \WHILE{$U^i > D_w$}
                    \STATE $k\leftarrow k+1$
                    \STATE $D_w \leftarrow D_w + \tilde{w}^k$
    \ENDWHILE
    \STATE Set $\tilde{\bi{x}^i}=\bi{x}^k$.
  \ENDFOR
  \STATE {\bfseries Output}: Equal weight particle ensemble $\{\tilde{\bi{x}}^i\}_{i=1}^J$.
\end{algorithmic}
\label{alg: systematic}
\end{algorithm}

\section{Sequential Monte Carlo Implementations}\label{sec: smc implement}

In Algorithm \ref{alg: nf smc tpcn} we give the pseudocode for the normalizing flow preconditioned SMC implementation, used as a benchmark for comparing the performance of the SKT samplers. The SMC implementation without NF preconditioning follows the same structure as Algorithm \ref{alg: nf smc tpcn}, without the NF fits such that the tpCN iterations are performed in the original data space. Similarly to the SKT samplers, we perform diminishing adaptation of the tpCN step size and reference measure mean at each temperature level.

\begin{algorithm}
   \caption{Flow Preconditioned Sequential Monte Carlo}
\begin{algorithmic}[1]
   \STATE {\bfseries Input:} Set of $J$ samples from the prior $\{\bi{x}_0^i\sim\pi_0(\bi{x})\}_{i=1}^J$, data $\bi{y}$, observation covariance $\Gamma$, target fractional ESS $\tau$, number of tpCN iterations to perform at each temperature level $M$, initial tpCN step size $\rho$, target tpCN acceptance rate $\alpha^{\star}$, tpCN autocorrelation threshold $\tau_\mathrm{corr}$.
   \STATE Set $\beta_0=0$ and iteration counter $n=0$.
   \WHILE{$\beta_n<1$ \do}
        \STATE Solve for target inverse temperature $\beta_{n+1}$ in Equation \ref{eqn:ess criterion}.
        \STATE $w_n(\bi{x}_n^i)\leftarrow\pi(\bi{y}|\bi{x}_n^i)^{\beta_{n+1}}/\pi(\bi{y}|\bi{x}_n^i)^{\beta_{n}},\quad i\in\{1,\ldots,J\}$.
        \IF{$\beta_{n+1}=1$}
            \STATE $n^* \leftarrow n+1$
        \ENDIF
        \STATE Fit NF map, $\bi{z}=f_n(\bi{x})$ to current particle locations $\{\bi{x}_n^i\}_{i=1}^{J}$.
        \STATE Obtain latent space particle locations $\{\bi{z}_n^i=f_n(\bi{x}_n^i)\}_{i=1}^J$.
        \STATE Resample weighted particles $\{\bi{z}_n^i, w_n(\bi{x}_n^i=f_n^{-1}(\bi{z}_n^i))\}$ using systematic resampling to give equal weight ensemble $\{\bi{z}_{n+1}^i\}_{i=1}^J$.
        \STATE Fit the multivariate $t$-distribution, $t_{\nu_s}(\mu_s, \mathcal{C}_s)$ to the latent space particle ensemble $\{\bi{z}^{i}_{n+1}\}_{i=1}^J$ with an EM algorithm. Set component-wise autocorrelations $\hat{\rho}_0(j)=1\:\forall j$.
        \FOR{$m=1,\ldots,M$}
                \FOR{$i=1, \ldots , J$}
                \STATE Update particle state $\bi{z}_{n+1}^i$ using Algorithm \ref{alg: tpCN} in NF latent space.
            \ENDFOR
            \STATE $\log\rho \leftarrow \log\rho + (\langle\alpha\rangle-\alpha^\star)/m$
            \STATE $\bi{\mu}_s \leftarrow \bi{\mu}_s + (\langle\bi{z}_{n+1}\rangle - \bi{\mu_s}) / m$
            \STATE Calculate component-wise autocorrelations $\hat{\rho}_m(j)$.
            \IF{$\prod_{l=1}^m\hat{\rho}_l(j)<\tau_{\mathrm{corr}}\:\forall j$}
                \STATE End tpCN iterations.
            \ENDIF
      \ENDFOR
      \STATE Map particle ensemble back to the original data space $\{\bi{x}_{n+1}^i=f^{-1}_n(\bi{z}_{n+1}^i)\}_{i=1}^{J}$
      \STATE $n \leftarrow n+1$
    \ENDWHILE
   \STATE {\bfseries Output}: Converged particle ensemble $\{\bi{x}_{n^*}^i\}_{i=1}^J$.
\end{algorithmic}
\label{alg: nf smc tpcn}
\end{algorithm}



\section{Field and Source Term Reconstructions}\label{sec:field recon}

\begin{figure}[ht]
    \centering
    \includegraphics[width=\textwidth]{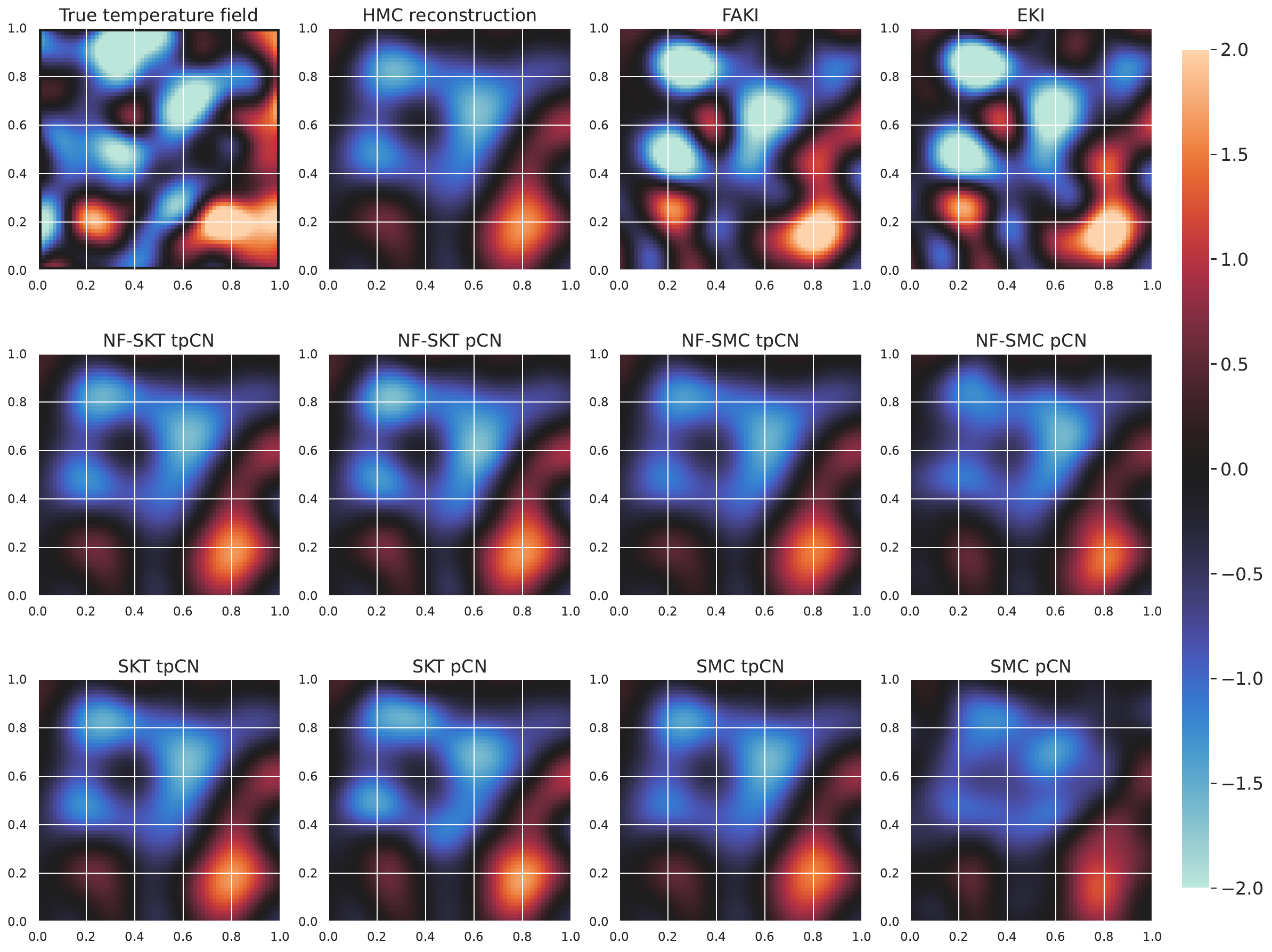}
    \caption{True initial temperature field $u(\bi{x}, t=0)$ for the heat equation example (Section \ref{subsec: heat}), plotted alongside the initial field reconstructions obtained using samples from HMC, which is treated as the target posterior predictive mean, and from the final particle ensembles of the FAKI, EKI, NF-SKT, NF-SMC, SKT and SMC algorithms (where results are shown for an ensemble size $J=10d$, and for a single random seed initialization). Results for the adaptation algorithms are shown as applied to tpCN and pCN.}
    \label{fig:heat initial recon}
\end{figure}

\begin{figure}[ht]
    \centering
    \includegraphics[width=\textwidth]{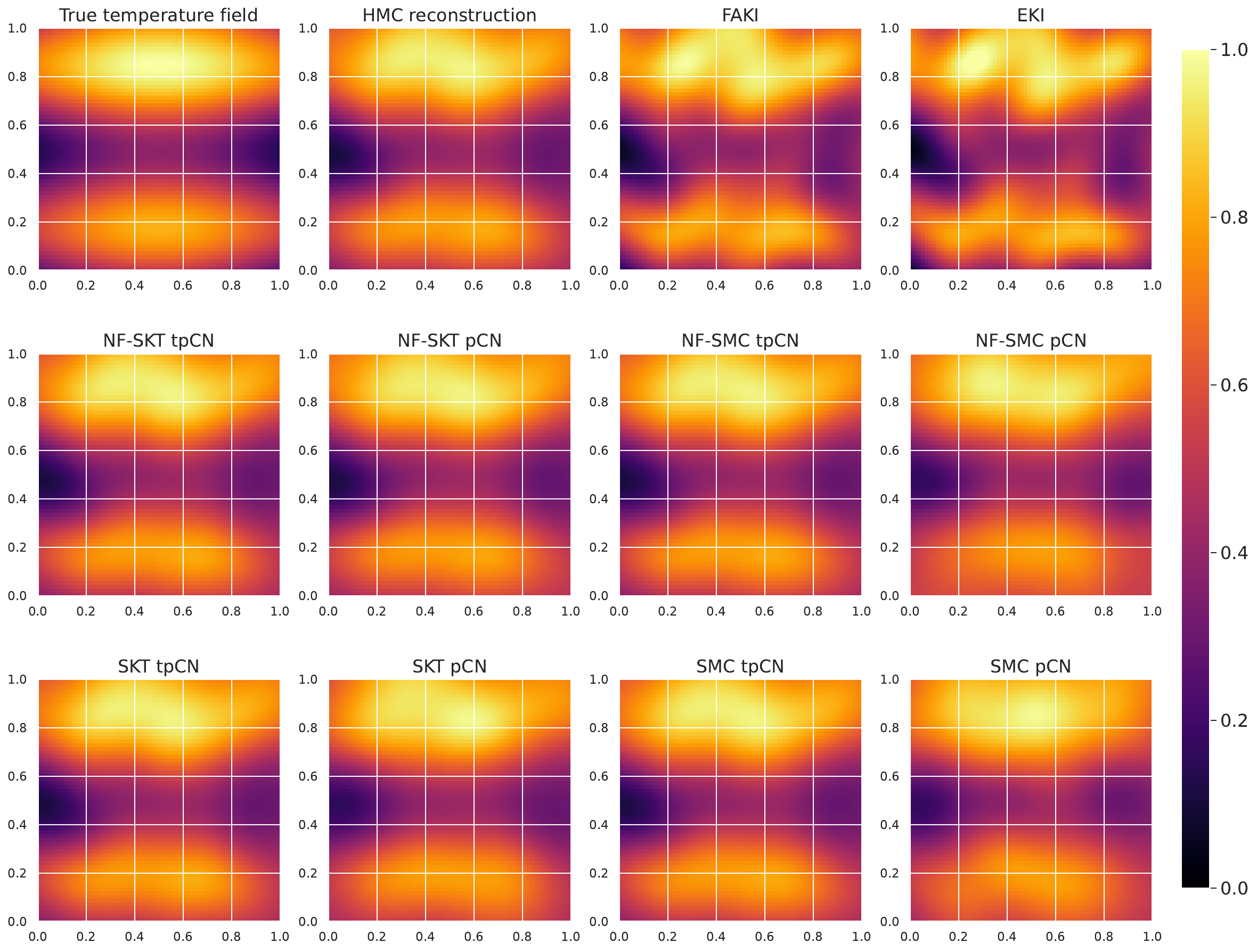}
    \caption{True subsurface mass density field $\varrho(\bi{x})$ for the gravity survey example (Section \ref{subsec: gravity}), plotted alongside the density field reconstructions obtained using samples from HMC, which is treated as the target posterior predictive mean, and from the final particle ensembles of the FAKI, EKI, NF-SKT, NF-SMC, SKT and SMC algorithms (where results are shown for an ensemble size $J=10d$, and for a single random seed initialization). Results for the adaptation algorithms are shown as applied to tpCN and pCN.}
    \label{fig:gravity initial recon}
\end{figure}

\begin{figure}[ht]
    \centering
    \includegraphics[width=\textwidth]{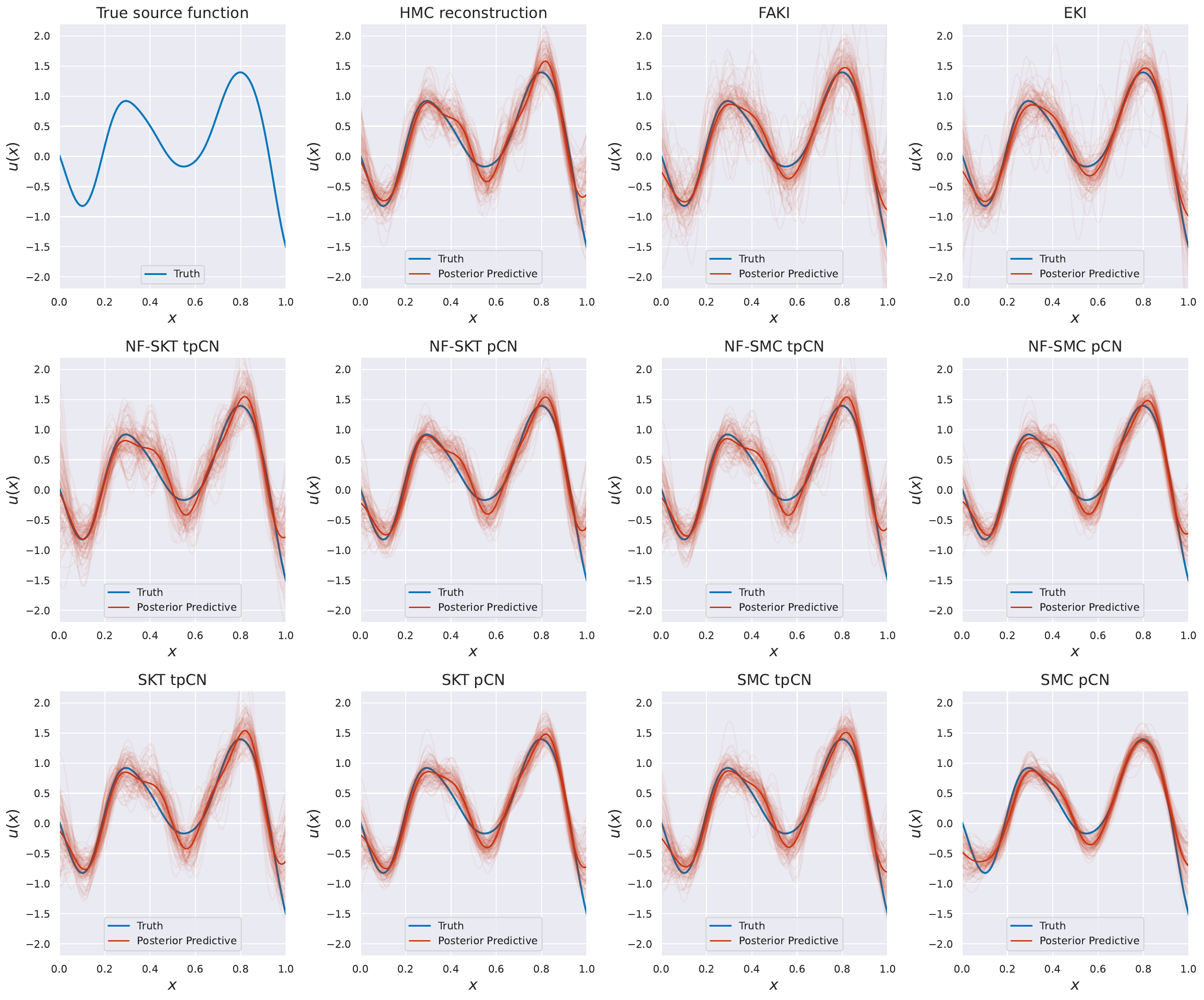}
    \caption{True source function $u(x)$ for the reaction-diffusion example (Section \ref{subsec: reaction diffusion}), plotted alongside posterior predictive samples obtained with HMC, which are treated as true posterior predictive samples, and from the final particle ensembles of the FAKI, EKI, NF-SKT, NF-SMC, SKT and SMC algorithms (where results are shown for an ensemble size $J=10d$, and for a single random seed initialization). For each algorithm we show 100 posterior predictive samples, with the dark red line showing the estimated posterior predictive mean in each case. Results for the adaptation algorithms are shown as applied to tpCN and pCN.}
    \label{fig:reaction diffusion recon}
\end{figure}

In Figure \ref{fig:heat initial recon} we show the true initial temperature field from the heat equation example in Section \ref{subsec: heat}, alongside the reconstructed initial field from HMC samples, and the field reconstructions obtained by evaluating the average over the final particle ensembles for each of the NF-SKT, NF-SMC, SKT and SMC adaptation algorithms as applied to tpCN and pCN, for a single random seed initialization with an ensemble size $J=10d$. We also show reconstructed fields obtained with the final FAKI and EKI ensembles. In Figure \ref{fig:gravity initial recon} we similarly show the true and reconstructed subsurface density fields for the gravity survey example in Section \ref{subsec: gravity}, and in Figure \ref{fig:reaction diffusion recon} we show the true and reconstructed source functions, $u(x)$ for the reaction-diffusion example in Section \ref{subsec: reaction diffusion}.

\section{Converged Ensemble Corner Plots}\label{sec: corner plots}

\begin{figure}[ht]
    \centering
    \includegraphics[width=0.48\textwidth]{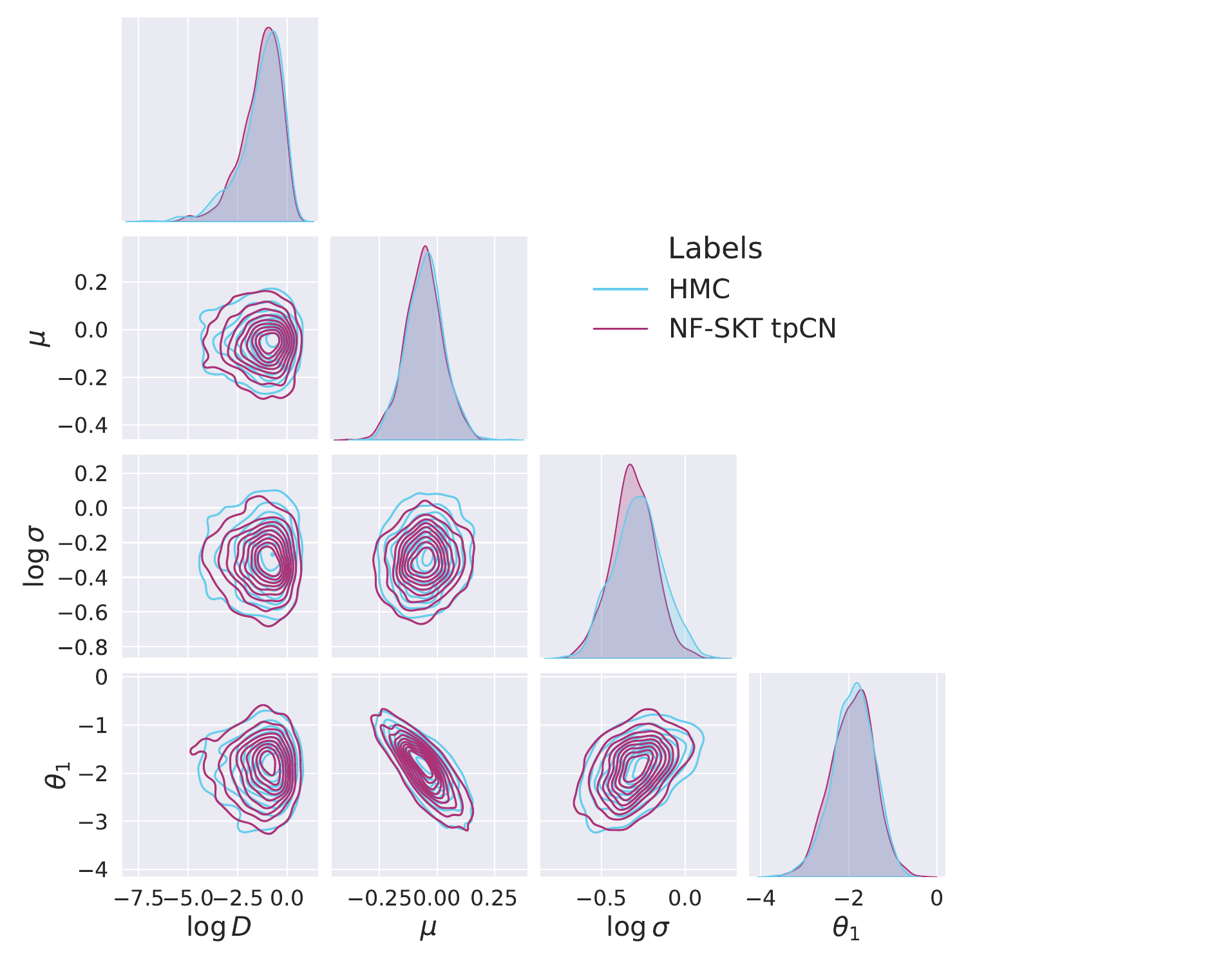}
    \includegraphics[width=0.48\textwidth]{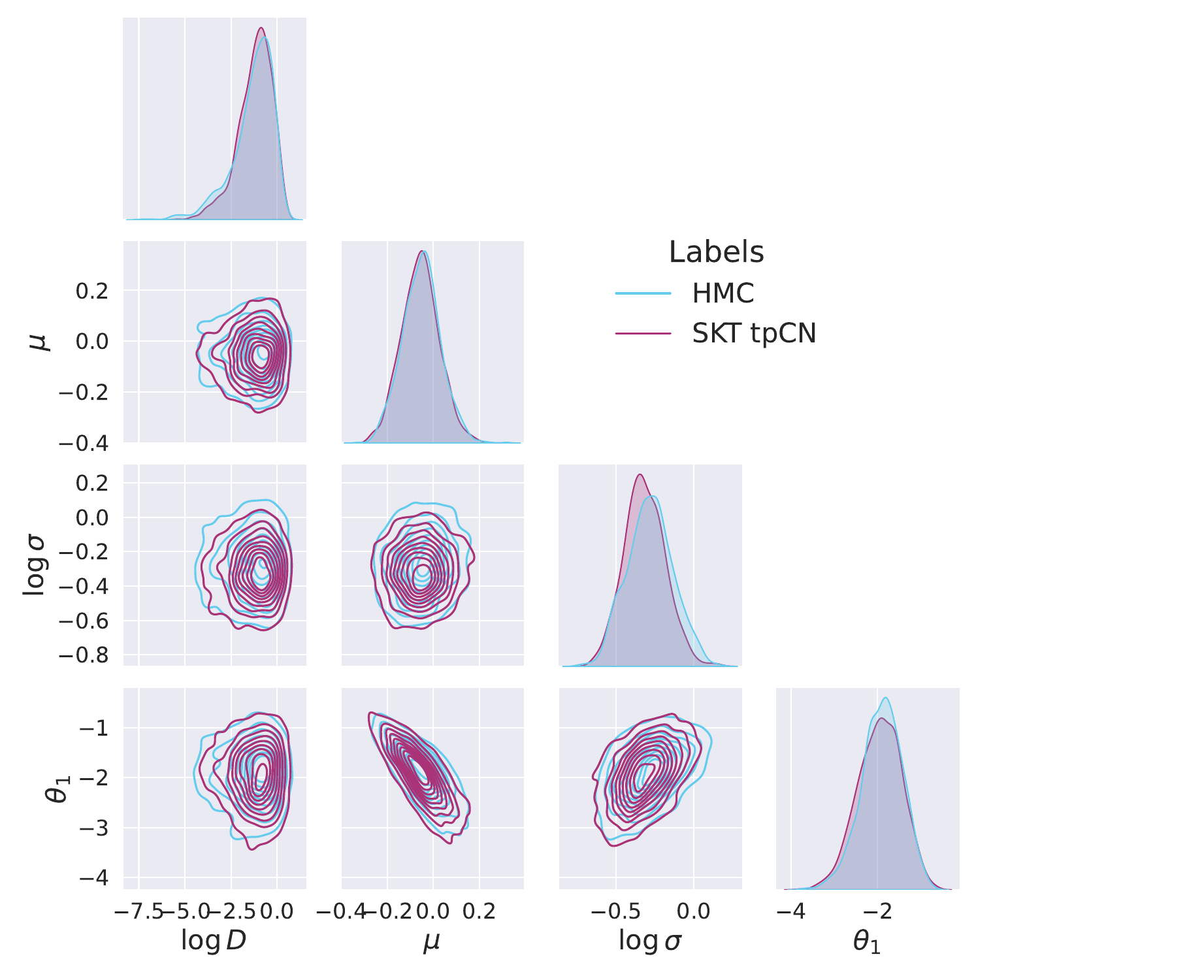}
    \includegraphics[width=0.48\textwidth]{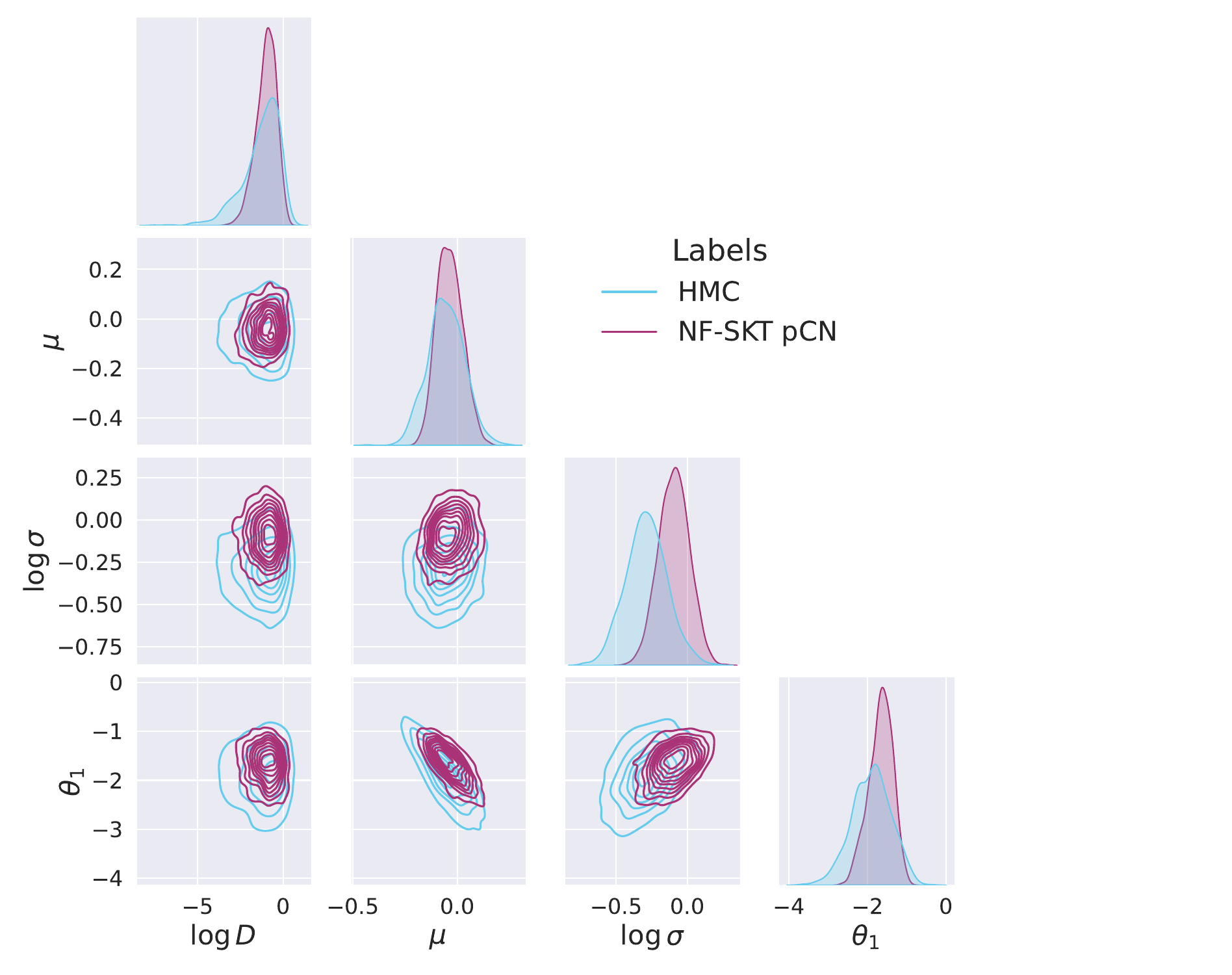}
    \includegraphics[width=0.48\textwidth]{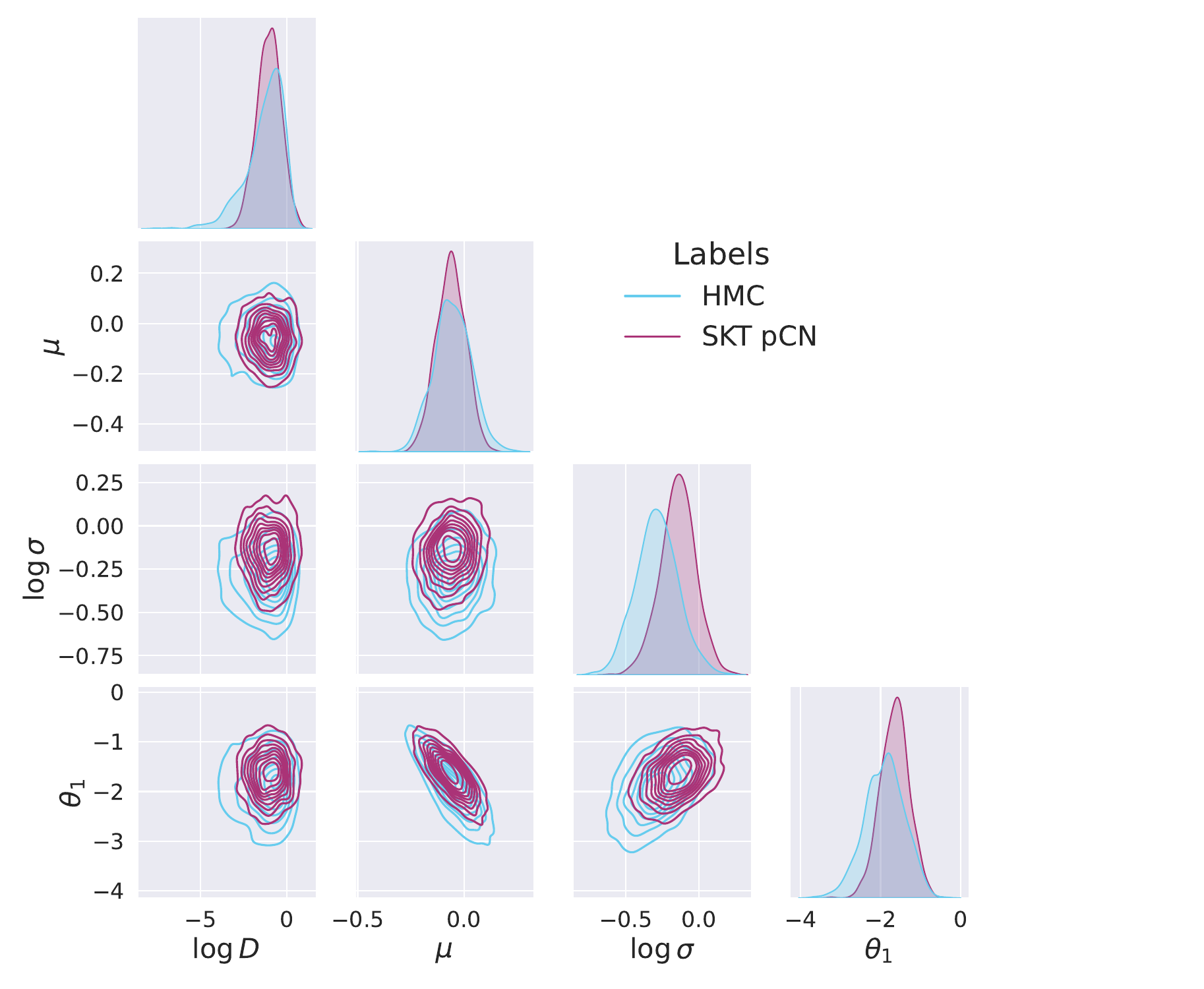}
    \caption{Corner plots showing the final particle ensemble for the first 4 dimensions, obtained using NF-SKT and SKT adaptation applied to the tpCN and pCN samplers, plotted alongside reference HMC samples for the heat equation example. The ensemble size was $J=10d$, where $d=103$.}
    \label{fig:heat eqn nf-SKT SKT corner}
\end{figure}

\begin{figure}[ht]
    \centering
    \includegraphics[width=0.48\textwidth]{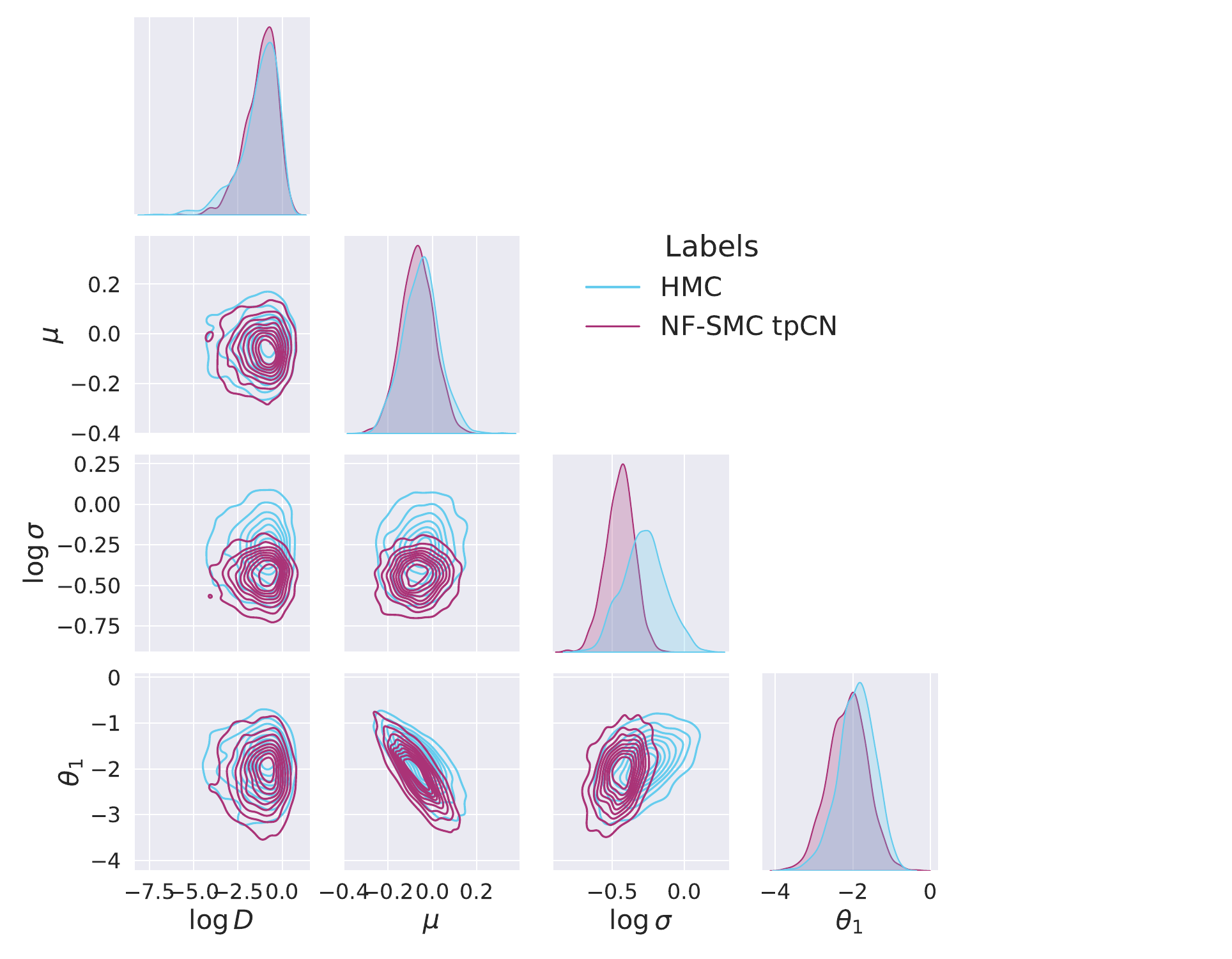}
    \includegraphics[width=0.48\textwidth]{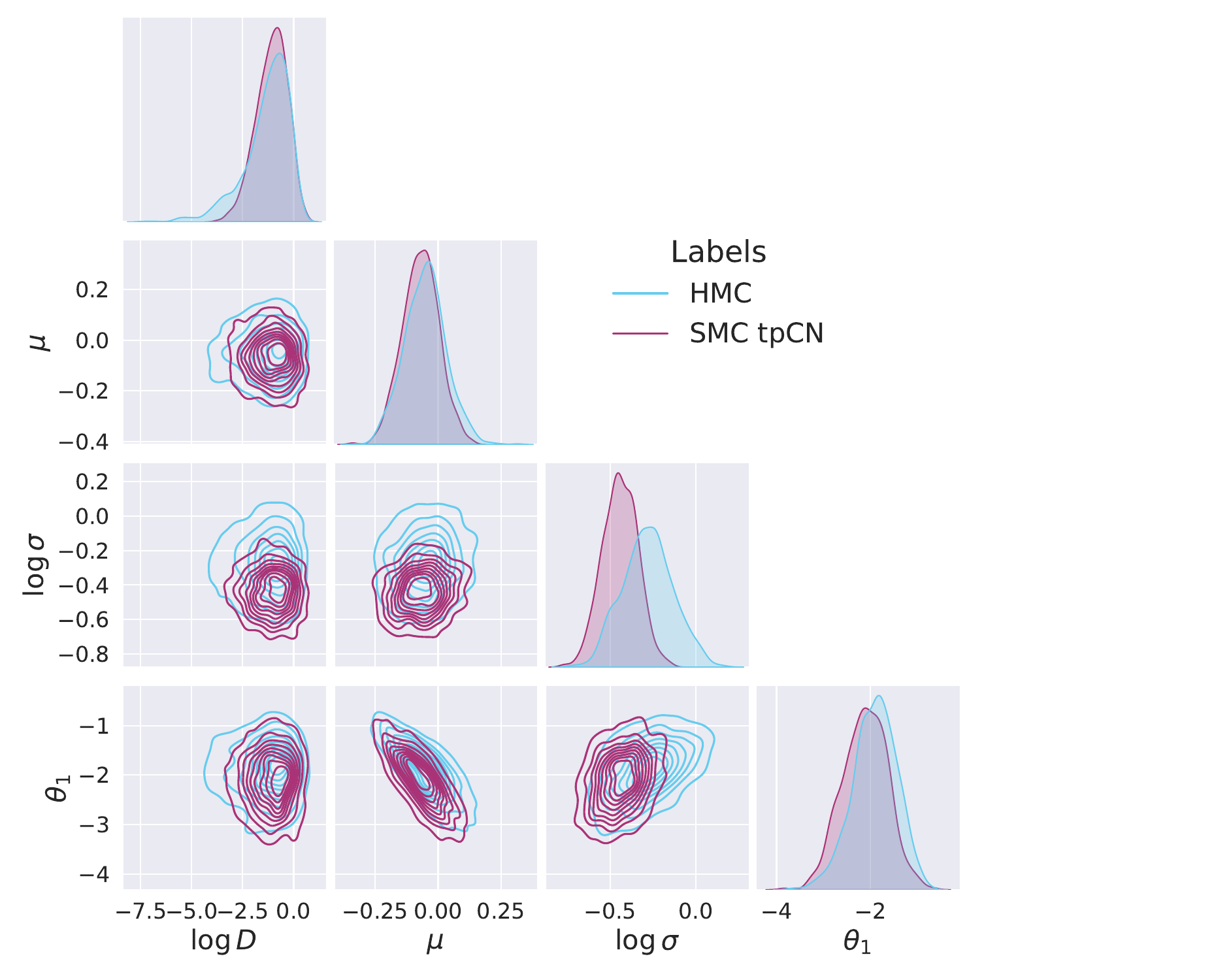}
    \includegraphics[width=0.48\textwidth]{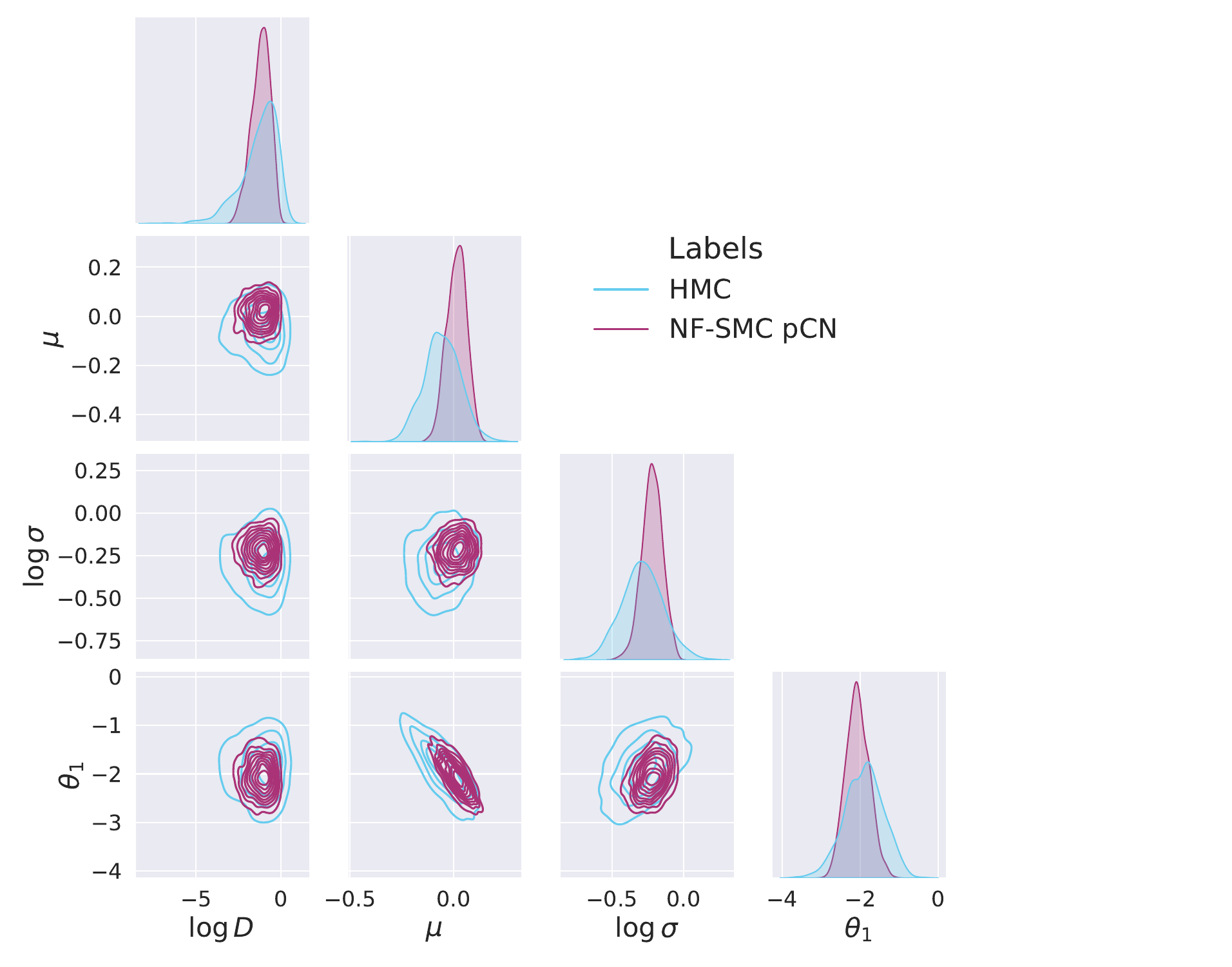}
    \includegraphics[width=0.48\textwidth]{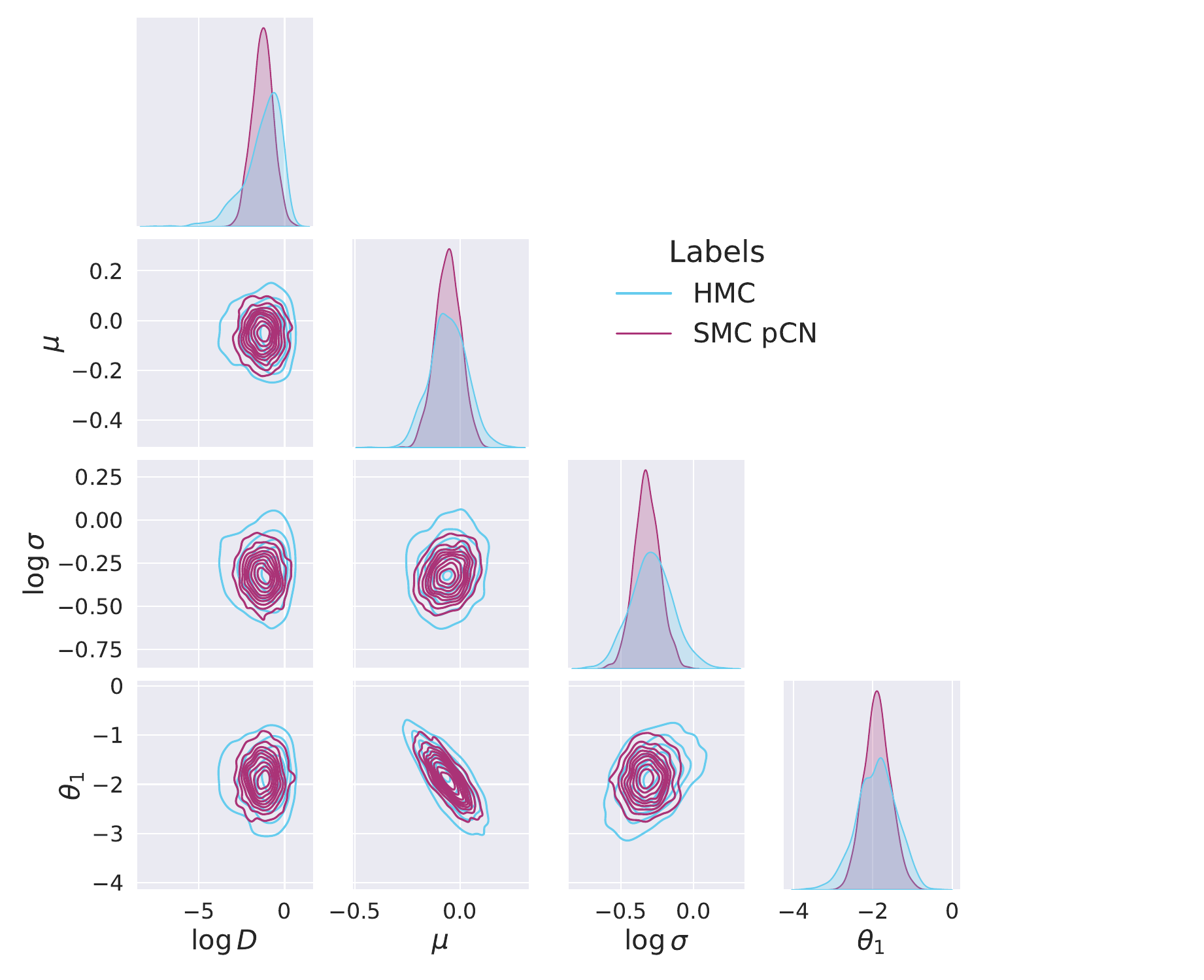}
    \caption{Corner plots showing the final particle ensemble for the first 4 dimensions, obtained using NF-SMC and SMC adaptation applied to the tpCN and pCN samplers, plotted alongside reference HMC samples for the heat equation example. The ensemble size was $J=10d$, where $d=103$.}
    \label{fig:heat eqn nf-smc smc corner}
\end{figure}

\begin{figure}[ht]
    \centering
    \includegraphics[width=0.48\textwidth]{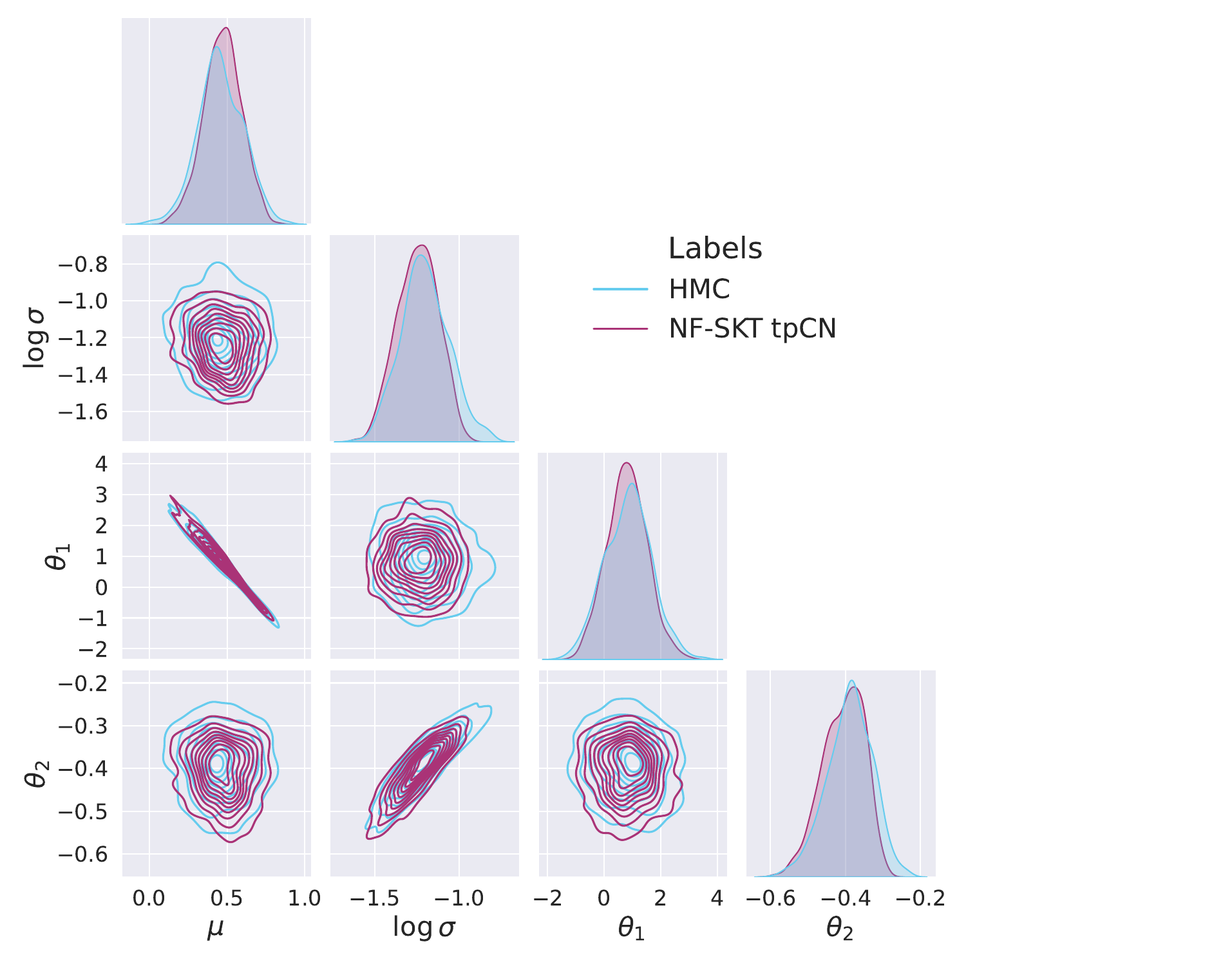}
    \includegraphics[width=0.48\textwidth]{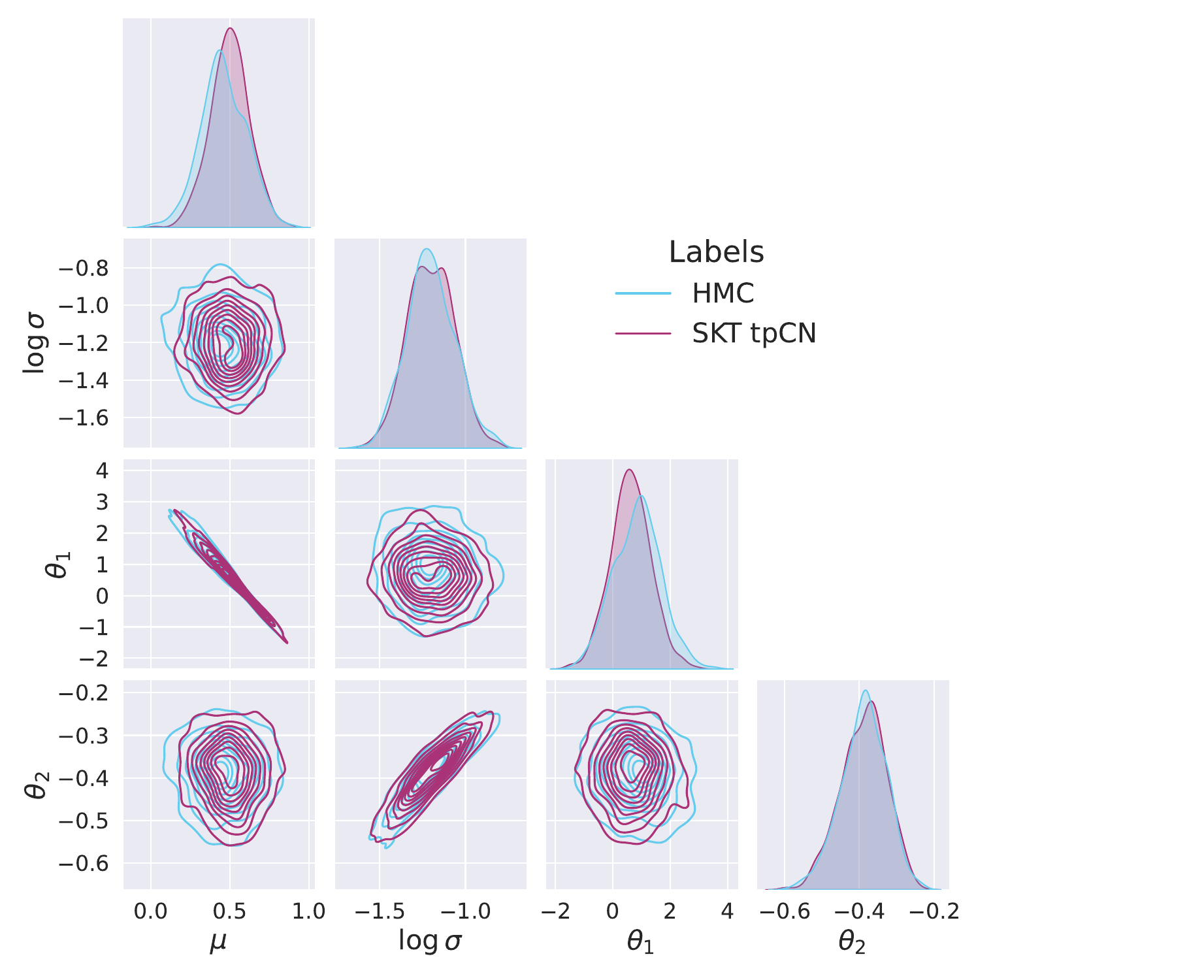}
    \includegraphics[width=0.48\textwidth]{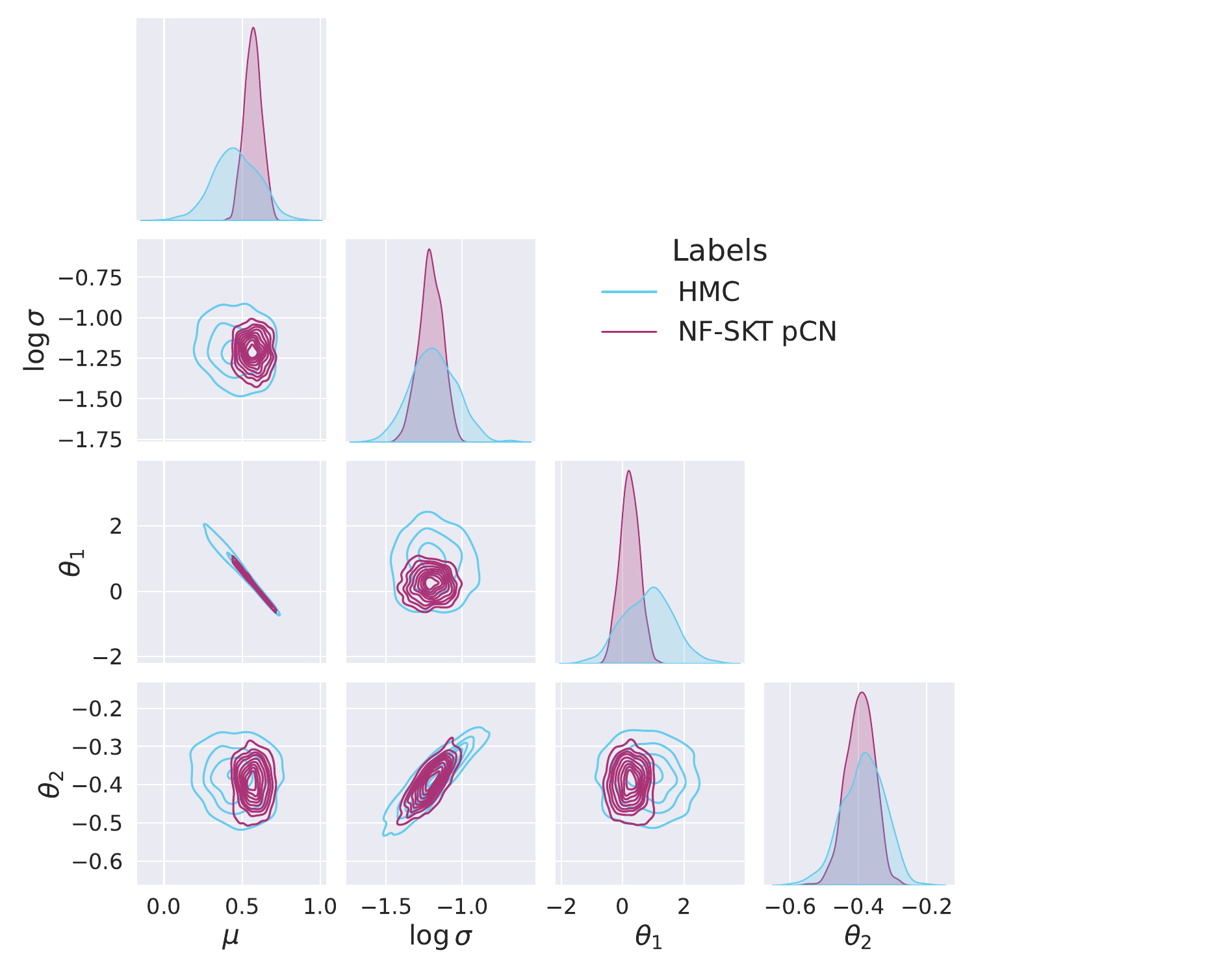}
    \includegraphics[width=0.48\textwidth]{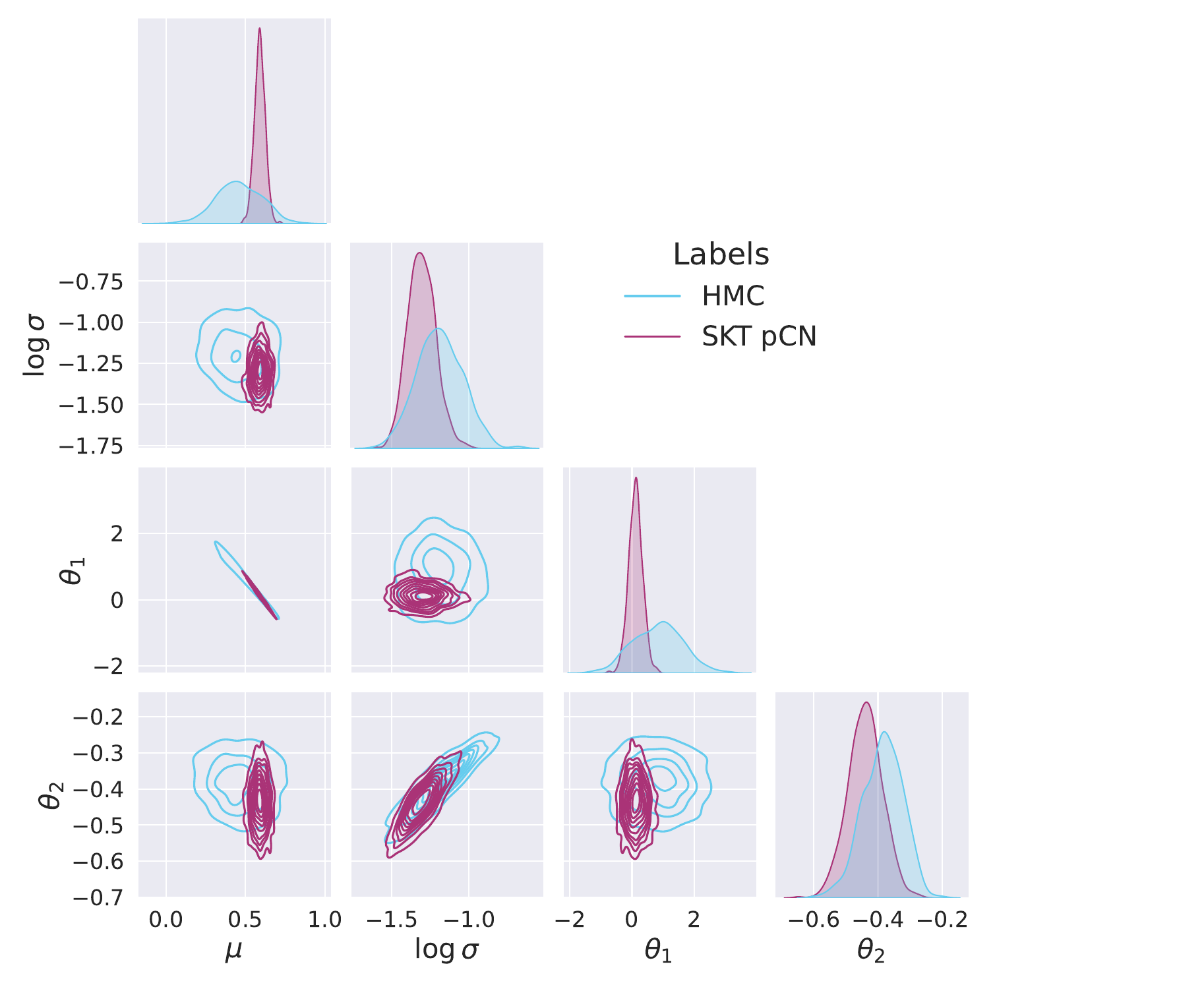}
    \caption{Corner plots showing the final particle ensemble for the first 4 dimensions, obtained using NF-SKT and SKT adaptation applied to the tpCN and pCN samplers, plotted alongside reference HMC samples for the gravity survey example. The ensemble size was $J=10d$, where $d=62$.}
    \label{fig:gravity nf-SKT SKT corner}
\end{figure}

\begin{figure}[ht]
    \centering
    \includegraphics[width=0.48\textwidth]{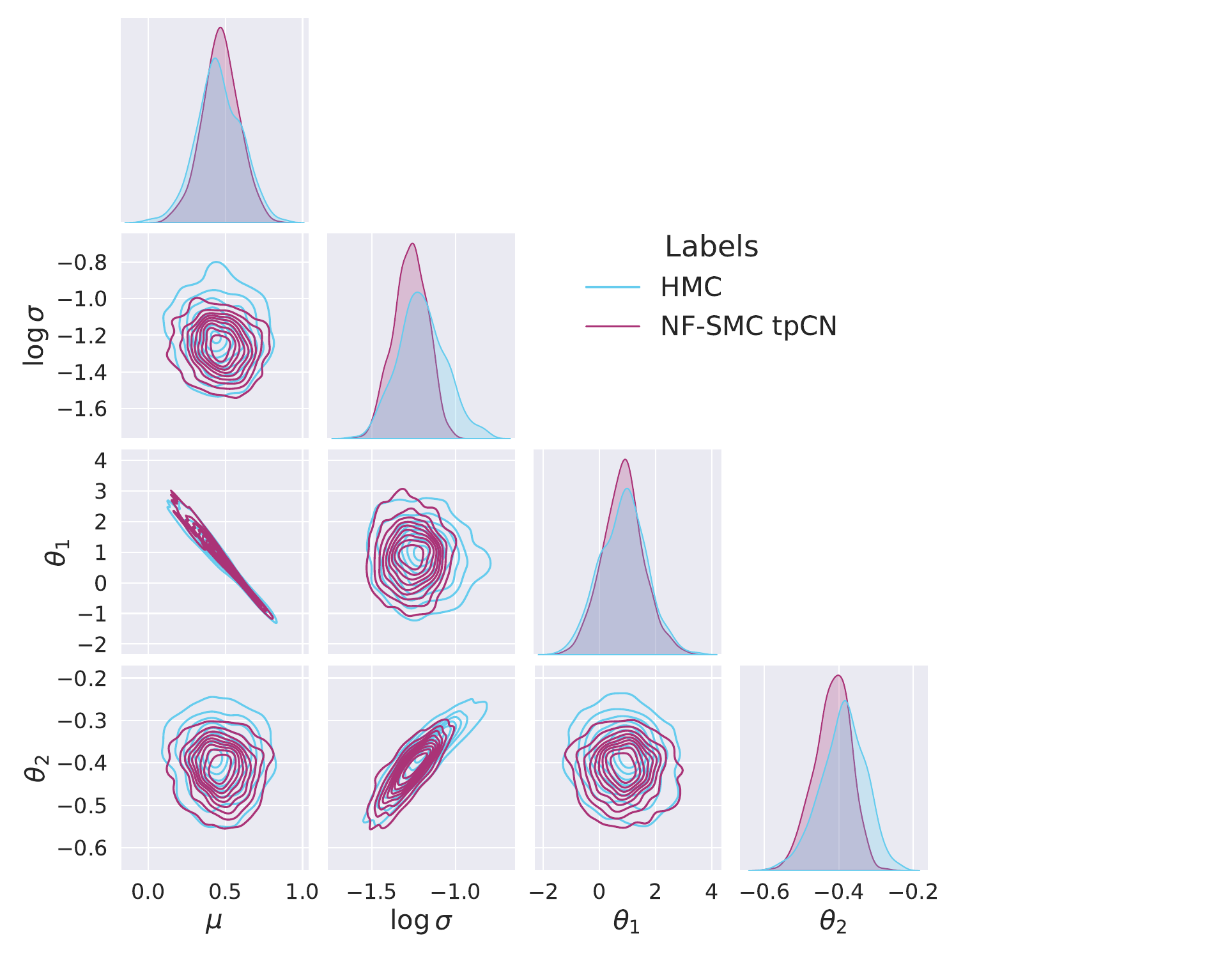}
    \includegraphics[width=0.48\textwidth]{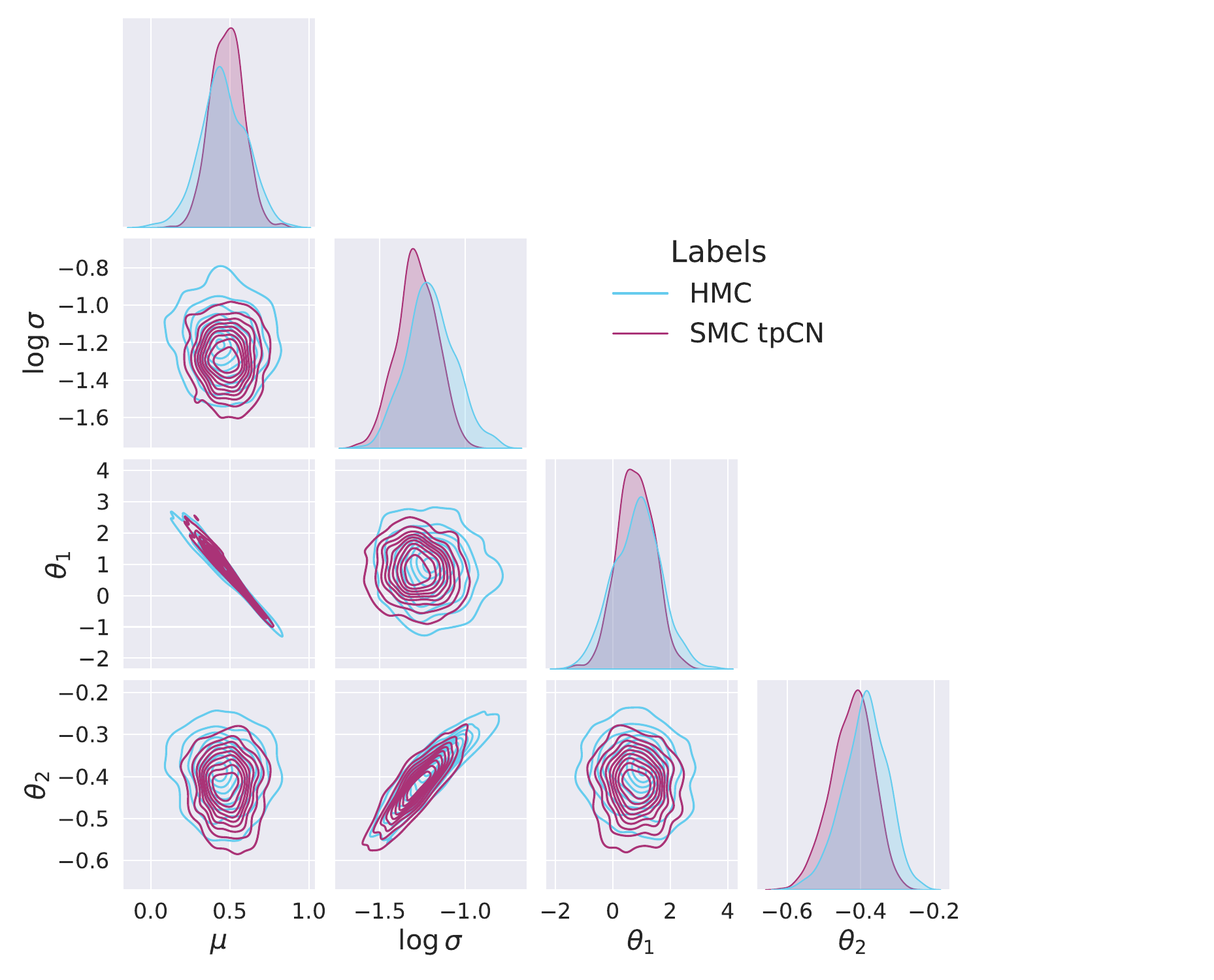}
    \includegraphics[width=0.48\textwidth]{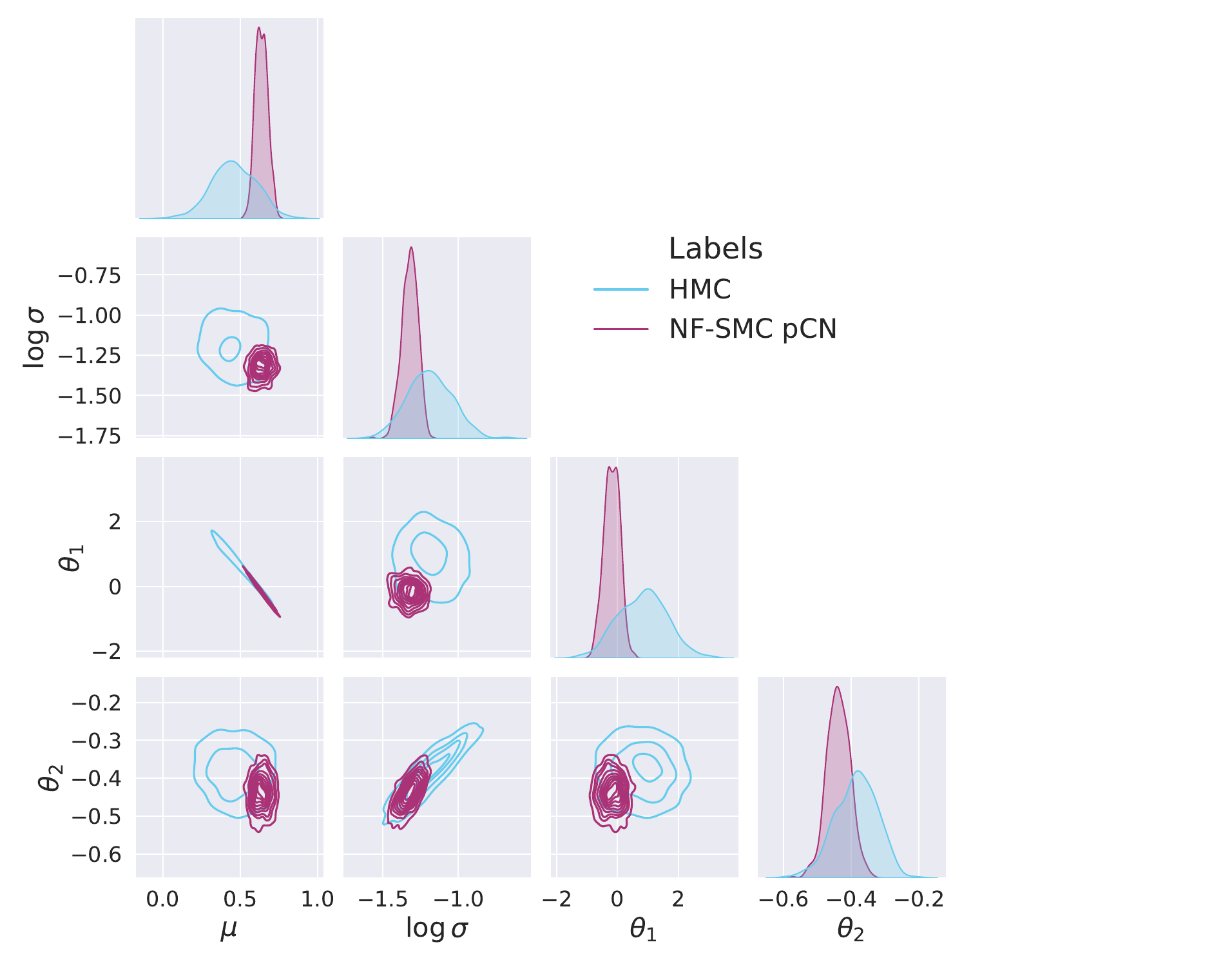}
    \includegraphics[width=0.48\textwidth]{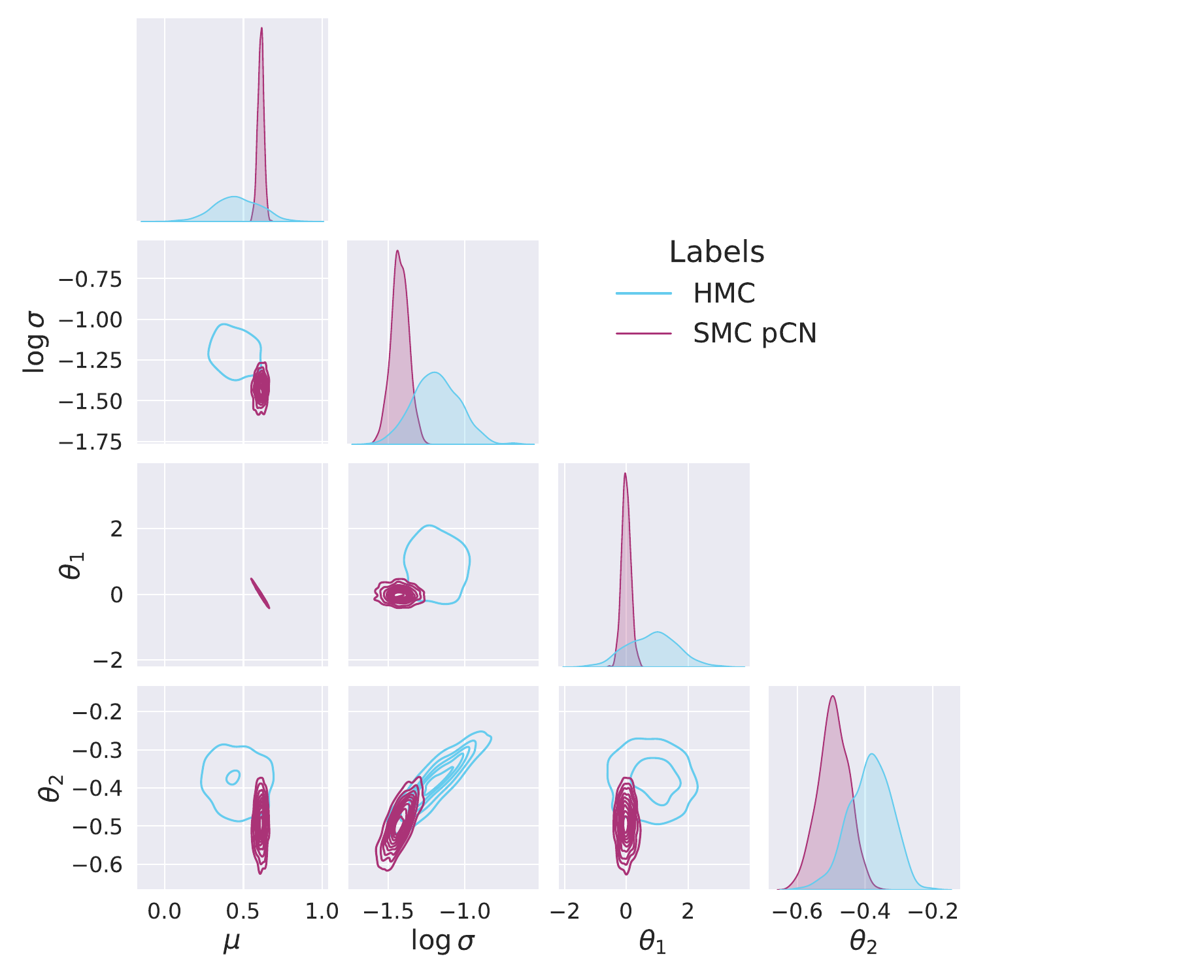}
    \caption{Corner plots showing the final particle ensemble for the first 4 dimensions, obtained using NF-SMC and SMC adaptation applied to the tpCN and pCN samplers, plotted alongside reference HMC samples for the gravity survey example. The ensemble size was $J=10d$, where $d=62$.}
    \label{fig:gravity nf-smc smc corner}
\end{figure}

\begin{figure}[ht]
    \centering
    \includegraphics[width=0.48\textwidth]{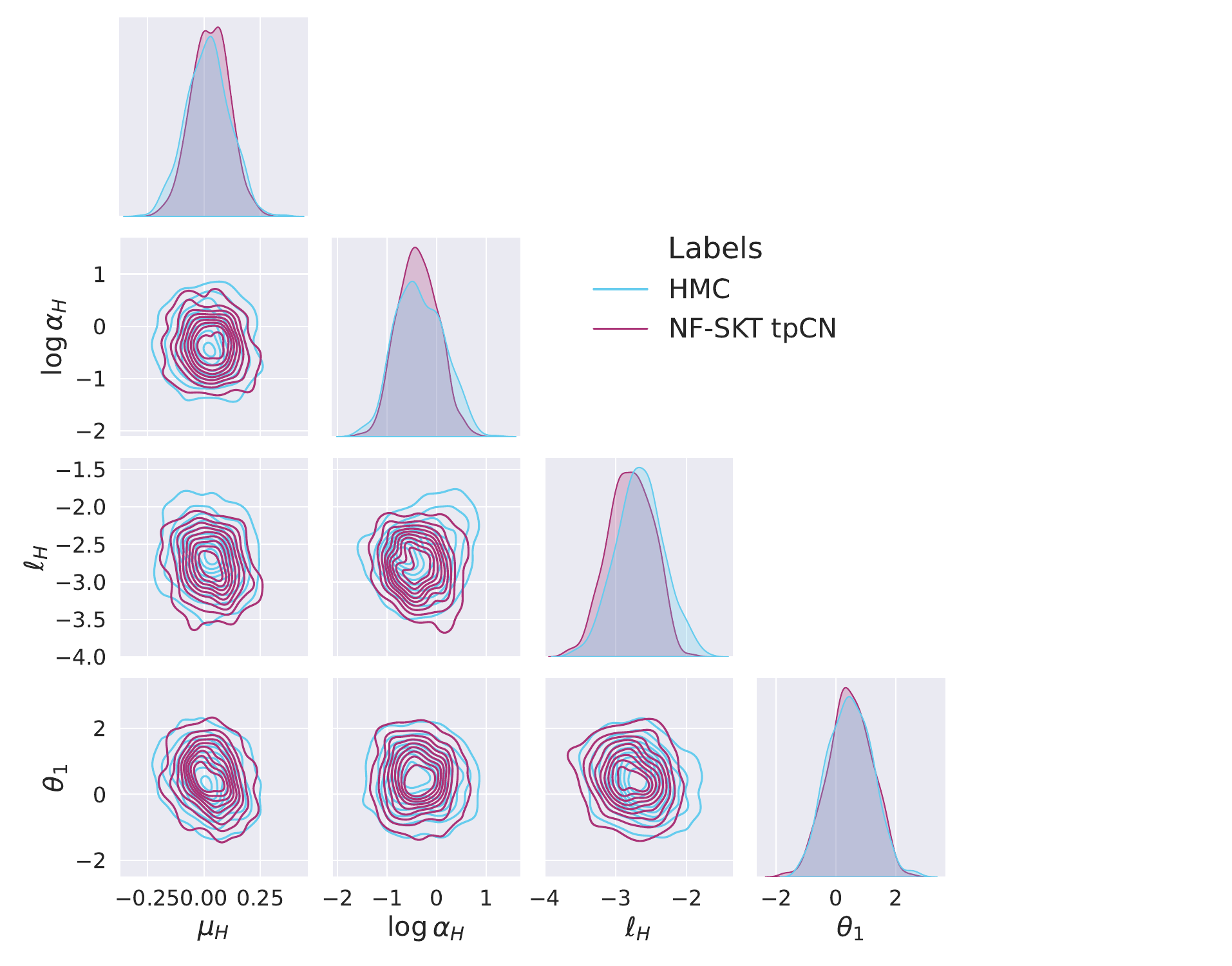}
    \includegraphics[width=0.48\textwidth]{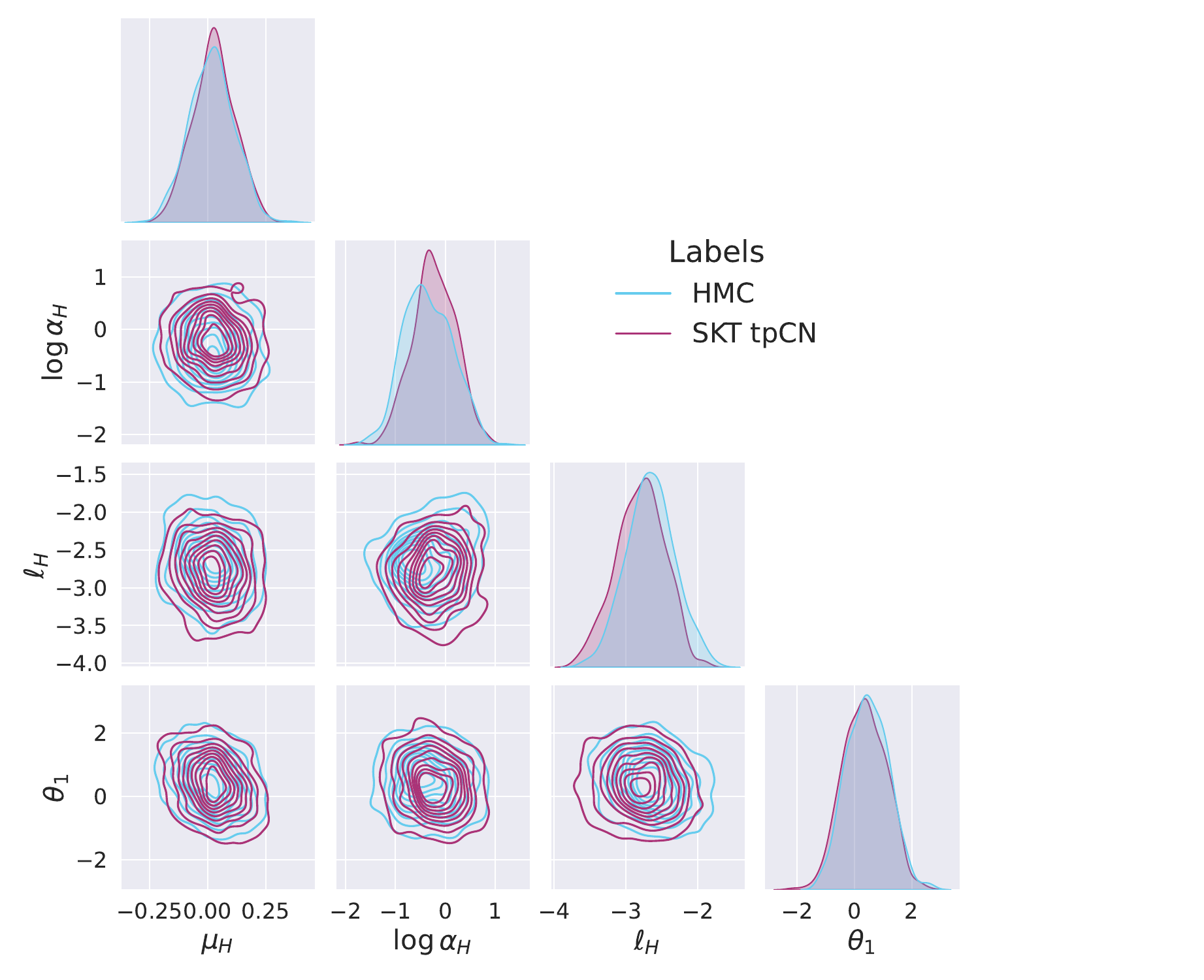}
    \includegraphics[width=0.48\textwidth]{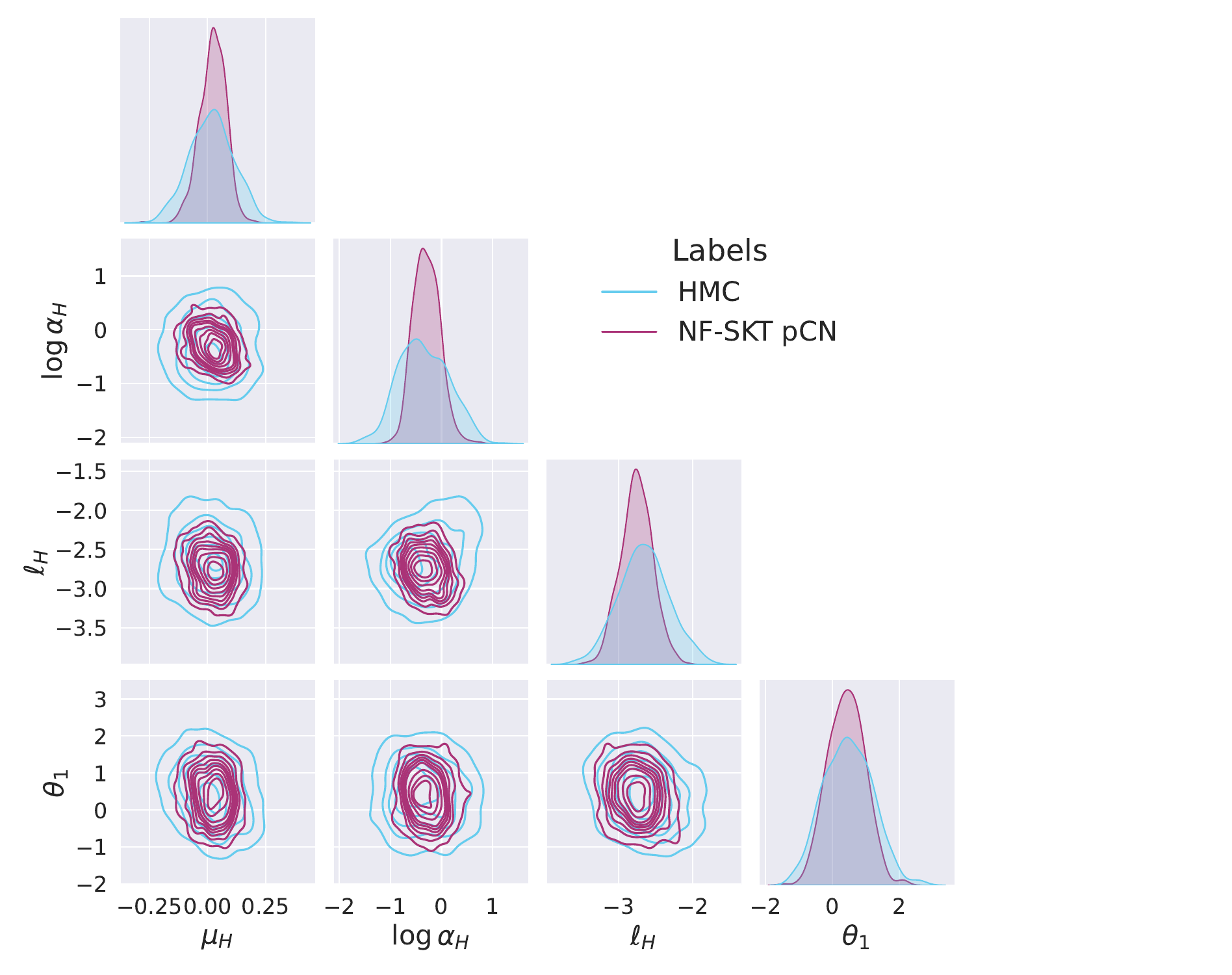}
    \includegraphics[width=0.48\textwidth]{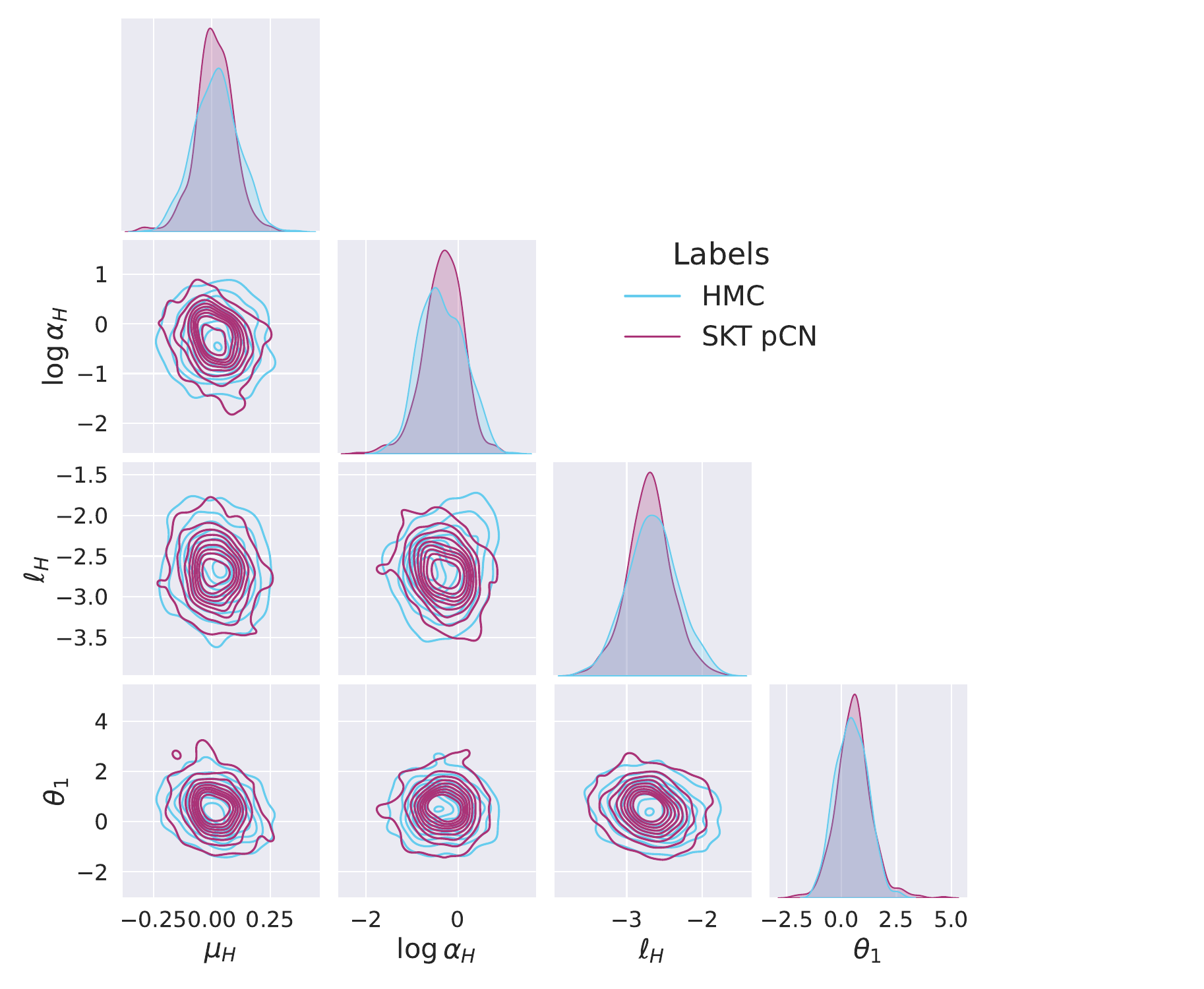}
    \caption{Corner plots showing the final particle ensemble for the first 4 dimensions, obtained using NF-SKT and SKT adaptation applied to the tpCN and pCN samplers, plotted alongside reference HMC samples for the reaction-diffusion example. The ensemble size was $J=10d$, where $d=53$.}
    \label{fig:rd nf-SKT SKT corner}
\end{figure}

\begin{figure}[ht]
    \centering
    \includegraphics[width=0.48\textwidth]{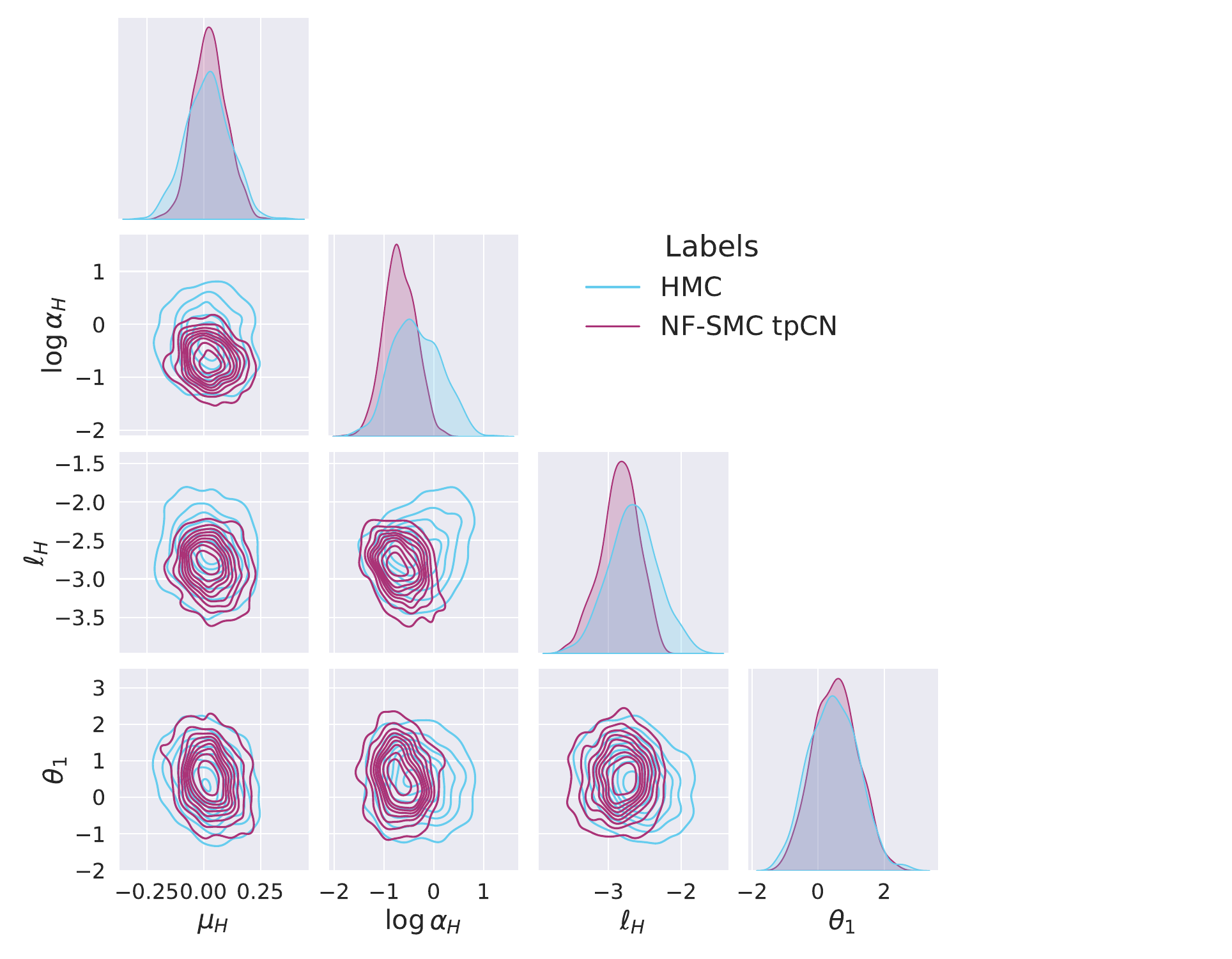}
    \includegraphics[width=0.48\textwidth]{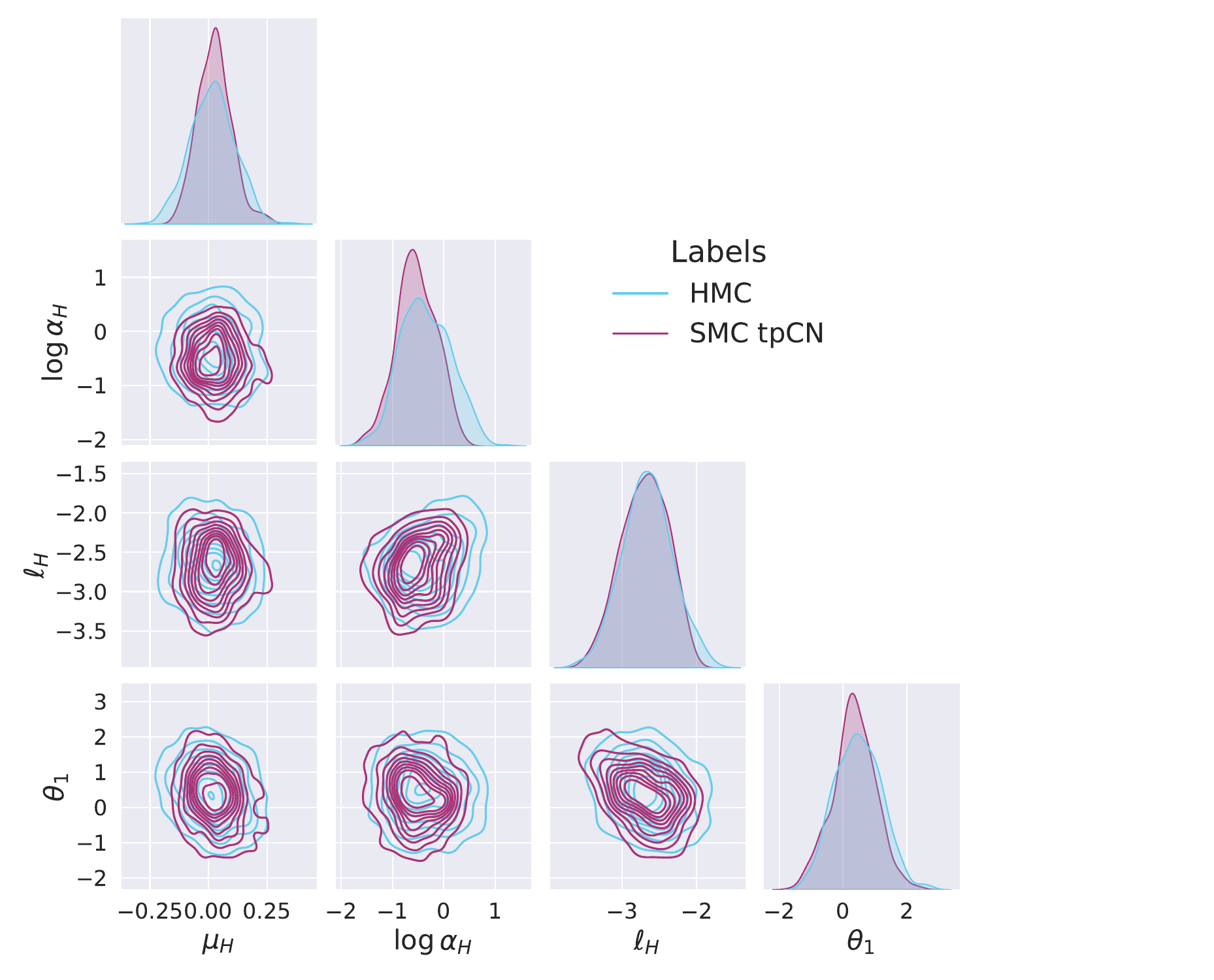}
    \includegraphics[width=0.48\textwidth]{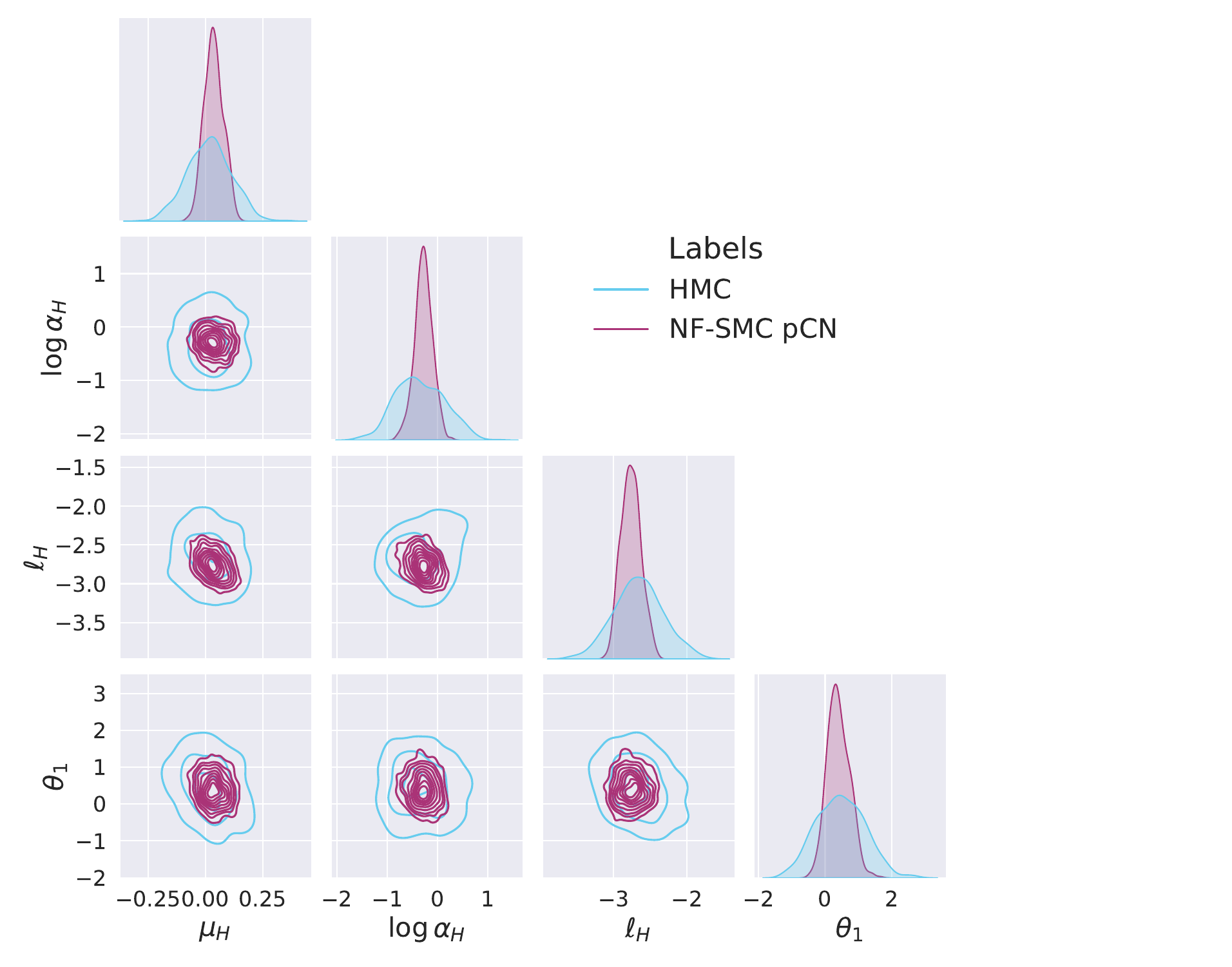}
    \includegraphics[width=0.48\textwidth]{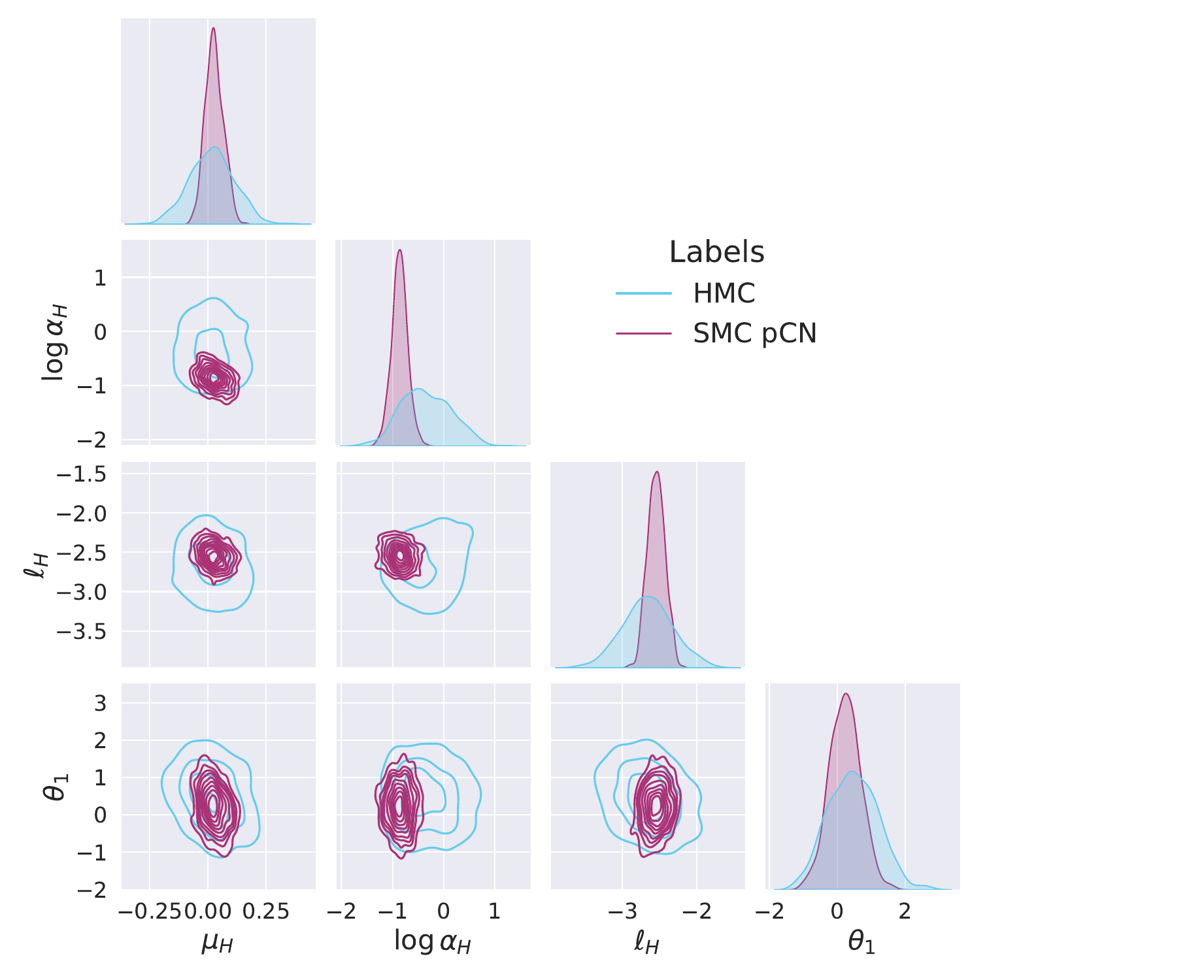}
    \caption{Corner plots showing the final particle ensemble for the first 4 dimensions, obtained using NF-SMC and SMC adaptation applied to the tpCN and pCN samplers, plotted alongside reference HMC samples for the reaction-diffusion example. The ensemble size was $J=10d$, where $d=53$.}
    \label{fig:rd nf-smc smc corner}
\end{figure}

In this section we provide corner plots showing the final particle ensembles over the first 4 dimensions, obtained by running each adaptation method on the tpCN and pCN samplers, plotted alongside reference HMC samples. All the ensembles shown in this section were obtained using an ensemble size of $J=10d$ and selecting the number of sampling iterations adaptively using an autocorrelation threshold of $\tau_\mathrm{corr}=0.1$. The figures included in this section are as follows:
\begin{itemize}
    \item Figure \ref{fig:heat eqn nf-SKT SKT corner}: Corner plot showing the final ensembles for the heat equation example, using NF-SKT and SKT adaptation applied to the tpCN and pCN samplers.
    \item Figure \ref{fig:heat eqn nf-smc smc corner}: Corner plot showing the final ensembles for the heat equation example, using NF-SMC and SMC adaptation applied to the tpCN and pCN samplers.
    \item Figure \ref{fig:gravity nf-SKT SKT corner}: Corner plot showing the final ensembles for the gravity survey example, using NF-SKT and SKT adaptation applied to the tpCN and pCN samplers.
    \item Figure \ref{fig:gravity nf-smc smc corner}: Corner plot showing the final ensembles for the gravity survey example, using NF-SMC and SMC adaptation applied to the tpCN and pCN samplers.
    \item Figure \ref{fig:rd nf-SKT SKT corner}: Corner plot showing the final ensembles for the reaction-diffusion example, using NF-SKT and SKT adaptation applied to the tpCN and pCN samplers.
    \item Figure \ref{fig:rd nf-smc smc corner}: Corner plot showing the final ensembles for the reaction-diffusion example, using NF-SMC and SMC adaptation applied to the tpCN and pCN samplers.
\end{itemize}

\section{Flow Annealed Kalman Inversion and Ensemble Kalman Inversion Ablation}\label{subsec: faki and eki}

\begin{figure}[ht]
    \centering
    \includegraphics[width=0.48\textwidth]{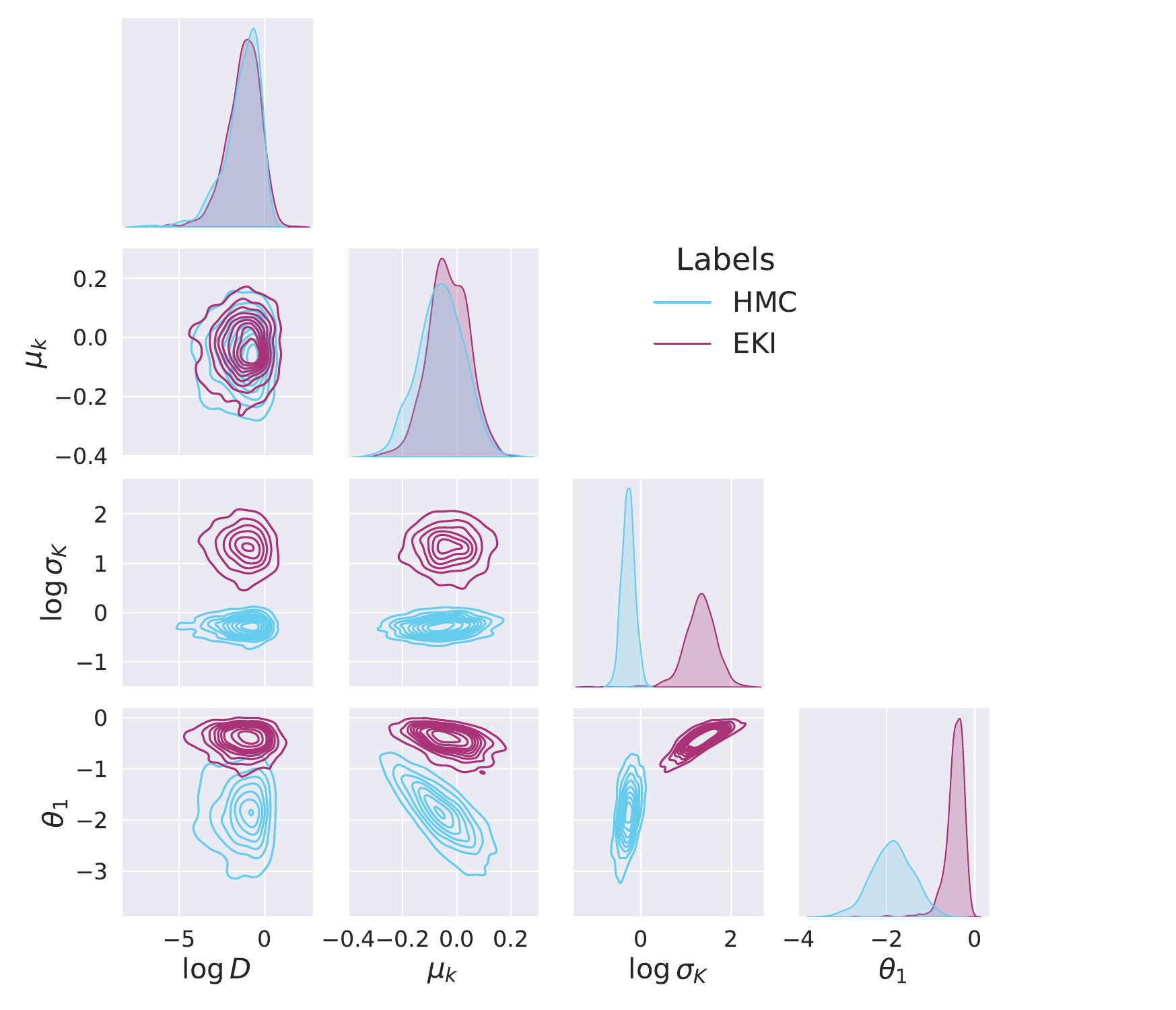}
    \includegraphics[width=0.48\textwidth]{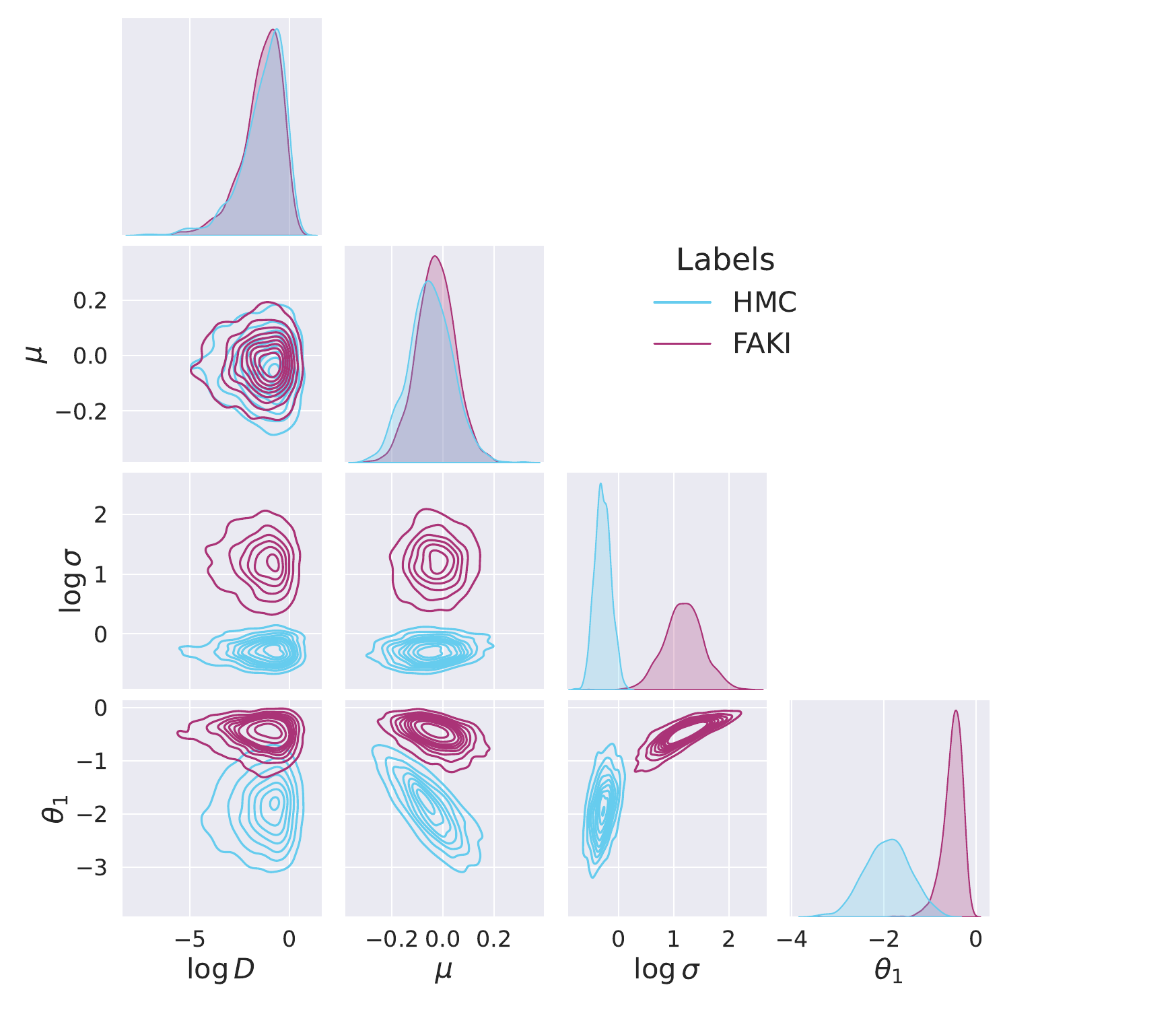}
    \caption{Corner plots showing the final particle ensemble for the first 4 dimensions, obtained using EKI (left panel) and FAKI (right panel), plotted alongside reference HMC samples for the heat equation example. The ensemble size was $J=10d$, where $d=103$.}
    \label{fig:heat eqn faki eki}
\end{figure}

\begin{figure}[ht]
    \centering
    \includegraphics[width=0.48\textwidth]{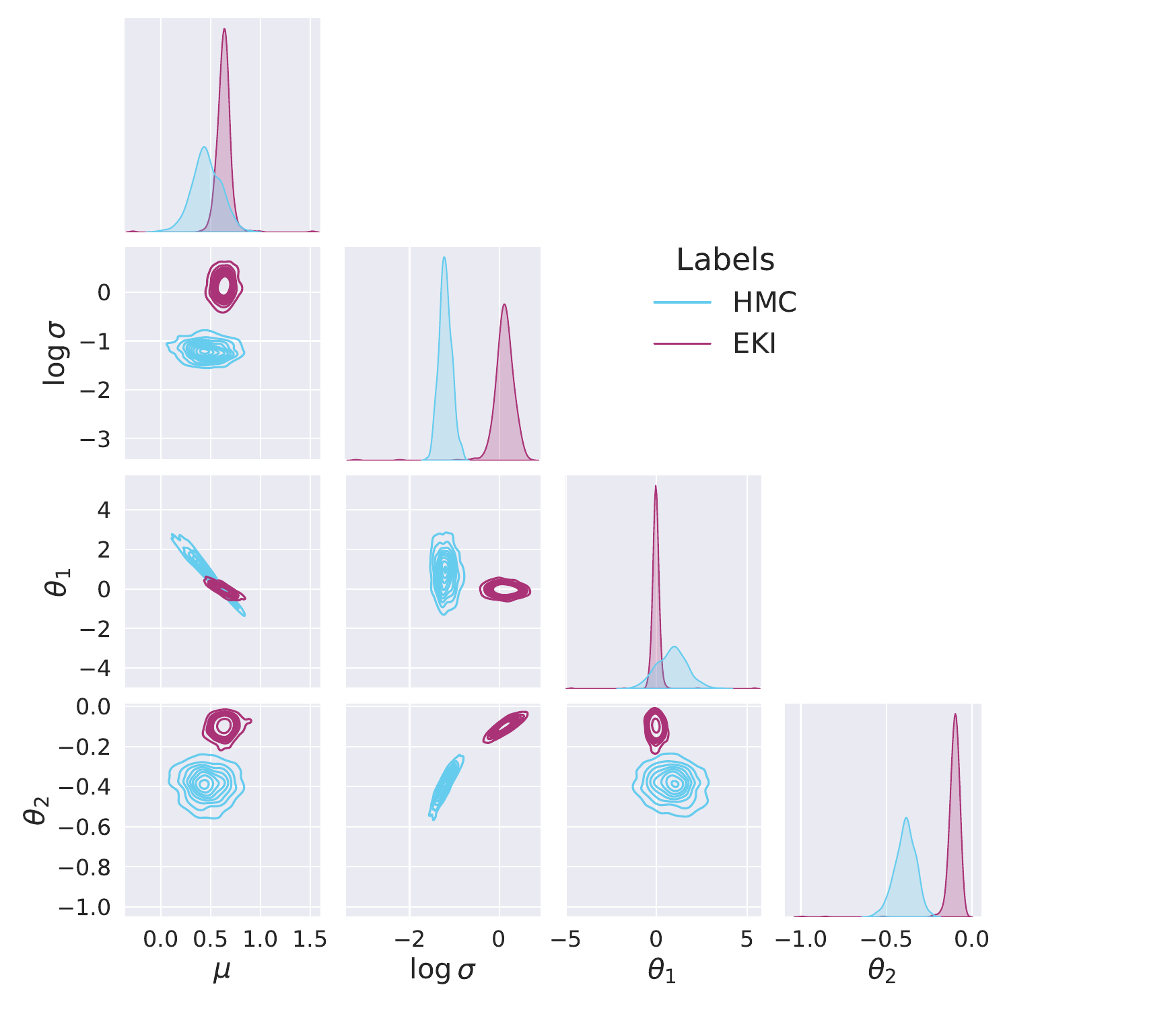}
    \includegraphics[width=0.48\textwidth]{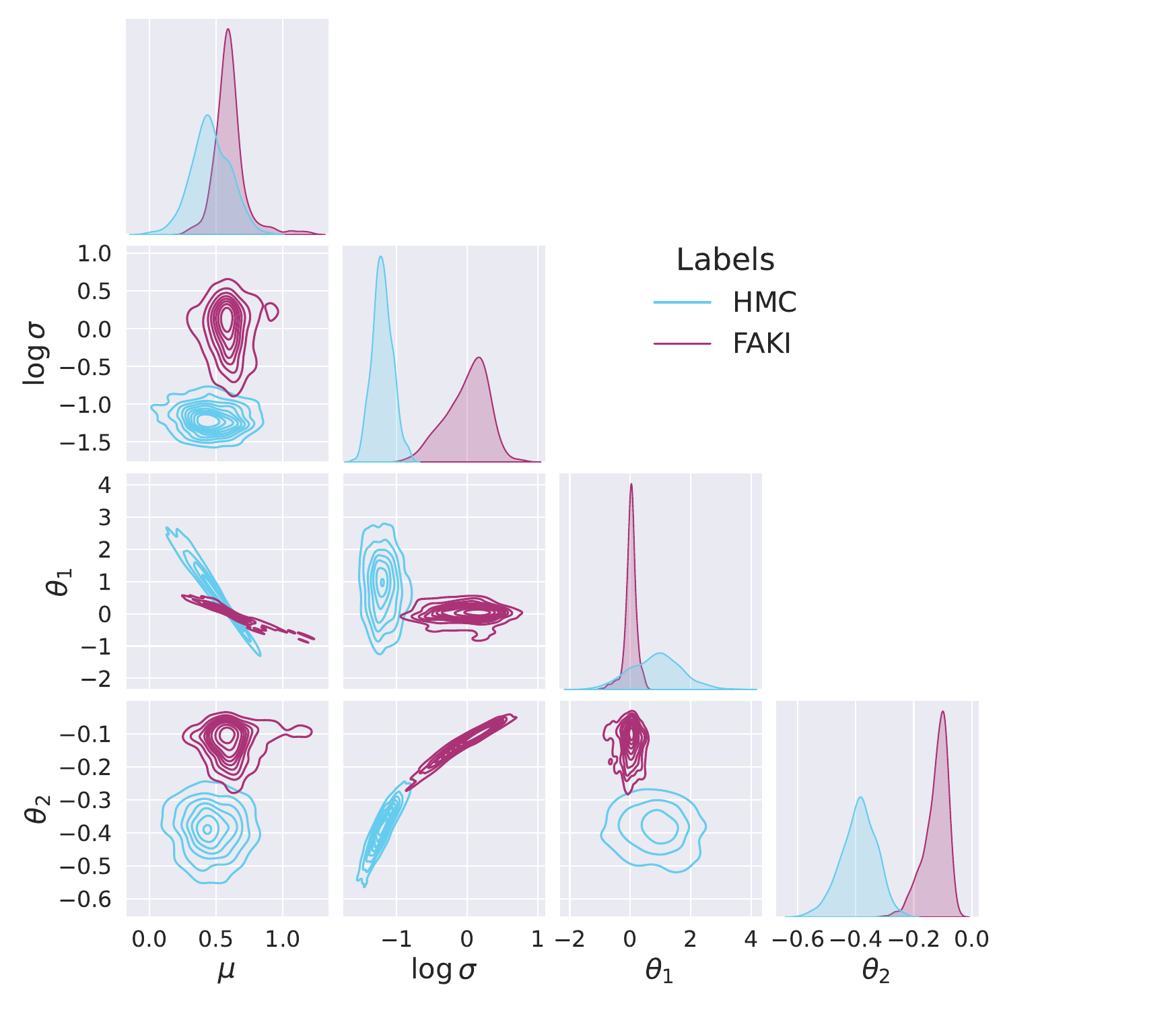}
    \caption{Corner plots showing the final particle ensemble for the first 4 dimensions, obtained using EKI (left panel) and FAKI (right panel), plotted alongside reference HMC samples for the gravity survey example. The ensemble size was $J=10d$, where $d=62$.}
    \label{fig:gravity faki eki}
\end{figure}

\begin{figure}[ht]
    \centering
    \includegraphics[width=0.48\textwidth]{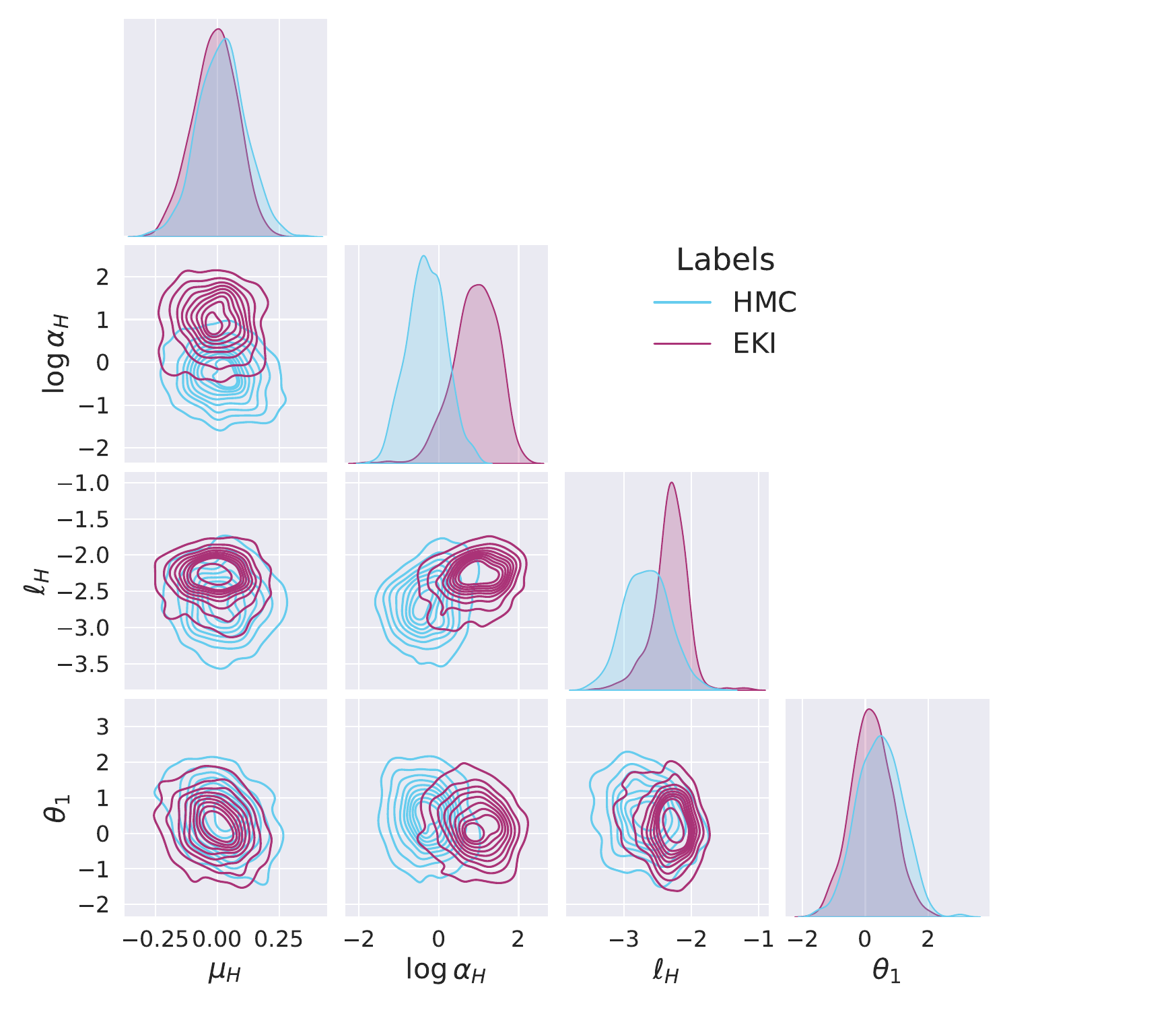}
    \includegraphics[width=0.48\textwidth]{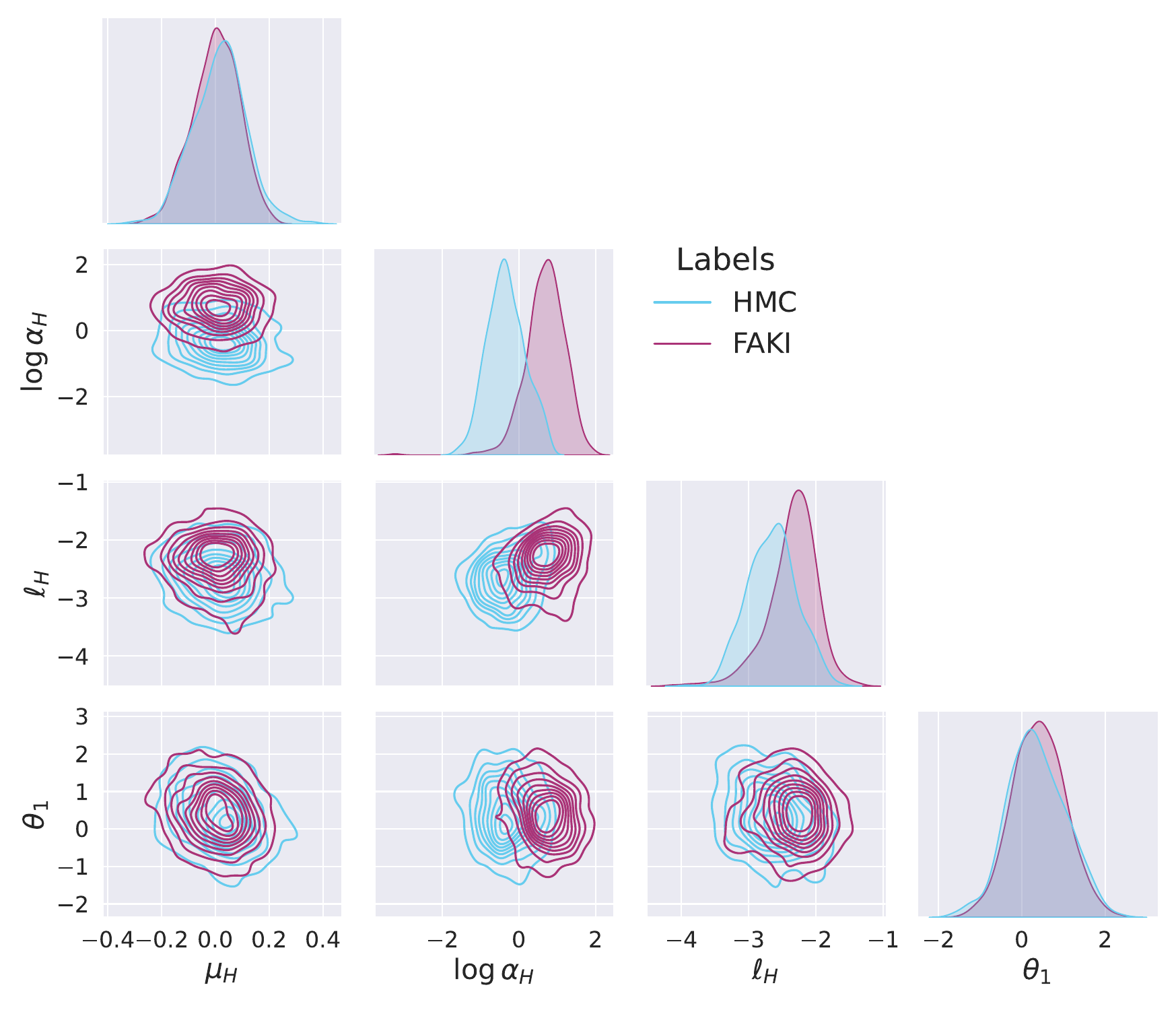}
    \caption{Corner plots showing the final particle ensemble for the first 4 dimensions, obtained using EKI (left panel) and FAKI (right panel), plotted alongside reference HMC samples for the reaction-diffusion example. The ensemble size was $J=10d$, where $d=53$.}
    \label{fig:reaction diffusion eki}
\end{figure}

In Figures \ref{fig:heat eqn faki eki}, \ref{fig:gravity faki eki} and \ref{fig:reaction diffusion eki} we show corner plots of the recovered particle distributions from running EKI and FAKI on the heat equation, gravity survey and reaction-diffusion examples respectively. For all plots we also show the sample distributions from our reference HMC samples. We use an ensemble size of $J=10d$ throughout for EKI and FAKI, set $\tau=0.5$ for temperature level adaptation, and show the particle distributions over the first 4 dimensions for illustrative purposes. In Table \ref{tab:eki faki summary} we state the average number of iterations and bias-squared results for EKI and FAKI over 10 runs for each of the numerical examples.

From these corner plots, we can immediately see that the particle distributions from EKI and FAKI are strongly offset from the reference HMC sample distributions, which manifests in the high bias results reported in Table \ref{tab:eki faki summary}. For our numerical examples, we break the core assumptions underlying EKI. This means that in moving from the prior to the first annealed target, the updated ensemble will not be correctly distributed according to the annealed target. When updating the particles for the next temperature level, we do not have the correct effective prior ensemble, meaning these errors will accumulate as we move from the prior to the posterior. For FAKI, we still have these problems, given that the NF maps do not address any errors arising due to nonlinearity. If the particle ensemble is not correctly distributed at a given temperature level, the NF will not Gaussianize the correct effective prior, meaning we lose the additional benefits from mapping the particle ensemble to a Gaussian latent space at each iteration. These results all demonstrate the importance of using the sampling iterations in SMC to correct the EKI and FAKI updates in order to obtain reliable estimates for posterior moments. 

\begin{table}
    \centering
    \begin{tabular}{c|c|c|c}
        \hhline{====}
         Experiment&  $N_\beta$&  $\langle b_1^2\rangle$& $\langle b_2^2\rangle$\\
         \hhline{=|=|=|=}
         Heat Equation FAKI&  $26.2\pm0.6$&  $1.72\pm 0.12$& $3.25\pm0.64$\\
         Heat Equation EKI&  $37.4\pm 1.5$&  $2.07\pm0.13$& $5.36\pm 1.01$\\
         \hhline{=|=|=|=}
         Gravity Survey FAKI&  $29.9\pm 0.7$&  $3.02\pm0.32$& $1.12\pm 0.03$\\
         Gravity Survey EKI&  $54.9\pm 8.5$&  $3.23\pm0.26$& $1.13\pm0.03$\\
         \hhline{=|=|=|=}
         Reaction-Diffusion FAKI&  $35.3\pm 3.2$&  $0.17\pm0.02$& $0.07\pm0.01$\\
         Reaction-Diffusion EKI&  $34.1\pm 4.4$&  $0.21\pm0.02$& $0.13\pm0.03$\\
         \hhline{====}
    \end{tabular}
    \caption{Number of iterations ($N_\beta$) and bias-squared results for the final ensembles obtained using FAKI and EKI on the heat equation, gravity survery and reaction-diffusion examples. The ensemble size was $J=10d$, where $d=103$ for the heat equation example, $d=62$ for the gravity survey example and $d=53$ for the reaction-diffusion example.}
    \label{tab:eki faki summary}
\end{table}

\section{Ensemble Kalman Sampler}\label{sec: eks}

The idea of exploiting the EKI ensemble structure for preconditioning in sampling has previously been used in the Ensemble Kalman Sampler (EKS) \cite{garbuno2020interacting}. Assuming that we have a Gaussian prior such that $\bi{x}\sim\mathcal{N}(0, \Gamma_0)$, EKS iterates over ensemble updates for the particle $j$ given by,
\begin{multline}
    \hat{\bi{x}}_{n+1}^j=\bi{x}_n^j-\frac{\Delta t_n}{J}\sum_{k=1}^J \left(\mathcal{F}(\bi{x}^k_n)-\bar{\mathcal{F}}_n\right)^\intercal\Gamma^{-1}\left(\mathcal{F}(\bi{x}^j_n)-\bi{y}\right)\bi{x}^k_n+\frac{d+1}{J}\left(\bi{x}_n^j-\bar{\bi{x}}_n\right)\\-\Delta t_n C_n^{\bi{x},\bi{x}}\Gamma_0^{-1}\hat{\bi{x}}_n^j,
    \label{eqn: eks update 1}
\end{multline}
\begin{equation}
    \bi{x}^j_{n+1}=\hat{\bi{x}}_{n}^j+\sqrt{2\Delta t_nC_n^{\bi{x},\bi{x}}}\bi{\xi}_n^j,
    \label{eqn: eks update 2}
\end{equation}
where $\bar{\mathcal{F}}_n=J^{-1}\sum_{k=1}^J\mathcal{F}(\bi{x}_n^k)$, $\bar{\bi{x}}_n=J^{-1}\sum_{k=1}^J\bi{x}_n^k$, $\bi{\xi}_n^j\sim\mathcal{N}(0,I_{d})$ and
\begin{equation}
    C_n^{\bi{x},\bi{x}}=\frac{1}{J}\sum_{k=1}^{J}\left(\bi{x}_n^k-\bar{\bi{x}}_n\right)\otimes\left(\bi{x}_n^k-\bar{\bi{x}}_n\right).
\end{equation}
Following \cite{garbuno2020interacting} the time step may be chosen adaptively such that
\begin{equation}
    \Delta t_n = \frac{\Delta t_0}{\norm{D_n}_F+\epsilon},
\end{equation}
where $\norm{\cdot}_F$ denotes the Frobenius norm, $\Delta t_0=1$ and $\epsilon=10^{-5}$. The matrix $D_n\in\mathbb{R}^{J\times J}$ is defined as
\begin{equation}
    D_n=\frac{1}{J}(F_n-\bar{F}_n)\Gamma^{-1}(F_n-Y)^\intercal,
\end{equation}
where $F_n\in\mathbb{R}^{J\times n_y}$ is a matrix where row $k$ contains the vector $\mathcal{F}(\bi{x}_n^k)$, $\bar{F}_n\in\mathbb{R}^{J\times n_y}$ is a matrix where every row contains the vector $\bar{\mathcal{F}}_n$, and $Y\in\mathbb{R}^{J\times n_y}$ is a matrix where every row contains the data vector $\bi{y}$.

The EKS ensemble will converge to an approximation of the posterior, which is only exact for linear forward models. In contrast to EKI, the prior is explicitly accounted for in the update equations, and noise is added in parameter space as opposed to data space. We ran EKS on each of our heat equation, gravity survey and reaction-diffusion examples. Given the EKS updates assume a Gaussian prior, we began by fitting an NF to the prior samples before performing EKS updates in the NF latent space. The EKS sampler was run with an ensemble size of $J=100d$ for each problem. This very large ensemble size was required to ensure the numerical stability of the EKS updates over a large number of iterations. Using an ensemble size of e.g., $J=10d$ resulted in serious numerical instabilities that meant the update procedure failed within $\sim 10$ iterations.

\begin{figure}
     \centering
     \begin{subfigure}[b]{0.48\textwidth}
         \centering
         \includegraphics[width=\textwidth]{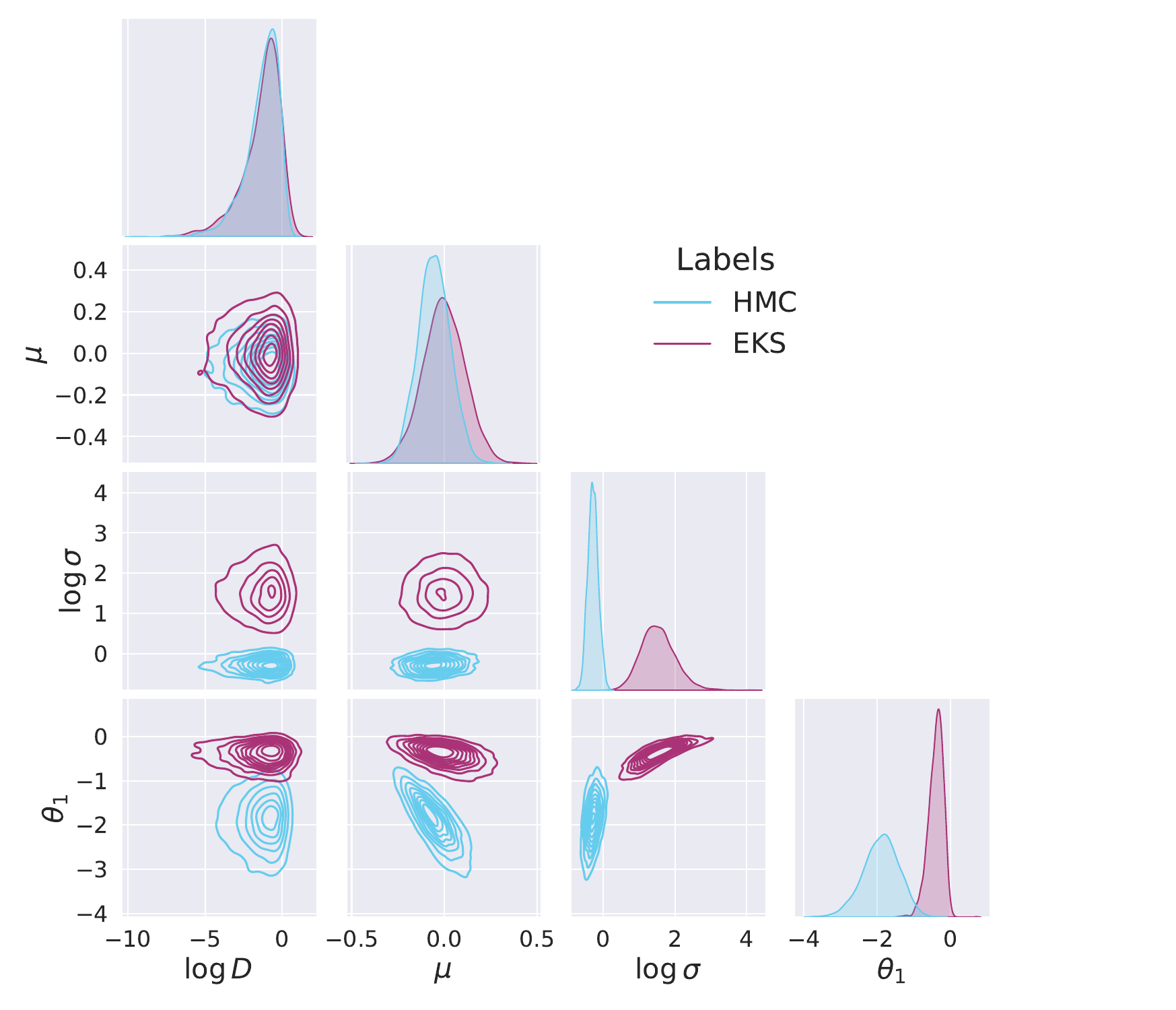}
         \caption{Heat Equation}
         \label{fig:heat eqn eks}
     \end{subfigure}
     \hfill
     \begin{subfigure}[b]{0.48\textwidth}
         \centering
         \includegraphics[width=\textwidth]{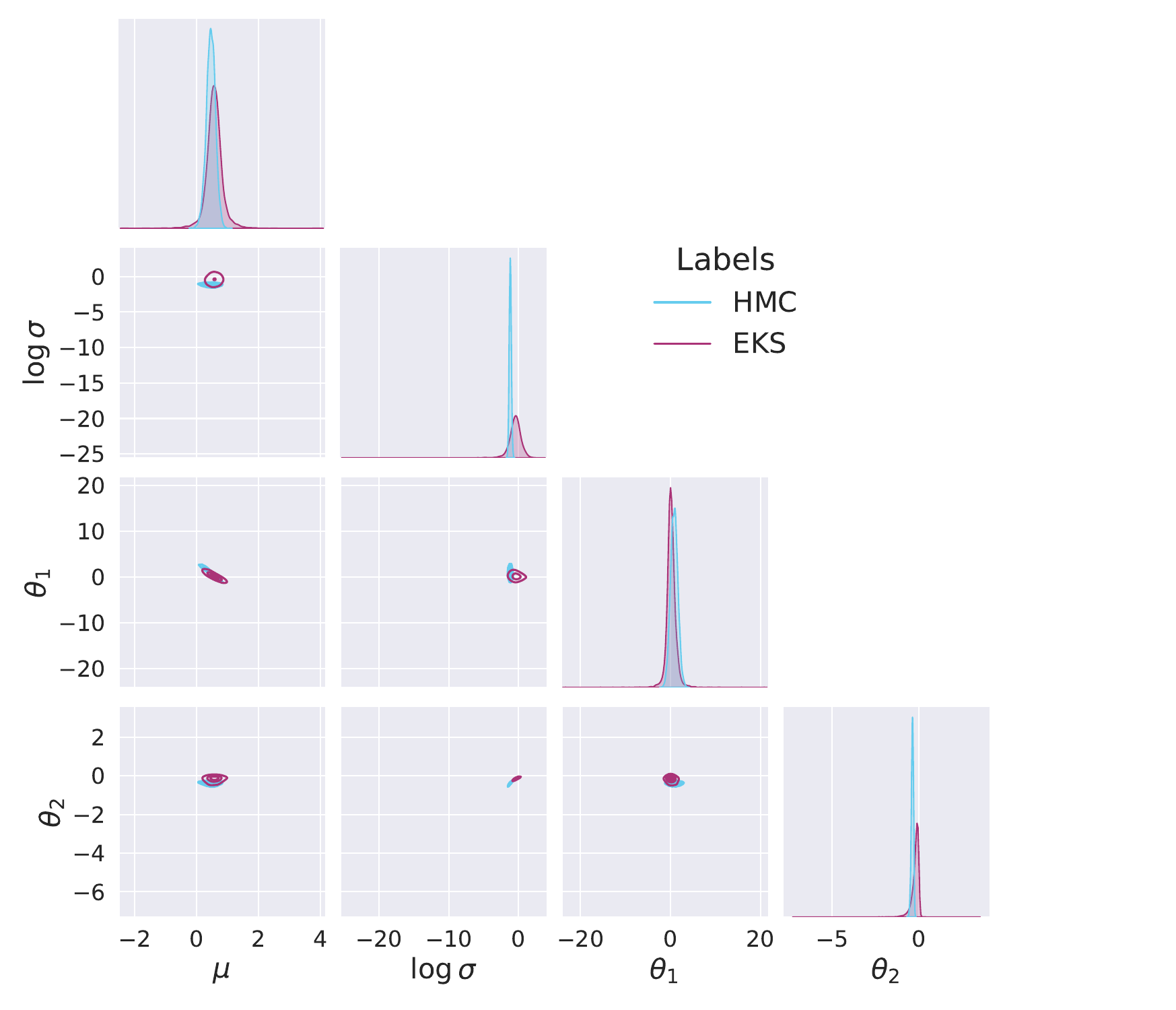}
         \caption{Gravity Survey}
         \label{fig:gravity eks}
     \end{subfigure}
     \hfill
     \begin{subfigure}[b]{0.48\textwidth}
         \centering
         \includegraphics[width=\textwidth]{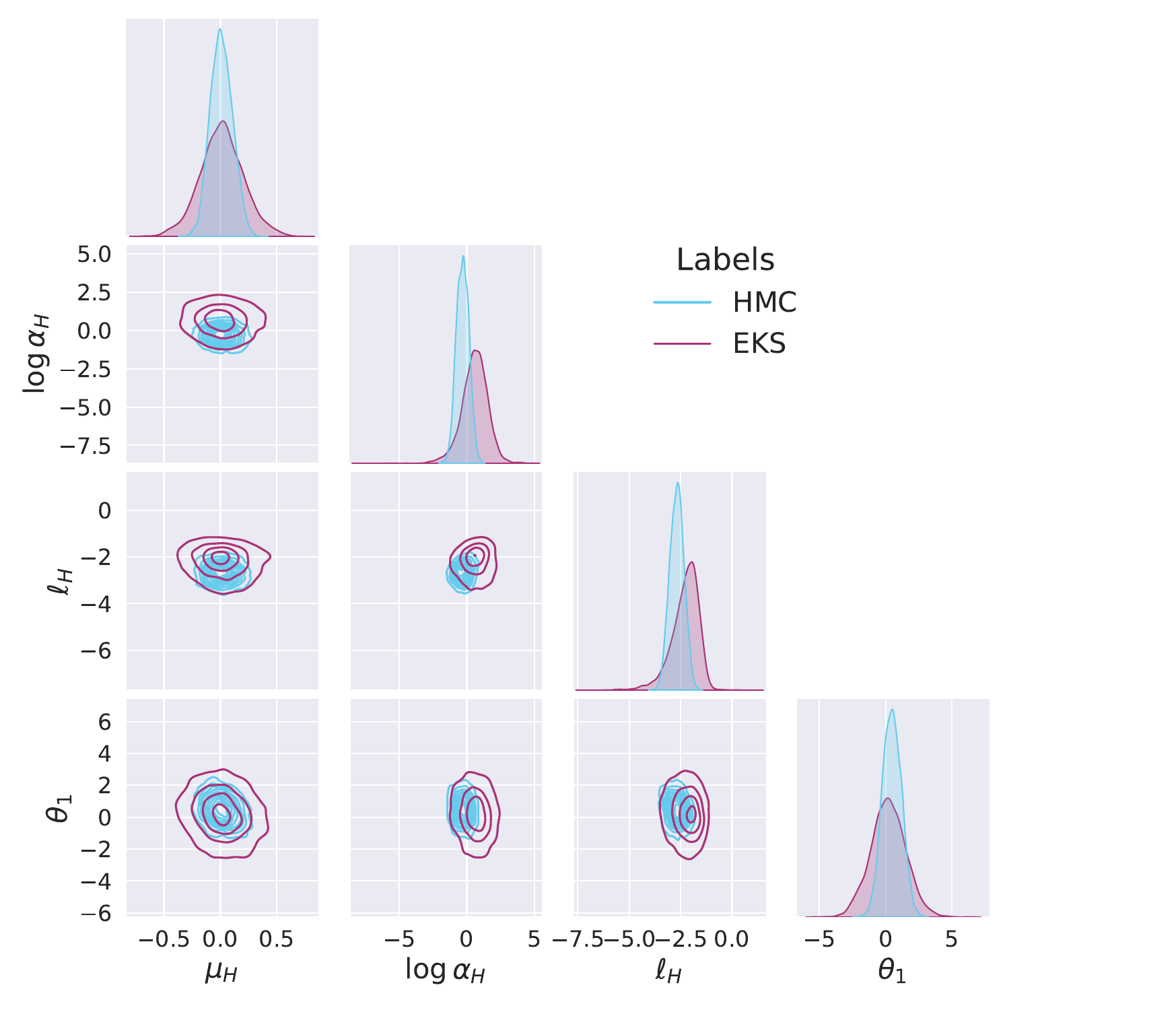}
         \caption{Reaction-Diffusion}
         \label{fig:rd eks}
     \end{subfigure}
        \caption{Corner plots showing the particle distributions over the first four dimensions obtained after 100 iterations of EKS, shown alongside reference samples from long runs of HMC. For all examples we use an ensemble size of $J=100d$, where $d=103$ for the heat equation example, $d=62$ for the gravity survey example, and $d=53$ for the reaction-diffusion example.}
        \label{fig:eks corner}
\end{figure}

In Figure \ref{fig:eks corner} we show corner plots comparing the ensemble distributions obtained with EKS after 100 iterations with reference samples obtained using HMC. For all three experiments EKS converges on a highly biased approximation to the posterior. In contrast, the tpCN sampler is able to preserve the exact target as its invariant measure, and when run with the SKT scheme is able to rapidly converge on low bias estimates of posterior moments.

\section*{References}
\bibliography{main}

\end{document}